%% file: sadeghi_vontobel_shams_trans_it_2007_subm1.tex
\def\statusstring{Submitted to IEEE Transactions on Information Theory, November 24, 2007.}

\documentclass[11pt,onecolumn]{latex_styles/IEEEtran} 

\usepackage{amsmath}
\usepackage{amscd}
\usepackage{epsfig}
\usepackage{amssymb}
\usepackage{theorem}
\usepackage{bbm}
\usepackage{amscd}
\usepackage[english]{babel}
\usepackage{amssymb}
\usepackage{cite}
\usepackage{dsfont}
\usepackage{./latex_styles/mydefs}

\renewcommand{\baselinestretch}{1.5}

\input{abbreviations1}

\renewcommand{\baselinestretch}{1.3}

\begin{document}

\input{title1}

\input{abstract1}

\input{introduction1}

\input{channel_models1}

\input{information_rates1}

\input{optimization_methods1}

\input{optimize_upper_bound1}

\input{optimize_diff_bound1}

\input{optimize_lower_bound1}

\input{results_partial1}

\input{results_fading1}

\input{conclusions1}

\input{appendices}

\input{acknowledgments1}

\end{document}

%% file: abbreviations1.tex
\newcommand{\suml}{\sum\limits}

\newcommand{\dint}[1]{{\operatorname{d}\!{#1}}}

\renewcommand{\proof}{\noindent \emph{Proof:} }
\newcommand{\proofend}{\hfill $\square$\\}

\newcommand{\vect}[1]{{\mathbf{#1}}}

\newcommand{\set}[1]{{\mathcal{#1}}}

\newcommand{\setB}{\set{B}}
\newcommand{\setBhat}{\set{\Bhat}}
\newcommand{\setShat}{\set{\Shat}}

\newcommand{\setS}{\set{S}}
\newcommand{\setX}{\set{X}}

\newcommand{\setY}{\set{Y}}

\newcommand{\defeq}{\triangleq}

\newcommand{\vS}{{\vect{S}}}
\newcommand{\vX}{{\vect{X}}}
\newcommand{\vY}{{\vect{Y}}}

\newcommand{\vb}{\vect{b}}
\newcommand{\vbhat}{\vect{\hat b}}

\newcommand{\vs}{{\vect{s}}}
\newcommand{\vshat}{{\vect{\shat}}}

\newcommand{\vx}{{\vect{x}}}

\newcommand{\vy}{{\vect{y}}}

\newcommand{\vyD}{{\vect{y}^{\underline{D}}}}

\newcommand{\diff}[1]{\frac{\partial}{\partial #1}}
\newcommand{\ddiff}[1]{\frac{\partial^2}{\partial {#1}^2}}
\newcommand{\fddiff}[2]{\frac{\partial^2}{\partial #1 \partial #2}}

\newcommand{\Phat}{\hat P}

\newcommand{\Ptilde}{\tilde P}
\newcommand{\Vtilde}{\tilde V}
\newcommand{\vtilde}{\tilde v}
\newcommand{\Vhat}{\hat V}
\newcommand{\vhat}{\hat v}
\newcommand{\What}{\hat W}
\newcommand{\Wtilde}{\tilde W}

\newcommand{\Ttilde}{\tilde T}

\newcommand{\Bhat}{\hat B}
\newcommand{\bhat}{\hat b}
\newcommand{\Shat}{\hat S}
\newcommand{\shat}{\hat s}
\newcommand{\shatL}{\hat s_{\mathrm p}}

\newcommand{\Iupper}{\overline{I}}
\newcommand{\Ilower}{\underline{I}}
\newcommand{\Iuldiff}{\Delta}
\newcommand{\Psiupper}{\overline{\Psi}}
\newcommand{\Psilower}{\underline{\Psi}}
\newcommand{\Psidiff}{\Psi_{\Delta}}

\newcommand{\sIN}{{\set I}_N}

\newcommand{\ellsum}{\sum_{\ell \in \sIN}}

\newcommand{\sumb}{\suml_{\vb}}

\newcommand{\sumxy}{\suml_{\vx,\,\vy}}

\newcommand{\sumx}{\suml_{\vx}}
\newcommand{\sumy}{\suml_{\vy}}
\newcommand{\sumbhat}{\suml_{\vbhat}}

\newtheorem{PreLemma}{{\textbf{Lemma}}}
  \newenvironment{Lemma}[1]%
    {\begin{PreLemma}[\textbf{#1}]}{\end{PreLemma}}

\newtheorem{PreTheorem}[PreLemma]{{\textbf{Theorem}}}
    {\begin{PreTheorem}[\textbf{#1}]}{\end{PreTheorem}}

\newtheorem{PreCorollary}[PreLemma]{{\textbf{Corollary}}}
    {\begin{PreCorollary}[\textbf{#1}]}{\end{PreCorollary}}

\newtheorem{PreDefinition}[PreLemma]{{\textbf{Definition}}}%
  \newenvironment{Definition}[1]%
    {\begin{PreDefinition}[\textbf{#1}]}{\hfill$\square$\end{PreDefinition}}

\newtheorem{PreAlgorithm}[PreLemma]{{\textbf{Algorithm}}}
  \newenvironment{Algorithm}[1]%
    {\begin{PreAlgorithm}[\textbf{#1}]\upshape}{\hfill$\square$\end{PreAlgorithm}}

\newtheorem{PreRemark}[PreLemma]{{\textbf{Remark}}}
  \newenvironment{Remark}[1]%
    {\begin{PreRemark}[\textbf{#1}]\upshape}{\hfill$\square$\end{PreRemark}}

\newtheorem{PreConjecture}[PreLemma]{{\textbf{Conjecture}}}
    {\begin{PreConjecture}[\textbf{#1}]\upshape}{\hfill$\square$\end{PreConjecture}}

\newtheorem{PreAssumption}[PreLemma]{{\textbf{Assumption}}}
    {\begin{PreAssumption}[\textbf{#1}]}{\hfill$\square$\end{PreAssumption}}

\newtheorem{PreExample}[PreLemma]{{\textbf{Example}}}
  \newenvironment{Example}[1]%
    {\begin{PreExample}[\textbf{#1}]\upshape}{\hfill$\square$\end{PreExample}}

\newenvironment{Proof}%
  {\proof}{\proofend}

\def\BibTeX{{\rm B\kern-.05em{\sc i\kern-.025em b}\kern-.08em
    T\kern-.1667em\lower.7ex\hbox{E}\kern-.125emX}}

\newcounter{theoremcntr}
{\begin{trivlist}\item[]\refstepcounter{theoremcntr}%
{\bfseries Theorem~\thetheoremcntr%
  \ifthenelse{\equal{#1}{}}{}{ (#1)}.
}}%
{\hfill$\QED$\end{trivlist}}

\newcounter{rulcntr}
{\begin{trivlist}\item[]\refstepcounter{rulcntr}%
 {\bfseries Rule~\therulcntr%
  \ifthenelse{\equal{#1}{}}{}{ (#1)}.
}}%
{\hfill$\QED$\end{trivlist}}

\newcommand{\good}{{\mathrm{g}}}
\newcommand{\bad}{{\mathrm{b}}}
\newcommand{\pgood}{p_{\mathrm{g}}}
\newcommand{\pbad}{p_{\mathrm{b}}}
\newcommand{\epsgood}{\epsilon_{\mathrm{g}}}
\newcommand{\epsbad}{\epsilon_{\mathrm{b}}}

\newcommand{\Mhat}{\hat{M}}

\newcommand{\condcs}{}
\newcommand{\condvs}{}

%% file: title1.tex

\title{Optimization of Information Rate Upper and Lower Bounds for Channels
  with Memory%
  \footnotemark${}^{*}$%
  \thanks{${}^{*}$%
    \statusstring{} Preliminary results of this paper were presented at the
    IEEE International Symposium on Information Theory (ISIT), Nice, France,
    June 2007.  The work of Parastoo Sadeghi was partly supported under
    Australian Research Council's Discovery Projects funding scheme (project
    number DP0773898).}  }

\author{Parastoo Sadeghi%
        \footnotemark${}^{\dag}$%
        \thanks{${}^{\dag}$%
          Parastoo Sadeghi (contact author) and Ramtin Shams are with the
          Department of Information Engineering, Research School of
          Information Sciences and Engineering, The Australian National
          University, Canberra 0200 ACT, Australia. Emails:
          \texttt{parastoo.sadeghi@anu.edu.au};
          \texttt{ramtin.shams@anu.edu.au}},
        Pascal O.~Vontobel%
        \footnotemark${}^{\ddag}$%
        \thanks{${}^{\ddag}$%
          Pascal O.~Vontobel is with Hewlett-Packard Laboratories, Palo Alto,
          CA 94304, USA. Email: \texttt{pascal.vontobel@ieee.org}},
        and Ramtin Shams%
        \footnotemark${}^{\dag}$}

\maketitle

%% file: abstract1.tex

\begin{abstract}
  We consider the problem of minimizing upper bounds and maximizing lower
  bounds on information rates of stationary and ergodic discrete-time channels
  with memory. The channels we consider can have a finite number of states,
  such as partial response channels, or they can have an infinite state-space,
  such as time-varying fading channels.

  We optimize recently-proposed information rate bounds for such channels,
  which make use of auxiliary finite-state machine channels (FSMCs). Our main
  contribution in this paper is to provide iterative expectation-maximization
  (EM) type algorithms to optimize the parameters of the auxiliary FSMC to
  tighten these bounds. We provide an explicit, iterative algorithm that
  improves the upper bound at each iteration. We also provide an effective
  method for iteratively optimizing the lower bound. To demonstrate the
  effectiveness of our algorithms, we provide several examples of partial
  response and fading channels, where the proposed optimization techniques
  significantly tighten the initial upper and lower bounds. Finally, we
  compare our results with an improved variation of the \emph{simplex} local
  optimization algorithm, called \emph{Soblex}. This comparison shows that our
  proposed algorithms are superior to the Soblex method, both in terms of
  robustness in finding the tightest bounds and in computational efficiency.

  Interestingly, from a channel coding/decoding perspective, optimizing the
  lower bound is related to increasing the achievable mismatched information
  rate, \ie, the information rate of a communication system where the decoder
  at the receiver is matched to the auxiliary channel, and not to the original
  channel.
\end{abstract}


\newpage

%% file: introduction1.tex

\section{Introduction}
\label{sec:introduction:1}

\subsection{Motivation and Background}

Channels with memory are common in practical communication systems. The
partial response channel is an important example of a channel with memory with
applications in magnetic and optical recording, as well as in communications
over band-limited channels with inter-symbol interference
(ISI)~\cite{Proakis:00:1}. The time-varying multipath fading channel in
wireless communication systems is another example of a channel with
memory~\cite{Biglieri1998}. Although the information rate of such channels is
formulated (we assume them to be stationary and ergodic), the direct
computation of the information rate (under the assumption of no channel state
information (CSI) at the receiver or the transmitter) has remained an open
problem~\cite{Arnold:Loeliger:Vontobel:Kavcic:Zeng:06:1}.

Partial response channels can be closely modeled by finite-state machine
channels (FSMCs)~\cite{Gallager:68}. Stochastic and numerical strategies have
been proposed in the literature for efficient computation of information rates
for FSMCs~\cite{Arnold:Loeliger:01:1,
  Arnold:Loeliger:Vontobel:Kavcic:Zeng:06:1, Sharma:Singh:01:1,
  Pfister:Soriaga:Siegel:01:1}. The numerical estimate of the information rate
converges under mild conditions with probability one to the true value when
the length of the channel input and output sequences goes to infinity. Upper
and lower bounds on the capacity of FSMCs have been proposed
in~\cite{Vontobel:Arnold:01:1, Kavcic2005,
  Vontobel:Kavcic:Arnold:Loeliger:04:1:subm}, where the upper bound
in~\cite{Vontobel:Arnold:01:1} is based on Lagrange duality, the upper bound
in~\cite{Kavcic2005} is based on the FSMC feedback capacity, and the lower
bound in~\cite{Vontobel:Kavcic:Arnold:Loeliger:04:1:subm} is obtained by
numerically optimizing the parameters of the Markov input source to the
channel. From a practical viewpoint, information rates, the capacity, and the
capacity-achieving input distribution of FSMCs with not too many states is
numerically computable.  However, for more complex partial response channels
with longer memories, the large number of states in the FSMC prohibits
efficient computation of information rates.  Physical multipath fading
channels are channels with an infinite (continuous-valued) state space.
Therefore, direct application of the techniques in~\cite{Arnold:Loeliger:01:1,
  Arnold:Loeliger:Vontobel:Kavcic:Zeng:06:1, Sharma:Singh:01:1,
  Pfister:Soriaga:Siegel:01:1} for computing information rates for fading
channels is not possible.

For the case of stationary and ergodic channels with memory that are
non-finite-state or where the number of states is large, we would still like
to efficiently (stochastically) compute upper and lower bounds on the
information rate. Such upper and lower bounds were proposed
in~\cite{Arnold:Loeliger:Vontobel:02:1,
  Arnold:Loeliger:Vontobel:Kavcic:Zeng:06:1} based on the introduction of an
auxiliary FSMC. The bounds are generally applicable to finite-state and
non-finite-state channels with memory that are stationary and ergodic. The
lower bound in~\cite{Arnold:Loeliger:Vontobel:02:1,
  Arnold:Loeliger:Vontobel:Kavcic:Zeng:06:1} is a special case of the
generalized mutual information (GMI) lower bound for mismatched
decoding~\cite{Ganti:Lapidoth:Telatar:00:1}. In other words, the lower bound
signifies achievable information rates when the receiver is equipped with a
decoding algorithm which is matched to the auxiliary channel model and hence,
usually mismatched to the original channel (over which the actual
communication takes place).

The maximum number of states and branches in the auxiliary FSMC model is
dictated by the computational complexity of running the
Bahl-Cocke-Jelinek-Raviv (BCJR) algorithm~\cite{BCJR1974} or the mismatched
decoding budget. The number of auxiliary FSMC states and branches, along with
the FSMC state transition and output probabilities, affect the tightness of
the bounds. Therefore, for a given auxiliary FSMC trellis section, it is
desirable to optimize the remaining auxiliary FSMC parameters to obtain the
tightest information rate upper and lower bounds, which is the topic of this
paper.

\subsection{Contributions and Organization}

In this paper, we optimize the parameters of the auxiliary FSMC model in order
to tighten the information rate upper and lower bounds that were introduced
in~\cite{Arnold:Loeliger:Vontobel:02:1,
  Arnold:Loeliger:Vontobel:Kavcic:Zeng:06:1} for channels with memory.
Actually, the lower bounds that we use in the present paper are slightly more
general than the lower bounds in~\cite{Arnold:Loeliger:Vontobel:02:1,
  Arnold:Loeliger:Vontobel:Kavcic:Zeng:06:1}. They are also more general than
the lower bounds that were optimized in the preliminary
version~\cite{Sadeghi:Vontobel:Shams:07:1} of this paper.

The original and auxiliary channels are reviewed in
\secref{sec:channel:models:1} and the information rate bounds under
consideration are reviewed in \secref{sec:information:rate:bound:1}. In our
approach, we assume that we are given a fixed trellis section of an auxiliary
FSMC and that we optimize its remaining parameters (the state transition and
output probabilities). The general optimization idea is shown in
\secref{sec:optimization:methods1}.  Briefly speaking, we replace the
optimization of a function by a succession of surrogate-function
optimizations, an approach that can be seen as as a variation of the
expectation-maximization (EM) algorithm~\cite{dempster:laird:rubin:1977}.

Five main contributions of the paper are summarized as follows.

\begin{enumerate}

\item In \secref{sec:optimize:upper:bound:1}, we propose an iterative
  procedure for the minimization of the upper bound. For this purpose, we
  devise an easily optimizable surrogate function. We establish that this
  surrogate function is never below the original upper bound and that by
  minimizing it, we ensure non-increasing upper bounds in each iteration.

\item In \secref{sec:optimize:diff:bound:1}, we propose a similar iterative
  procedure for the minimization of the difference between the upper bound and
  a specialized version of the lower bound. The parameters of the auxiliary
  FSMC that minimize this difference can be used as the initial point for the
  optimization of upper and lower bounds, resulting in quicker convergence or
  tighter bounds.


\item In \secref{sec:optimize:lower:bound:1}, we propose an iterative
  procedure for the maximization of the lower bound by devising an easily
  optimizable surrogate function. The important property of this surrogate
  function is that at the current point in the auxiliary FSMC parameter space, the
  function value and its gradient agree with the lower bound function value
  and its gradient, respectively. However, we were not able to establish
  analytically that this surrogate function is never above the lower bound,
  which makes the approach rather heuristic. Note though that our surrogate
  function of choice will have a parameter which enables one to control the
  ``aggressiveness'' of the optimization step. Adaptively setting this
  parameter allows one to have a non-decreasing lower bound after every step.

\item In \secref{sec:num:partial:response}, we apply our optimization
  techniques to several partial response channels and observe that they result
  in noticeably tighter upper and lower bounds. We analyze the convergence
  properties and the numerical tolerance of the proposed algorithms. Moreover,
  we compare our optimization results with those obtained using an improved
  version of the \emph{simplex} local optimization
  algorithm~\cite{press:flannery:teukolsky:vetterling:numerical:Recipes:in:c}.
  The improved simplex method is called \emph{Soblex} and was recently
  proposed in~\cite{shams:kennedy:sadeghi:hartley:iccv:07}. This comparison
  shows the superiority of our algorithms in terms of computational
  efficiency, tightness, and reliability of the optimized bounds.

\item In \secref{sec:num:fading}, we apply our optimization techniques to
  time-varying fading channels. We also propose a tight lower bound for the
  conditional entropy of the original Gauss-Markov fading channel, which is
  required for computing the upper bound. Compared to the widely used perfect
  CSI upper bound, we have obtained significantly tighter upper bounds, which
  together with the optimized lower bound, successfully bound the range of
  fading channel information rates.
\end{enumerate}

Some of the proofs have been relegated to the appendices at the end of the
paper.

\subsection{Notations}
\label{sec:notation:1}

The following general notations will be used. Other special notations will be
introduced later in the paper.
\begin{itemize}

\item Alphabet sets will be denoted by calligraphic characters, such as
  $\set{A}$.

\item Random variables will be denoted by upper-case characters (\eg $X$),
  while their realizations will be denoted by lower-case characters (\eg
  $x$).

\item Random vectors will be denoted by upper-case boldface characters
  (\eg $\vX$), while their realizations will be denoted by lower-case
  boldface characters (\eg $\vx$).

\item Sequences like $\ldots, x_{-1}, x_0, x_1, \ldots$ will be denoted by $\{
  x_{\ell} \}$. If $j \geq i$, we will use the short-hand ${\vx}_i^j$ to
  denote the vector $\left[ x_i, x_{i+1}, \ldots , x_{j-1}, x_j \right]$.

\item All logarithms are natural logarithms (base $e$); therefore, all
  entropies and mutual informations will be measured in nats. The only
  exceptions are figures and their corresponding discussions, where the
  information rates will be presented in bits per channel use (bits / channel
  use).

\item All channel input and output alphabets are assumed to be finite. Unless
  stated otherwise, we only deal with probability mass functions (pmfs) and
  conditional probability mass functions in this paper.

\item In order to avoid cluttering the summation signs, we will use the
  following conventions. Summations like $\sum_{x}$, $\sum_{y}$, $\sum_{s}$,
  and $\sum_{b}$ will implicitly mean $\sum_{x \in \setX}$, $\sum_{y \in
    \setY}$, $\sum_{s \in \setS}$, and $\sum_{b \in \setB}$, respectively.
  Summations like $\sum_{\vx}$ and $\sum_{\vy}$ will implicitly mean
  $\sum_{\vx \in \set{X}_{-N+1}^{N}}$ and $\sum_{\vy \in \set{Y}_{-N+1}^{N}}$.
  Summations like $\sum_{\vs}$ and $\sum_{\vb}$ will be over all valid channel
  state sequences and valid channel branch sequences,
  respectively.\footnote{The notion of ``valid channel state sequences'' and
    ``valid channel branch sequences'' will be introduced later on.}
\end{itemize}

%% file: channel_models1.tex

\section{Source and Channel Models}
\label{sec:channel:models:1}

Before presenting source and channel definitions, it is useful to define the
following index set.

\begin{figure}
  \begin{center}
    \epsfig{file=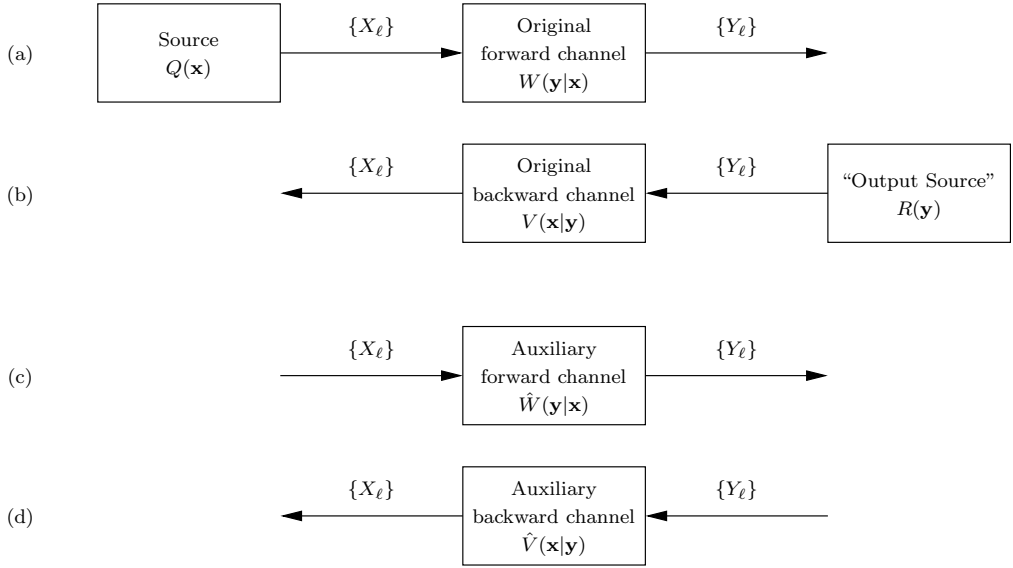,
            width=0.75\linewidth}
  \end{center}
  \caption{Block diagrams of (a) the source and the original (forward)
    channel, (b) the ``output source'' and the original backward channel, (c) the
    auxiliary forward channel, and (d) the auxiliary backward channel under
    consideration.}
  \label{fig:source:channel:prob:models:1}
\end{figure}

\begin{Definition}{The Index Set}
  \label{def:notation:indices:1}

  We assume $N$ to be a positive integer. We define the following index set
  \begin{alignat*}{2}
    \sIN
      &\defeq
         [-N+1,N]
     &&= \{ -N+1, \ldots, N \}.
  \end{alignat*}
  Observe that the size of the set is $|{\sIN}| = 2N$. Note that in our
  results, we are mainly interested in the limit $N \to \infty$
\end{Definition}

Consider the block diagram in
Fig.~\ref{fig:source:channel:prob:models:1}(a) which shows a source
and a channel, in the following also called the original forward
channel. We assume that both the source and the channel are fixed
and that we would like to find tight upper and lower bounds on the
information rate of this source/channel pair. Our upper bounding
technique will use a so-called auxiliary forward channel (see
Fig.~\ref{fig:source:channel:prob:models:1}(c)) whose input/output
alphabets match the input/output alphabets of the original forward
channel. The tightening of the upper bound will be achieved by
optimizing the parameters of the auxiliary forward channel.
Similarly, our lower bounding technique will use a so-called
auxiliary backward channel (see
Fig.~\ref{fig:source:channel:prob:models:1}(d)) whose input/output
alphabets match the input/output alphabets of the original backward
channel (see Fig.~\ref{fig:source:channel:prob:models:1}(b)) that is
associated to the original forward channel. The tightening of the
lower bound will be achieved by optimizing the parameters of the
auxiliary backward channel.

\subsection{\textbf{Source Model}}
\label{sec:source:model:1}

\begin{Definition}{Source}
  \label{def:source:model:1}

  In this paper we only consider sources that are discrete-time, stationary,
  and ergodic and that produce a sequence $\ldots, X_{-1}, X_0, X_1, \ldots$,
  where $X_{\ell} \in \setX$ for all $\ell \in {\mathbb
    Z}$. We assume that the alphabet
  $\setX$ is finite and that for any $i \leq j$ the probability of observing
  $\vx_{i}^{j}$ is $P_{\vX_{i}^{j}}(\vx_{i}^{j})$. In the following, we will
  use the abbreviation $Q(\vx_{i}^{j}) \defeq P_{\vX_{i}^{j}}(\vx_{i}^{j})$
  for any $i \leq j$.
\end{Definition}

\subsection{\textbf{Finite-State Machine Channels (FSMCs)}}
\label{sec:finite:state:channel:models:1}

Many of the original channels that we will consider can be described as
FSMCs.\footnote{Note that Gallager~\cite{Gallager:68} calls them finite-state
  channels (FSCs).}

\begin{Definition}{Finite-State Machine Channels (FSMCs)}
  \label{def:finite:state:channel:models:1}

  A time-independent, discrete-time finite-state machine
  channel~\cite{Gallager:68} has an input process $\ldots, X_{-1}, X_0, X_1,
  \ldots$, an output process $\ldots, Y_{-1}, Y_0, Y_1, \ldots$, and a state
  process $\ldots, S_{-1}, S_0, S_1, \ldots$, where $X_{\ell} \in \setX$,
  $Y_{\ell} \in \setY$, and $S_{\ell} \in \setS$ for all $\ell \in {\mathbb
    Z}$. We assume that the alphabets $\setX$, $\setY$, and $\setS$ are
  finite. Let us define the following finite windows of the FSMC states,
  inputs, and outputs as:
  \begin{align}
    \vs
      &\defeq
         {\vs}_{-N+1}^{N},
    \quad\quad
    \vx
       \defeq
         \vx_{-N+1}^{N},
    \quad\quad
    \vy
       \defeq
         \vy_{-N+1}^{N}.
  \end{align}
  Unless otherwise stated, we condition all FSMC-related probabilities on an
  initial state, such as $s_{-N}$. The joint FSMC conditional pmf decomposes
  as
  \begin{align}
    P_{\vS,\vY | \vX, S_{-N}}
      (\vs, \vy | \vx, s_{-N})
      &= \prod_{\ell \in \sIN}
           P_{S_{\ell},Y_{\ell} | S_{\ell-1}, X_{\ell}}
             (s_{\ell},y_{\ell} | s_{\ell-1}, x_{\ell}) \\
      &= \left(
           \prod_{\ell \in \sIN} P_{S_{\ell} | S_{\ell-1}, X_{\ell}}
                                 (s_{\ell} | s_{\ell-1}, x_{\ell})
         \right)
         \cdot
         \left(
           \prod_{\ell \in \sIN} P_{Y_{\ell} | S_{\ell-1}, X_{\ell},S_{\ell}}
                                 (y_{\ell} |  s_{\ell-1}, x_{\ell}, s_{\ell})
         \right),
  \end{align}
  where $P_{S_{\ell} | S_{\ell-1}, X_{\ell}}$ and $P_{Y_{\ell} | S_{\ell-1},
    X_{\ell},S_{\ell}}$ are referred to as the FSMC state transition
  probability and FSMC output probability, respectively, and are independent
  of the time index $\ell$. In the following, we will use the notation
  \begin{align}
    W(s_{\ell}| s_{\ell-1}, x_{\ell})
      &\defeq
         P_{S_{\ell} | S_{\ell-1}, X_{\ell}}(s_{\ell} | s_{\ell-1}, x_{\ell}), \\
    W(y_{\ell} | s_{\ell-1}, x_{\ell}, s_{\ell})
      &\defeq
         P_{Y_{\ell} | S_{\ell-1}, X_{\ell},S_{\ell}}
           (y_{\ell} |  s_{\ell-1}, x_{\ell}, s_{\ell}), \\
    W(\vs,\vy |\vx,s_{-N})
      &\defeq
         P_{\vS,\vY |\vX, S_{-N}}
           (\vs,\vy |\vx, s_{-N}), \\
    W(\vy |\vx,s_{-N})
      &\defeq
         P_{\vY |\vX, S_{-N}}
           (\vy |\vx, s_{-N})
       = \sum_{\vs}
           W(\vs,\vy |\vx,s_{-N}).
  \end{align}
\end{Definition}

\begin{figure}
  \begin{center}
    \epsfig{file=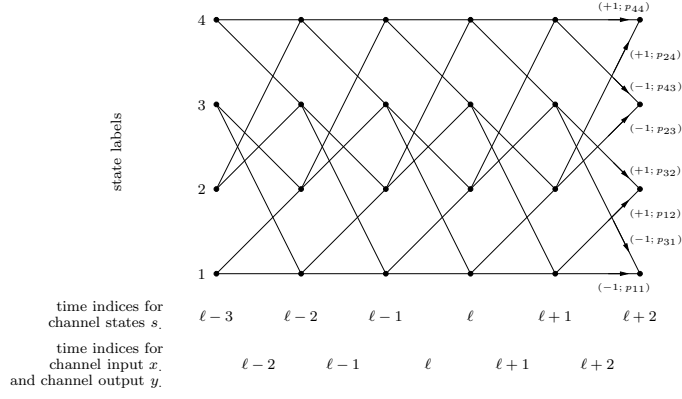, width=0.5\linewidth}
  \end{center}
  \caption{Trellis representation of the finite-state machine behind an
    FSMC. Because of the assumed time-invariance, the branch labels in every
    trellis section are the same. Therefore, we have shown these branch labels
    only in one trellis section.}
  \label{fig:trellis:channel:notation:1}
\end{figure}

\begin{figure}
  \begin{center}
    \epsfig{file=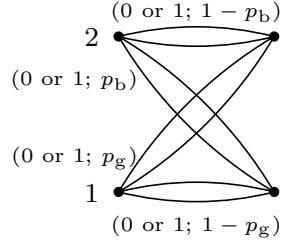, width=0.2\linewidth}
  \end{center}
  \caption{Trellis section of the trellis representation of the finite-state
    machine behind a Gilbert-Elliott channel. State ``$1$'' corresponds to the
    bad state ``$\bad$'' and state ``$2$'' corresponds to the good state
    ``$\good$''.}
  \label{fig:trellis:section:gilbert:elliott:channel:1}
\end{figure}

\figref{fig:trellis:channel:notation:1} shows how the finite-state machine
behind a typical FSMC can be represented by a trellis. This specific FSMC has
$4$ states, namely $\setS = \{ 1, 2, 3, 4 \}$, and two input symbols, namely
$\setX = \{ -1, +1 \}$. To a branch going from state $i$ to state $j$ we
associate a label like $(x_{ij}; p_{ij})$: it shows the probability $p_{ij}$
with which this state transition is chosen in case that the channel input
symbol is $x_{ij}$.

\begin{Example}{Partial Response Channel}
  \label{ex:partial:response:channel:1}

  The trellis in \figref{fig:trellis:channel:notation:1} actually represents
  the trellis of the finite-state machine that is behind a so-called partial
  response channel. Such a channel can be described by
  \begin{align}
    \label{eq:fir:1}
    y_{\ell}
      &= \mu
         \left(
           \sum_{m=0}^{M}
             h_m x_{\ell-m}
             +
             n_{\ell}
         \right).
  \end{align}
  Here, $\mu$ is a quantization function that maps elements of $\mathbb{R}$ to
  $\setY$. Moreover, $\{ x_{\ell} \}$, $\{ y_{\ell} \}$, $\{ n_{\ell} \}$, $\{
  h_m \}$ represent the channel input process, the channel output process, an
  additive noise process, and the filter coefficients, respectively. Assuming
  the channel input alphabet to be $\setX = \{ -1, +1 \}$ and the memory
  length to be $M = 2$, we see that in \figref{fig:trellis:channel:notation:1}
  the state $S_{\ell} = 1$ corresponds to $(X_{\ell-1},X_{\ell}) = (-1,-1)$,
  the state $S_{\ell} = 2$ corresponds to $(X_{\ell-1},X_{\ell}) = (-1,+1)$,
  the state $S_{\ell} = 3$ corresponds to $(X_{\ell-1},X_{\ell}) = (+1,-1)$,
  and the state $S_{\ell} = 4$ corresponds to $(X_{\ell-1},X_{\ell}) =
  (+1,+1)$.
\end{Example}

\begin{Example}{Gilbert-Elliott Channel}
  \label{example:gilbert:elliott:channel:1}

  The Gilbert-Elliott channel~\cite{Gilbert1960, Elliott1963} has the state
  alphabet $\set{S} = \{ \good, \bad \}$, \ie, a ``good'' state and a ``bad''
  state, the input alphabet $\setX = \{ 0, 1 \}$, and the output alphabet $\{
  0,1 \}$. One defines $W(s_{\ell},y_{\ell} | s_{\ell-1},x_{\ell}) \defeq
  W(s_{\ell} | s_{\ell-1}) \cdot W(y_{\ell} | s_{\ell-1},x_{\ell})$ where
  $W(\bad | \good) = \pbad$ and $W(\good | \bad) = \pgood$, where $W(y_{\ell}
  | s_{\ell-1},x_{\ell})$ is a binary symmetric channel (BSC) with cross-over
  probability $\epsgood$ when $s_{\ell-1} = \good$, and where $W(y_{\ell} |
  s_{\ell-1},x_{\ell})$ is a BSC with cross-over probability $\epsbad$ when
  $s_{\ell-1} = \bad$. A trellis section of the trellis representation of the
  finite-state machine behind such a channel is shown
  in~\figref{fig:trellis:section:gilbert:elliott:channel:1}. (Here, $\pgood$,
  $\pbad$, $\epsgood$, and $\epsbad$ are arbitrary real numbers between $0$ and
  $1$, where usually $|\epsgood - 1/2| > |\epsbad - 1/2|$.)
\end{Example}

It is useful to introduce the FSMC branch random variable $B_{\ell} \defeq
(S_{\ell-1}, X_{\ell}, S_{\ell})$ and the set $\setB$ of all branches in a
trellis section. (Note that because the original FSMC is assumed to be
time-invariant, the set $\setB$ is also time-invariant.) The initial state of
a branch $b_{\ell} \in \setB$ at time index $\ell$ will then be denoted by
$s_{\ell-1}(b_{\ell})$, the channel input symbol by $x(b_{\ell})$, and the
final state by $s_{\ell}(b_{\ell})$. It can easily be seen that without loss
of generality, we can assume that for any triple $(s_{\ell-1}, x_{\ell},
s_{\ell})$ there is at most one branch in the trellis that starts in state
$s_{\ell-1}$, has input symbol $x_{\ell}$, and ends in state $s_{\ell}$.  In
that sense, in the following we will use the notation
\begin{align}
  W(y_{\ell} | b_{\ell})
    &\defeq
       P_{Y_{\ell} | B_{\ell}}
         (y_{\ell} |  b_{\ell})
     = P_{Y_{\ell} | S_{\ell-1}, X_{\ell},S_{\ell}}
         (y_{\ell} |  s_{\ell-1}, x_{\ell}, s_{\ell})
     = W(y_{\ell} |  s_{\ell-1}, x_{\ell}, s_{\ell})
\end{align}
for $b_{\ell} = (s_{\ell-1}, x_{\ell}, s_{\ell})$. Similarly to $\vs$, $\vx$,
and $\vy$, we also define $\vb \defeq {\vb}_{-N+1}^{N}$ which helps us define
\begin{align}
  W(\vb, \vy | \vx, s_{-N})
    &\defeq
       W(\vs, \vy | \vx, s_{-N}).
           \label{eq:from:state:and:symbol:seq:to:branch:seq:1}
\end{align}
In this and upcoming similar expressions, $\vx$ and $\vs$ are implicitly given
by $\vb$, \ie, $\vx = \vx(\vb)$ and $\vs = \vs(\vb)$. The sequence $\vb$ is
called a valid branch sequence if the ending state of $b_{\ell-1}$ equals the
starting state of $b_{\ell}$. Similarly, the sequence $\vs$ is called a valid
state sequence if there is a valid branch sequence $\vb$ such that $\vs =
\vs(\vb)$. As already mentioned in the introduction, summations like
$\sum_{\vs}$ and $\sum_{\vb}$ will always be over all valid channel state
sequences and all valid channel branch sequences, respectively.

In this paper, we consider only FSMCs that are indecomposable, as defined by
Gallager~\cite{Gallager:68}, \ie channels where, roughly speaking, the
influence of the initial state fades out with time for every possible channel
input sequence. Feeding such a channel with a stationary and ergodic source
results in input and output processes that are jointly stationary and ergodic.
Therefore, when feeding an indecomposable FSMC with a source as in
Def.~\ref{def:source:model:1} we will use the notation
\begin{align}
  (QW)(\vy|s_{-N})
    &\defeq
       \sumb
         Q
         \big(\vx
           \big|
           \vx_{-\infty}^{-N}
           \big.
         \big) \,
         W(\vb, \vy | \vx, s_{-N}).
\end{align}
Moreover, in order to be able to apply the results
of~\cite{LeGland:Mevel:00:2} later on, we will impose the following condition
on $W$: for all $b_{\ell} \in \setB$ and all $y_{\ell} \in \setY$, we require
that $W(y_{\ell} | b_{\ell})$ is strictly positive, \ie, we require that any
output symbol $y _{\ell} \in \setY$ can potentially be observed for any branch
$b_{\ell} \in \setB$.

\begin{Definition}{Data-Controllable FSMCs}
  \label{def:controlable}

  If an indecomposable finite-state machine channel can be taken from any
  state into any other state by a finite number of channel inputs, which do
  not depend on the current state, the channel is called {\em controllable}.
  (Referring to~\cite[p.~111 and p.~527]{Gallager:68}, we note that there are
  also decomposable channels that could be called controllable and for which
  the unconstrained capacity is uniquely defined. However, in the following we
  will not consider such channels because we deal exclusively with
  indecomposable channels.)
\end{Definition}

Clearly, the partial response channel in
Ex.~\ref{ex:partial:response:channel:1} is data-controllable, whereas the
Gilbert-Elliott channel (cf.~\ref{example:gilbert:elliott:channel:1}) is
\emph{not} data-controllable.

\subsection{\textbf{General Channels with Memory}}
\label{sec:general:channels:memory:1}

We will allow the original channel to be a more general stationary
discrete-time channel than an indecomposable FSMC, namely, we allow the
state-space size to be infinite, as long as the following requirement is
satisfied: it should be possible to approximate such a general channel to any
desired degree by an indecomposable FSMC.

\subsection{\textbf{Original Backward Channel Model}}
\label{sec:backward:channel:model:1}

Reversing the usual meaning of $\vX$ and $\vY$, \ie, looking at $\vX$ as being
the output of some channel which is fed by $\vY$ which in turn is produced by
some source, we arrive at the ``backward'' channel model in
Fig.~\ref{fig:source:channel:prob:models:1}(b). Note that the mutual
information $I(\vX;\vY)$ is a functional of the joint pmf of $\vX$ and $\vY$.
Therefore, if $\vX$ and $\vY$ have the same joint pmf in both the forward and
the backward channel setup then both channels will have the same information
rate. This can be achieved by setting the pmf of the ``output'' source to be
$R(\vy) \defeq (QW)(\vy)$ and the ``backward'' channel law to be $V(\vx|\vy) =
\frac{Q(\vx) W(\vy|\vx)}{(QW)(\vy)}$ because then $Q(\vx) W(\vy|\vx) = R(\vy)
V(\vx|\vy)$.

\subsection{\textbf{Auxiliary Forward Finite-State Machine Channels}}
\label{sec:forward:aux:finite:state:channel:models:1}

The information rate upper bound in \secref{sec:information:rate:bound:1} will
be based on an auxiliary forward channel law. By an auxiliary forward channel
law we will simply mean a conditional pmf on $\vy$ given $\vx$. In general, we
will denote this conditional pmf by $\What(\vy|\vx)$ and it fulfills the usual
properties of a conditional pmf, \ie, $\What(\vy|\vx) \geq 0$ for all $\vx$
and all $\vy$, and $\sum_{\vy} \What(\vy|\vx) = 1$ for all $\vx$.

In the following, we will focus on the case where the auxiliary forward
channel is an auxiliary forward finite-state machine channel (AF-FSMC).

\begin{Definition}{AF-FSMCs}
  \label{def:forward:aux:finite:state:channel:models:1}

  A time-independent and discrete-time AF-FSMC has an input process $\ldots,
  X_{-1}, X_0, X_1, \ldots$, an output process $\ldots, Y_{-1}, Y_0, Y_1,
  \ldots$, and a state process $\ldots$, $\Shat_{-1}$, $\Shat_0$, $\Shat_1$,
  $\ldots$, where $X_{\ell} \in \setX$, $Y_{\ell} \in \setY$, and
  $\Shat_{\ell} \in \setShat$ for all $\ell \in {\mathbb Z}$. We assume the
  set $\setShat$ to be finite. Let us define the following finite windows of
  the AF-FSMC states and branches as:
  \begin{align}
    \vshat
      &\defeq
         {\vshat}_{-N+1}^{N},
    \quad\quad
    \vbhat
       \defeq
         {\vbhat}_{-N+1}^{N}.
  \end{align}
  The AF-FSMC conditional pmf decomposes as
  \begin{align}
    \What(\vshat, \vy | \vx, \shat_{-N})
      &= \prod_{\ell \in \sIN}
           \What(\shat_{\ell}, y_{\ell} | \shat_{\ell-1}, x_{\ell})
       = \left(
           \prod_{\ell \in \sIN}
             \What(\shat_{\ell} | \shat_{\ell-1}, x_{\ell})
         \right)
         \cdot
         \left(
           \prod_{\ell \in \sIN}
             \What(y_{\ell} |  \shat_{\ell-1}, x_{\ell}, \shat_{\ell})
         \right),
         \label{eq:aux:FSMC:decompose:2}
  \end{align}
  where $\What(\shat_{\ell} | \shat_{\ell-1}, x_{\ell})$ and $\What(y_{\ell} |
  \shat_{\ell-1}, x_{\ell}, \shat_{\ell})$ are referred to as the AF-FSMC
  state transition probability and AF-FSMC output probability, respectively,
  and are independent of the time index $\ell$. The input-output conditional
  pmf will then be
  \begin{align}
    \What(\vy | \vx, \shat_{-N})
      &\defeq
         \sum_{\vshat}
           \What(\vshat, \vy | \vx, \shat_{-N}).
  \end{align}
\end{Definition}

It is useful to introduce the AF-FSMC branch (random) variable $\Bhat_{\ell}
\defeq (\Shat_{\ell-1}, X_{\ell}, \Shat_{\ell})$ and the set $\setBhat$ of all
branches in a trellis section. (Note that because the AF-FSMC is assumed to be
time-invariant, the set $\setBhat$ is also time-invariant.) The initial state
of a branch $\bhat_{\ell}$ at time index $\ell$ will then be denoted by
$\shat_{\ell-1}(\bhat_{\ell})$, the channel input symbol by $x(\bhat_{\ell})$,
and the final state by $\shat_{\ell}(\bhat_{\ell})$. In that sense, we will
often write $\What(y_{\ell} | \bhat_{\ell})$ instead of $\What(y_{\ell} |
\shat_{\ell-1}, x_{\ell}, \shat_{\ell})$, and $\What(\vbhat, \vy | \vx,
\shat_{-N})$ instead of $\What(\vshat, \vy | \vx, \shat_{-N})$ if $\vbhat =
(\vx, \vshat)$. As in \secref{sec:finite:state:channel:models:1}, without loss
of generality we can assume that for any triple $(\shat_{\ell-1}, x_{\ell},
\shat_{\ell})$ there is at most one branch in the trellis that starts in state
$\shat_{\ell-1}$, has input symbol $x_{\ell}$, and ends in state
$\shat_{\ell}$.

Similar to the original channel, we consider only AF-FSMCs that are
indecomposable and for which $\What(y_{\ell} | \bhat_{\ell})$ is strictly
positive for all $\bhat_{\ell} \in \setBhat$ and all $y_{\ell} \in \setY$.

\begin{Remark}{Induced pmfs}
  \label{remark:induced:pmfs:1}

  Of special interest is the case where $\vx$ is generated according to the
  standard source: the induced joint pmf of $\vx$, $\vshat$, and $\vy$ is then
  called $\Phat(\vx,\vshat,\vy | \shat_{-N})$ and equals $\Phat(\vx,\vshat,\vy
  | \shat_{-N}) \defeq Q(\vx) \What(\vshat, \vy|\vx, \shat_{-N})$. In that
  sense,
  \begin{align}
    \Phat(\vbhat,\vy | \shat_{-N})
      &\defeq
         \Phat(\vx,\vshat,\vy | \shat_{-N}), \\
    \Phat(\vx, \vy | \shat_{-N})
      &\defeq
         \sum_{\vshat}
           \Phat(\vx,\vshat,\vy | \shat_{-N}), \\
    \Phat(\vy | \shat_{-N})
      &\defeq
         \sum_{\vx, \, \vshat}
           \Phat(\vx, \vshat, \vy | \shat_{-N})
       = \sum_{\vx}
           \Phat(\vx, \vy | \shat_{-N})
       = (Q\What)(\vy | \shat_{-N}), \\
    \Phat(\vbhat | \vy, \shat_{-N})
      &\defeq
         \frac{\Phat(\vbhat, \vy | \shat_{-N})}
              {\Phat(\vy | \shat_{-N})}, \\
    \Phat(\vbhat | \vx, \vy, \shat_{-N})
      &\defeq
         \frac{\Phat(\vbhat, \vy | \shat_{-N})}
              {\sum\limits_{\vbhat': \,
                           \text{$\vbhat'$ is compatible with $\vx$}}
                 \Phat(\vbhat', \vy | \shat_{-N})},
  \end{align}
  where in the last expression we assume that $\vx$ is compatible with
  $\vbhat$.
\end{Remark}

\subsection{\textbf{Auxiliary Backward Finite-State Machine Channels}}
\label{sec:backward:aux:finite:state:channel:models:1}

The information rate lower bound in \secref{sec:information:rate:bound:1} will
be based on an auxiliary backward channel law. By an auxiliary backward
channel law we will simply mean a conditional pmf on $\vx$ given $\vy$. In
general, we will denote this conditional pmf by $\Vhat(\vx|\vy)$ and it
fulfills the usual properties of a conditional pmf, \ie, $\Vhat(\vx|\vy) \geq
0$ for all $\vx$ and all $\vy$, and $\sum_{\vx} \Vhat(\vx|\vy) = 1$ for all
$\vy$.

In the following, we will focus on the case where the auxiliary backward
channel is an auxiliary backward finite-state machine channel (AB-FSMC).

\begin{Definition}{AB-FSMCs}
  \label{def:backward:aux:finite:state:channel:models:1}

  A time-independent and discrete-time AB-FSMC has an input process $\ldots,
  Y_{-1}, Y_0, Y_1, \ldots$, an output process $\ldots, X_{-1}, X_0, X_1,
  \ldots$, and a state process $\ldots$, $\Shat_{-1}$, $\Shat_0$, $\Shat_1$,
  $\ldots$, where $Y_{\ell} \in \setY$, $X_{\ell} \in \setX$, and
  $\Shat_{\ell} \in \setShat$ for all $\ell \in {\mathbb Z}$. We assume the
  set $\setShat$ to be finite. Let us define the following finite windows of
  the AB-FSMC states and branches as:
  \begin{align}
    \vshat
      &\defeq
         {\vshat}_{-N+1}^{N},
    \quad\quad
    \vbhat
       \defeq
         {\vbhat}_{-N+1}^{N}.
  \end{align}
  The AB-FSMC conditional pmf is defined to be
  \begin{align}
    \Vhat(\vshat, \vx | \vy, \shat_{-N})
      &\defeq
         \frac{\vhat(\shat_{-N}, \vshat, \vx, \vy)}
              {\sum\limits_{\vshat', \vx'}
                 \vhat(\shat_{-N}, \vshat', \vx', \vy)}
         \label{eq:backward:auxiliary:channel:law:1}
  \end{align}
  with
  \begin{align}
    \vhat(\shat_{-N}, \vshat, \vx, \vy)
      &\defeq
         Q(\vx)
         \cdot
         \prod_{\ell \in \sIN}
           \vhat
             \left(
               \shat_{\ell-1}, x_{\ell}, \shat_{\ell}, \vy_{\ell-D_1}^{\ell+D_2}
             \right).
         \label{eq:backward:auxiliary:channel:law:1:part:2}
  \end{align}
  Here, $D_1$ and $D_2$ are some arbitrary non-negative integers\footnote{We
    will comment on the selection of suitable values for $D_1$ and $D_2$ at
    the end of \secref{sec:information:rate:bound:1} and in
    \secref{subsec:lower:optimization:idea}. In the expressions $\Vhat(\vshat,
    \vx | \vy, \shat_{-N})$ and $\vhat(\shat_{-N}, \vshat, \vx, \vy)$, it
    would actually be more precise to write $\vy_{-N+1-D_1}^{N+D_2}$ instead
    of $\vy$ ($ = \vy_{-N+1}^{N}$) in the case that $D_1 > 0$ and/or $D_2 >
    0$.  However, in order to keep the notation simple and because we are
    mostly interested in the limit $N \to \infty$, we will not do so.} and
  $\vhat(\shat_{\ell-1}, x_{\ell}, \shat_{\ell}, \vy_{\ell-D_1}^{\ell+D_2})$ is a
  non-negative function of $\bigl( (\shat_{\ell-1}, x_{\ell}, \shat_{\ell}),
  \, \vy_{\ell-D_1}^{\ell+D_2} \bigr) \in \setBhat \times \setY^{D_1+D_2+1}$.
  Note that the latter requirement on $\vhat(\shat_{\ell-1}, x_{\ell},
  \shat_{\ell}, \vy_{\ell-D_1}^{\ell+D_2})$ is sufficient for $\Vhat(\vshat, \vx |
  \vy, \shat_{-N})$ to represent a conditional pmf of $\vshat$ and $\vx$ given
  $\vy$ and $\shat_{-N}$.\footnote{The letter $\vhat$ was chosen to show the
    closeness of the $\vhat$ function to the $\Vhat$ function, yet to remind
    the reader that the $\vhat$ function is not properly normalized to be a
    conditional pmf of $\vshat$ and $\vx$ given $\vy$ and $\shat_{-N}$.}  The
  backward input-output conditional pmf will then be
  \begin{align}
    \Vhat(\vx | \vy, \shat_{-N})
      &\defeq
         \sum_{\vshat}
           \Vhat(\vshat, \vx | \vy, \shat_{-N}).
         \label{eq:backward:auxiliary:channel:law:1:part:3}
  \end{align}
\end{Definition}

In the following, we will also use
\begin{align}
  \vhat(\shat_{-N}, \vx, \vy)
    &\defeq
       \sum_{\vshat}
         \vhat(\shat_{-N}, \vshat, \vx, \vy), \\
  \Vhat(\vshat | \vx, \vy, \shat_{-N})
    &\defeq
       \frac{\Vhat(\vshat, \vx | \vy, \shat_{-N})}
            {\sum_{\vshat'}
               \Vhat(\vshat', \vx | \vy, \shat_{-N})}
     = \frac{\vhat(\shat_{-N}, \vshat, \vx, \vy)}
            {\sum_{\vshat'}
               \vhat(\shat_{-N}, \vshat', \vx, \vy)}
     = \frac{\vhat(\shat_{-N}, \vshat, \vx, \vy)}
            {\vhat(\shat_{-N}, \vx, \vy)}.
    \label{eq:capital:v:small:v:1}
\end{align}
It is useful to introduce the AB-FSMC branch (random) variable $\Bhat_{\ell}
\defeq (\Shat_{\ell-1}, X_{\ell}, \Shat_{\ell})$ and the set $\setBhat$ of all
branches in a trellis section. (Note that because the AB-FSMC is
assumed to be time-invariant, the set $\setBhat$ is also
time-invariant.) The initial state of a branch $\bhat_{\ell}$ at
time index $\ell$ will then be denoted by
$\shat_{\ell-1}(\bhat_{\ell})$, the (backward channel) output symbol
by $x(\bhat_{\ell})$, and the final state by
$\shat_{\ell}(\bhat_{\ell})$. In that sense, we will often write
$\vhat(\bhat_{\ell}, \vy_{\ell-D_1}^{\ell+D_2})$ instead of
$\vhat(\shat_{\ell-1}, x_{\ell}, \shat_{\ell},
\vy_{\ell-D_1}^{\ell+D_2})$, $\Vhat(\vbhat | \vy, \shat_{-N})$
instead of $\Vhat(\vshat, \vx | \vy, \shat_{-N})$ if $\vbhat = (\vx,
\vshat)$, and $\Vhat(\vbhat | \vx, \vy, \shat_{-N})$ instead of
$\Vhat(\vshat | \vx, \vy, \shat_{-N})$ if $\vbhat = (\vx, \vshat)$.
As in \secref{sec:finite:state:channel:models:1}, without loss of
generality we can assume that for any triple $(\shat_{\ell-1},
x_{\ell}, \shat_{\ell})$ there is at most one branch in the trellis
that starts in state $\shat_{\ell-1}$, has (backward channel) output
symbol $x_{\ell}$, and ends in state $\shat_{\ell}$.

Similar to the original channel, we consider only AB-FSMCs that are
indecomposable and for which $\vhat(\bhat_{\ell},
\vy_{\ell-D_1}^{\ell+D_2})$ is strictly positive for all
$\bhat_{\ell} \in \setBhat$ and all $\vy_{\ell-D_1}^{\ell+D_2} \in
\setY^{D_1+D_2+1}$.

\subsection{\textbf{Remarks}}

Let us briefly point out some notational differences with the
notation used in the paper on the generalized Blahut-Arimoto
algorithm~\cite{Vontobel:Kavcic:Arnold:Loeliger:04:1:subm}. Similar
to that paper, we are using $\vx$ to denote the source output /
channel input sequence and $\vy$ to denote the channel output
sequence.  However, in the current paper $\vs$ and $\vb$ denote the
state and branch sequence, respectively, in the trellis
representation of the \emph{channel}, whereas
in~\cite{Vontobel:Kavcic:Arnold:Loeliger:04:1:subm}, $\vs$ and $\vb$
were used to denote the state and branch sequence, respectively, in
the trellis representation of the \emph{source}. Finally, note that
in both papers the time indexing of the components of the input,
output, state, and branch sequences are done in the same manner,
however, $\vs$ is defined to be $\vs_{-N+1}^{N}$ and not
$\vs_{-N}^{N}$.

We have already mentioned that we will be mainly interested in the limit $N
\to \infty$. Because our setup is such that the limits of the expressions of
interest do not depend on the past $\vx_{-\infty}^{-N}$ and the initial states
$s_{-N}$ and $\shat_{-N}$ (this can be justified using results like
in~\cite{LeGland:Mevel:00:2}), in the following we will assume that
$\vx_{-\infty}^{-N}$, $s_{-N}$, and $\shat_{-N}$ are fixed to some suitable
and mutually compatible values. In order to simplify the notation in all the
upcoming formulas, we will omit the explicit conditioning on
$\vx_{-\infty}^{-N}$, $s_{-N}$, and $\shat_{-N}$, however, it should be kept
in mind that such a conditioning is still present.

In the following, a generic branch of an AF-FSMC / AB-FSMC will
often be denoted by $\bhat = (\shatL, x, \shat)$ instead of
$\bhat_{\ell} = (\shat_{\ell-1}, x_{\ell}, \shat_{\ell})$. (Here,
``p'' stands for previous.) Similarly, the generic output symbol
corresponding to this branch will often be denoted by $y$ instead of
$y_{\ell}$. Moreover, the generic $\vyD$ will be used instead of
$\vy_{\ell-D_1}^{\ell+D_2}$. In this manner, $\vhat(\bhat_{\ell},
\vy_{\ell-D_1}^{\ell+D_2})$ in an AB-FSMC will simply be denoted by
$\vhat(\bhat,\vyD)$. Such simplifications in notation are possible
because of our stationarity assumptions for sources and channels.

%% file: information_rates1.tex

\section{Information Rates and Their Upper and
             Lower Bounds}
\label{sec:information:rate:bound:1}

In the following, the source pmf and the original (forward) channel law will
be assumed to be fixed, and we will only vary the auxiliary forward and
auxiliary backward channels. Therefore, in order to simplify the notation, $Q$
and $W$ will not appear as arguments of information rates, upper and lower
bound formulas, etc. (Note that a fixed source pmf and a fixed original
forward channel law imply, according to our comments in
\secref{sec:backward:channel:model:1}, that the ``output'' source pmf and the
original backward channel law are also fixed.)

\begin{Definition}{Information Rate}
  \label{def:information:rate:1}

  For the type of sources and channels that were considered in
  \secref{sec:channel:models:1}, the information rate is given by
  \begin{align}
    I^{(N)}
       \defeq
         \frac{1}{2N}
           I(\vX;\vY \condvs)
      &\defeq
         \frac{1}{2N}
           \sumx
             \sumy
               Q(\vx \condvs)
                 W(\vy|\vx \condcs)
                   \log
                     \left(
                       \frac{W(\vy|\vx \condcs)}
                            {(QW)(\vy \condvs)}
                     \right),
             \label{eq:mut:inf:rate:3}
  \end{align}
  with asymptotic version $I \defeq \lim_{N \to \infty} I^{(N)}$.
\end{Definition}

Note that for any finite $N$, the information rate $I^{(N)}$ depends on the
choice of $\vx_{-\infty}^{-N}$ and $s_{-N}$, however, we will not mention
these quantities explicitly as arguments of $I^{(N)}$. Similarly, we will not
mention them in upcoming functionals, along with not mentioning the initial
state $\shat_{-N}$ of an auxiliary channel.

\begin{Definition}{Information Rate Upper Bound Based on an AF-FSMC}
  \label{def:information:rate:upper:1}

  Using an AF-FSMC as defined in
  Def.~\ref{def:forward:aux:finite:state:channel:models:1}, an upper bound on
  the information rate is given by
  \begin{align}
    \label{eq:information:rate:upper:1}
    \Iupper^{(N)}(\What \condcs)
      &\defeq
         \frac{1}{2N}
         \sumxy
           Q(\vx \condvs)
             W(\vy|\vx \condcs)
             \log
               \left(
                 \frac{W(\vy|\vx \condcs)}
                      {(Q\What)(\vy \condvs)}
               \right),
  \end{align}
  with asymptotic version $\Iupper(\What) \defeq \lim_{N \to \infty}
  \Iupper^{(N)}(\What \condcs)$. This upper bound was also used
  in~\cite{Arnold:Loeliger:Vontobel:02:1,
    Arnold:Loeliger:Vontobel:Kavcic:Zeng:06:1} and we refer the interested
  reader to these papers for some historical comments.
\end{Definition}

The fact that $\Iupper^{(N)}(\What \condcs)$ is indeed an upper bound on
$I^{(N)}$ can easily be verified by writing the difference
$\Iupper^{(N)}(\What \condcs) - I^{(N)}$ as a Kullback-Leibler (KL)
divergence~\cite[p.~18]{Cover:Thomas:91} and by using the well-known fact that
the KL divergence is never negative, \ie
  \begin{align}
    \Iupper^{(N)}(\What \condcs)-I^{(N)}
        &= D_{\vy}
             \left(
                 (QW)(\vy \condvs)
               \,\left\lVert\,
                 (Q\What)(\vy \condvs)
               \right.
             \right)
         \ge 0.
  \end{align}
  It is noted that for computing the upper bound $\Iupper^{(N)}(\What)$, an
  analytical or numerical evaluation method for the conditional entropy
\begin{align}
  \label{eq:upper:analytical}
  H^{(N)}
    &\defeq
       -
       \frac{1}{2N}
       \sumxy
         Q(\vx \condvs)
           W(\vy|\vx \condcs)
           \log
             \bigl(
               W(\vy|\vx \condcs)
             \bigr)
\end{align}
of the output process given the input process in the original channel is
required. (The asymptotic version of $H^{(N)}$ will be called $H$.)
Alternatively, a lower bound on $H^{(N)}$ can be used to obtain an upper bound
on $\Iupper^{(N)}(\What \condcs)$. Later in \secref{sec:num:fading}, we prove
a tight lower bound on $H^{(N)}$ for Gauss-Markov fading channels.

Let us briefly mention that the expression in
~\eqref{eq:information:rate:upper:1} is still a valid information rate upper
bound also if $(Q\What)(\vy)$ is replaced by some arbitrary pmf over $\vy$.
However, we will not pursue this more general information rate upper bound any
further in this paper.

\begin{Definition}{Information Rate Lower Bound Based on an AB-FSMC}
  \label{def:information:rate:lower:1}

  Using an AB-FSMC as defined in
  Def.~\ref{def:backward:aux:finite:state:channel:models:1}, a lower bound on
  the information rate is given by
  \begin{align}
    \label{eq:information:rate:lower:1}
    \Ilower^{(N)}(\vhat \condcs)
      &\defeq
         \frac{1}{2N}
         \sumxy
           Q(\vx \condvs)
             W(\vy|\vx \condcs)
             \log
               \left(
                 \frac{\Vhat(\vx|\vy \condcs)}
                      {Q(\vx \condvs)}
               \right),
  \end{align}
  with asymptotic version $\Ilower(\vhat) \defeq \lim_{N \to \infty}
  \Ilower^{(N)}(\vhat \condcs)$, and where $\Vhat$ is implicitly defined by
  $\vhat$ as shown
  in~\eqref{eq:backward:auxiliary:channel:law:1}-\eqref{eq:backward:auxiliary:channel:law:1:part:3}.
\end{Definition}

Again, that $\Ilower^{(N)}(\vhat \condcs)$ is a lower bound on $I^{(N)}$ can
easily be verified by writing the difference $I^{(N)} - \Ilower^{(N)}(\vhat
\condcs)$ as a KL divergence, \ie,
\begin{align}
  I^{(N)}- \Ilower^{(N)}(\vhat \condcs)
      &= D_{\vx,\vy}
           \left(
             \Big.
               Q(\vx \condvs) W(\vy|\vx \condcs)
             \,\Big\lVert\,
               (QW)(\vy \condvs) \Vhat(\vx|\vy \condcs)\right)
       \geq
         0.
\end{align}
On the side, let us remark that $I^{(N)}- \Ilower^{(N)}(\vhat \condcs)$ can
also be written as
\begin{align}
  I^{(N)}- \Ilower^{(N)}(\vhat \condcs)
      &= \sum_{\vy}
           (QW)(\vy \condvs) \
           D_{\vx}
           \left(
             \left.
               \frac{Q(\vx \condvs) W(\vy|\vx \condcs)}
                    {(QW)(\vy \condvs)}
             \,\right\lVert\,
               \Vhat(\vx|\vy \condcs)\right).
\end{align}

The lower bound defined in \eqref{eq:information:rate:lower:1} is linked with
the generalized mutual information (GMI) which is defined
as~\cite{Ganti_JIT2000}
\begin{align}
  \label{EqGMI}
  I^{(N)}_{\mathrm{GMI}}
    &\defeq
       \frac{1}{2N}
       \sup_{a^{(N)}
      \geqslant 0}
       \sum_{\vx,\vy}
         Q(\vx \condvs)
         W(\vy|\vx \condcs)
         \log
           \frac{e^{-a^{(N)}d^{(N)}(\vx,\vy)}}
                {\sum_{\vx'}Q(\vx' \condvs)e^{-a^{(N)}d^{(N)}(\vx',\vy)}},
\end{align}
where $d^{(N)}(\vx,\vy)$ is a mapping from $\setX^{2N} \times \setY^{2N}$ to
$\mathbb{R}$. Under suitable time-invariance conditions on $d^{(N)}$, this GMI
has the following interesting meaning: it is a reliable communication rate
that is achievable under mismatched decoding, \ie, when the decoder uses the
decoding metric $d^{(N)}(\vx,\vy)$ instead of the ML decoding metric $-
\frac{1}{2N} \log W(\vy|\vx \condcs)$. (Maximizing the GMI over all possible
sources gives a lower bound on what is known as the mismatch capacity under
the decoding metric $d^{(N)}(\vx,\vy)$, $N \to \infty$.) Now, setting $a^{(N)}
\defeq 2N$ and
\begin{align}
  d^{(N)}(\vx,\vy)
    &\defeq
       - \frac{1}{2N}
           \log
             \left(
               \frac{f(\vy) \, \Vhat(\vx|\vy \condcs)}
                    {Q(\vx)}
             \right),
    \label{eq:gmi:distance:1}
\end{align}
where $f(\vy)$ is an arbitrary positive function over $\vy$, we see
that the mutual information rate lower bound in
Def.~\ref{def:information:rate:lower:1} is a special case of the
GMI. This means that $\Ilower^{(N)}(\vhat \condcs)$ is a reliable
communication rate that is achievable under mismatched decoding with
the decoding metric $d^{(N)}(\vx,\vy)$ as
in~\eqref{eq:gmi:distance:1}, \ie, a decoder that is matched to the
AB-FSMC and mismatched to the original channel.\footnote{Note that
if there is a function $f(\vy)$ such that $Q(\vx)
  = \sum_{\vy} f(\vy) \Vhat(\vx|\vy)$ for all $\vx$, then $f(\vy)
  \Vhat(\vx|\vy \condcs) / Q(\vx)$ can be seen a conditional pmf of $\vy$
  given $\vx$. In this special case, the mismatched decoder is the ML decoder
  that is matched to the auxiliary channel.}

\begin{Remark}{First Special Case of Information Rate Lower Bound}
  \label{remark:special:case:information:rate:lower:bound:1}

  A special case of $\Ilower^{(N)}(\vhat)$ is obtained by setting
  $\Vhat(\vx|\vy) \defeq Q(\vx) \What(\vy|\vx) / (Q\What)(\vy)$ where $\What$
  is some arbitrary AF-FSMC law. The lower bound then reads
  \begin{align}
    \label{eq:information:rate:lower:1:special}
   \left.
       \Ilower^{(N)}(\Vhat)
    \right|_{\Vhat(\vx | \vy) = \frac{Q(\vx) \What(\vy|\vx)}{(Q\What)(\vy)}}
      &\defeq \frac{1}{2N}
         \sumxy
           Q(\vx \condvs) W(\vy|\vx)
           \log
             \left(
               \frac{\What(\vy|\vx)}
                    {(Q\What)(\vy)}
             \right).
  \end{align}
  This lower bound was also used in~\cite{Arnold:Loeliger:Vontobel:02:1,
    Arnold:Loeliger:Vontobel:Kavcic:Zeng:06:1} and we refer the interested
  reader to these papers for some historical comments. Please note that this
  specialized lower bound is the lower bound that was optimized in the
  preliminary version of this paper~\cite{Sadeghi:Vontobel:Shams:07:1}; this
  is in contrast to the present paper where we will optimize the more general
  lower bound that was given in Def.~\ref{def:information:rate:lower:1}.
\end{Remark}

Ideally, we would like to define the difference of the information
rate upper bound for any AF-FSMC law $\What$ and the information
rate lower bound for any AB-FSMC law $\Vhat$. However, in order to
obtain something tractable for optimization purposes, it turns out
to be expedient to use a $\Vhat$ that is implicitly defined through
$\What$ as in~\eqref{eq:information:rate:lower:1:special}.

\begin{Definition}{Difference Function}
  \label{def:information:rate:diff:1}

  Let $\What(\vy|\vx)$ be the law of some AF-FSMC. The \emph{difference function} is
  defined to be
  \begin{align}
    \label{eq:information:rate:diff:1}
    \Iuldiff^{(N)}(\What \condcs)
       \defeq
         \Iupper(\What)
         -
         \left.
           \Ilower^{(N)}(\Vhat)
         \right|_{\Vhat(\vx| \vy) = \frac{Q(\vx) \What(\vy|\vx)}{(Q\What)(\vy)}}
      &= \frac{1}{2N}
         \sumxy
           Q(\vx \condvs) W(\vy|\vx \condcs)
           \log
             \left(
               \frac{W(\vy|\vx \condcs)}
                    {\What(\vy|\vx \condcs)}
             \right) \\
      &= \frac{1}{2N}
           D_{\vx,\vy}
            \left(
                Q(\vx \condvs)
                W(\vy|\vx \condcs)
              \,\left\lVert\,
                Q(\vx \condvs)
                \What(\vy|\vx \condcs)
              \right.
            \right),
  \end{align}
  with asymptotic version $\Iuldiff(\What) \defeq \lim_{N \to \infty}
  \Iuldiff^{(N)}(\What \condcs)$.
\end{Definition}

Minimizing this difference function would not be significant if we could
directly find the global minimum of the upper bound and the global maximum of
the lower bound. Moreover, a minimized difference between the bounds does not
necessarily mean that the individual upper and lower bounds are optimized.
Nevertheless, we will see that the minimization of the difference function can
give a useful initialization point for the iterative optimization of the upper
and lower bounds. Such an initialization can result in faster-converging
iterative algorithms or tighter upper and/or lower bounds. This is especially
the case for partial response FIR channels, where the AF-FSMC parameters that
minimize the difference function are found in closed form. For fading
channels, using the parameters of the optimized difference function for
initializing the maximization of the lower bound often yields better lower
bounds compared to using other initialization methods. However, using the
parameters of the optimized difference function for initializing the
minimization of the upper bound bound usually does not yield better bounds
compared to using other initialization methods. We refer the reader
to~\secref{sec:num:partial:response} and~\secref{sec:num:fading} for more
details.

Let us conclude this section with yet another special case of the information
rate lower bound.

\begin{Remark}{Second Special Case of Information Rate Lower Bound}
  \label{remark:special:case:information:rate:lower:bound:2}

  In Def.~\ref{def:backward:aux:finite:state:channel:models:1} the only
  requirement on \\ $\vhat( \shat_{\ell-1}, x_{\ell}, \shat_{\ell},
  \vy_{\ell-D_1}^{\ell+D_2})$ was that it is non-negative. Imposing
  additionally the condition that $\sum_{\shat_{\ell}} \vhat( \shat_{\ell-1},
  x_{\ell}, \shat_{\ell}, \vy_{\ell-D_1}^{\ell+D_2}) = 1$ for all
  $\shat_{\ell-1}$, all $x_{\ell}$, and all $\vy_{\ell-D_1}^{\ell+D_2}$, one
  can verify that $\sum_{\vshat, \vx} \vhat(\vshat, \vx, \vy) = 1$ for all
  $\vy$ and all $\shat_{-N}$, see
  App.~\ref{app:proof:remark:special:case:information:rate:lower:bound:2}.
  Therefore, the denominator in~\eqref{eq:backward:auxiliary:channel:law:1} is
  $1$ which means that $\Vhat(\vshat, \vx | \vy) = \vhat(\vshat, \vx, \vy)$.
  Finally, this implies $\Vhat(\vx | \vy) = \vhat(\vx, \vy)$, and so the lower
  bound~\eqref{eq:information:rate:lower:1} reads
  \begin{align}
    \label{eq:information:rate:lower:2}
    \Ilower^{(N)}(\vhat)
      &\defeq
         \frac{1}{2N}
         \sumxy
           Q(\vx)
             W(\vy|\vx)
             \log
               \left(
                 \frac{\vhat(\vx, \vy)}
                      {Q(\vx \condvs)}
               \right).
  \end{align}
\end{Remark}

Let us briefly comment on the number of parameters of AB-FSMCs. From
Def.~\ref{def:backward:aux:finite:state:channel:models:1} it follows that the
number of parameters is $|\setBhat| \times |\setY|^{D_1+D_2+1}$. It is clear
that larger $D_1$ and $D_2$ lead to better lower bounds but also to the need
of estimating and storing more parameters. Given the exponential growth in
$D_1 + D_2$, it is obviously desirable to choose $D_1$ and $D_2$ as large as
needed, yet as small as possible. Empirical evidence shows that $D_1 = 0$ and
$D_2 = 0$ (the smallest possible choice) is sufficient for the
Def.~\ref{def:information:rate:lower:1} lower bound to give good results.
However, non-zero values of $D_1$ and $D_2$ are usually needed for the
Rem.~\ref{remark:special:case:information:rate:lower:bound:2} lower bound to
work well. (Given typical sizes of $|\setY|$, a positive sum of $D_1$ and
$D_2$ yields a \emph{significant} increase in AB-FSMC model parameters.)
Although we will see in \secref{sec:optimize:lower:bound:1} that it is more
difficult to get a handle on the optimization of the
Def.~\ref{def:information:rate:lower:1} lower bound, which is in contrast to
the nice analytical properties of the optimization of the
Rem.~\ref{remark:special:case:information:rate:lower:bound:2} lower bound, the
Def.~\ref{def:information:rate:lower:1} lower bound will be the lower bound of
choice. Indeed, all examples in Secs.~\ref{sec:num:partial:response}
and~\ref{sec:num:fading} will use it with $D_1 = 0$ and $D_2 = 0$.

%% file: optimization_methods1.tex

\section{Optimization Methods and Definitions}
\label{sec:optimization:methods1}

The first main objective of the present paper is the minimization of the
information rate upper bound that was presented in
\secref{sec:information:rate:bound:1}. This optimization takes place over the
set of all AF-FSMCs that have the same trellis section $\setBhat$, \ie, for a
given $\setBhat$ we will optimize over all possible $\bigl\{ \What(\shat |
\shatL, x) \bigr\} \cup \bigl\{ \What(y | \bhat) \bigr\}$.\footnote{In the
  following, we assume that the trellis section $\setBhat$ is such that in the
  relative interior of the set of all possible settings of $\bigl\{
  \What(\shat | \shatL, x) \bigr\} \cup \bigl\{ \What(y | \bhat) \bigr\}$, the
  conditions in \secref{sec:forward:aux:finite:state:channel:models:1} that
  were imposed on AF-FSMCs are fulfilled.} The second main objective of the
present paper is the maximization of the information rate lower
bound that was presented in \secref{sec:information:rate:bound:1}.
This time, the optimization takes place over the set of all AB-FSMCs
that have the same trellis section $\setBhat$, \ie, for a given
$\setBhat$ we will optimize over all possible $\bigl\{
\vhat(\bhat,\vyD) \bigr\}$.\footnote{In the following, we
  assume that the trellis section $\setBhat$ is such that in the relative
  interior of the set of all possible settings of $\bigl\{ \vhat(\bhat,\vyD)
  \bigr\}$, the conditions in
  \secref{sec:backward:aux:finite:state:channel:models:1} that were imposed on
  AB-FSMCs are fulfilled.}

Note that the direct optimization of the information rate upper and lower
bounds is in general intractable. This is mainly due to the fact that
$(Q\What)(\vy \condvs)$, $\What(\vy|\vx \condcs)$, and $\Vhat(\vx|\vy)$ in the
logarithms are not readily decomposable into suitable products. This, in turn,
prevents their optimization in a mathematically tractable manner.

As a way out of this problem, we will use iterative approaches which after
every iteration yield better bounds. As we will see, each of these iterations
can be seen as the optimization of a suitably chosen surrogate function. The
rest of this section is devoted to the presentation of the general ideas; the
mathematical details and the proofs will be treated in later sections.

Although it very often makes sense that the type of the AF-FSMC is chosen such
that it matches the type of the original channel, e.g., that a controllable
AF-FSMC is chosen if the original channel is a controllable FSMC, please note
that such a restriction in choice is not required for the optimization
algorithms that we are about to present.

\subsection{\textbf{Optimization Approach for the Information Rate
                                 Upper Bound}}
\label{subsec:upper:optimization:idea}

\begin{figure}
  \begin{center}
    \epsfig{file=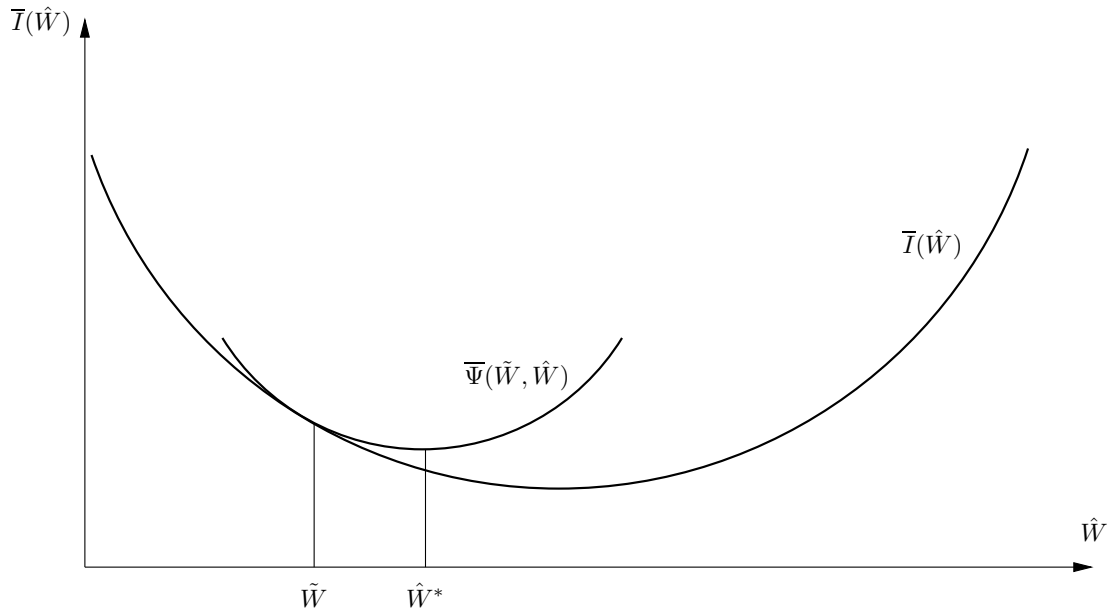, width=0.8\linewidth}
  \end{center}
  \caption{Iterative minimization of the upper bound using a surrogate
    function.}
  \label{fig:em:idea:upper:1}
\end{figure}

The underlying idea for optimizing the upper bound is as follows (see
Fig.~\ref{fig:em:idea:upper:1}):
\begin{itemize}

\item Set the iteration number to $t=0$. Start with an initial AF-FSMC model
  with channel law $\What^{\langle 0\rangle}$.

\item Assume that at the current iteration we have found an AF-FSMC model with
  channel law $\What^{\langle t\rangle}$ with the corresponding
  information rate upper bound $\Iupper(\What^{\langle t\rangle})$.

\item At the next iteration we would like to find a ``better'' AF-FSMC model
  with channel law $\What^{\langle t+1\rangle}$, which results in a tighter
  upper bound. To this end, we introduce a surrogate function
  $\Psiupper(\What^{\langle t\rangle}, \What)$ which locally approximates
  $\Iupper(\What)$ around $\What = \What^{\langle t\rangle}$. More explicitly,
  the surrogate function is chosen such that $\Psiupper(\What^{\langle
    t\rangle}, \What^{\langle t\rangle}) = \Iupper(\What^{\langle t\rangle})$
  and such that it is never below the upper bound, \ie,
  $\Psiupper(\What^{\langle t\rangle}, \What) \geq \Iupper(\What)$ for all
  $\What$.

\item Let us assume for the moment that we can find such a surrogate function
  $\Psiupper(\What^{\langle t\rangle}, \What)$ that can be easily minimized
  and let us call $\What^{\langle t+1\rangle}$ the channel law that achieves
  the minimum of $\Psiupper(\What^{\langle t\rangle}, \What)$ over $\What$
  (one such function is given in Section~\ref{sec:optimize:upper:bound:1}).

\item Using this approach, we obtain a new channel law $\What^{\langle
    t+1\rangle}$ that does not increase the upper bound, \ie,
  $\Iupper(\What^{\langle t+1\rangle}) \le \Psiupper(\What^{\langle t\rangle},
  \What^{\langle t+1\rangle}) \le \Psiupper(\What^{\langle t\rangle},
  \What^{\langle t\rangle}) = \Iupper(\What^{\langle t\rangle})$.

\item We increase $t$ by one and then repeat the above procedure until some
  termination criterion is met.

\end{itemize}

It is important to note that unlike the idealistic situation depicted in
Fig.~\ref{fig:em:idea:upper:1}, $\Iupper(\What)$ is in general not a unimodal
function of $\What$ and therefore, $\Iupper(\What)$ can have multiple local
minima. Later in \secref{sec:optimize:upper:bound:1}, we will study the
convergence properties of the surrogate function for the upper bound.

We will use the following nomenclature: a surrogate function that is never
below a certain function will be called a \emph{never-below} surrogate
function. On the other hand, a surrogate function that is never above a
certain function will be called a \emph{never-above} surrogate function.

\subsection{\textbf{Optimization Approach for the Difference Function}}
\label{subsec:diff:optimization:idea}

The underlying idea for minimizing the difference function is very much
similar to minimizing the information rate upper bound. Of course, we will
need a different class of never-below surrogate functions: the surrogate
function at the $t$-th iteration will be called $\Psidiff(\What^{\langle
  t\rangle}, \What)$.

\subsection{\textbf{Optimization Approach for the Information Rate
                                 Lower Bound}}
\label{subsec:lower:optimization:idea}

Plugging the definition~\eqref{eq:backward:auxiliary:channel:law:1} of $\Vhat$
in terms of $\vhat$ into the information rate lower bound
\eqref{eq:information:rate:lower:1}, we obtain
\begin{align}
  \Ilower^{(N)}(\vhat \condcs)
    &\defeq
       \frac{1}{2N}
       \sumxy
         Q(\vx \condvs)
           W(\vy|\vx \condcs)
           \log
             \left(
               \frac{\sum_{\vshat} \vhat(\vshat, \vx, \vy)}
                    {Q(\vx \condvs) \,
                     \sum_{\vshat', \vx'}
                       \vhat(\vshat', \vx', \vy)}
             \right).
    \label{eq:information:rate:lower:1:small:v}
\end{align}

In the second special case of the information rate lower bound,
cf.~Rem.~\ref{remark:special:case:information:rate:lower:bound:2},
the summation in the denominator of the logarithm
in~\eqref{eq:information:rate:lower:1:small:v} is always one and
therefore only the numerator of the logarithm depends on $\vhat$.
Maximizing $\Ilower^{(N)}(\vhat \condcs)$ is then very similar to
minimizing the information rate upper bound and the difference
function. This similarity comes from the fact that in both the
\emph{negative} upper bound and the \emph{negative} difference
function, only the numerator of the logarithm depends on the
auxiliary FSMC parameterization. Correspondingly, we are using a
never-above surrogate function for the present optimization. Our
practical experience shows that non-zero values for $D_1$ and $D_2$
must be used in order to obtain tight lower bounds. Given that the
size of the output alphabet is usually not too small, this means
that the AB-FSMC has many parameters, which is somewhat undesirable.

In the general setup of the lower bound, \ie, when we are not assuming the
restriction of Rem.~\ref{remark:special:case:information:rate:lower:bound:2},
the choice $D_1 = 0$ and $D_2 = 0$ seems to be sufficient. In this case
however, the numerator \emph{and} the denominator of the logarithm
in~\eqref{eq:information:rate:lower:1:small:v} depend on $\vhat$ and we have
not been able to derive a surrogate function for which we can analytically
guarantee that it is never above the lower bound. Therefore, instead of using
a surrogate function $\Psilower(\vhat^{\langle t\rangle}, \vhat)$ for which we
can prove that it is never above $\Ilower^{(N)}(\vhat)$, we will use a
surrogate function $\Psilower(\vhat^{\langle t\rangle}, \vhat)$ for which we
can prove the following two properties: firstly, $\Psilower(\vhat^{\langle
  t\rangle}, \vhat)$ equals $\Ilower(\vhat)$ when $\vhat = \vhat^{\langle
  t\rangle}$, and secondly, the gradient of $\Psilower(\vhat^{\langle
  t\rangle}, \vhat)$ w.r.t.~$\vhat$ equals the gradient of $\Ilower(\vhat)$
w.r.t.~$\vhat$ when $\vhat = \vhat^{\langle t\rangle}$. Such a surrogate
function guarantees that we will be moving in a good direction, however, it
does not guarantee that we obtain a non-decreasing lower bound after each
iteration. Note though that our surrogate function of choice will have a
parameter $\gamma$ which enables one to control the ``aggressiveness'' of the
optimization step. Adaptively setting this $\gamma$ parameter allows one to
have a non-decreasing lower bound after every step.

\subsection{\textbf{Some Key Quantities for Iterative Optimizations}}
\label{subsec:key:quantity:optimization}

The following quantities will be repeatedly used for iterative optimizations
in later sections. These quantities are evaluated at the currently found
AF-FSMC channel law $\Wtilde \defeq \What^{\langle t\rangle}$ or the currently
found AB-FSMC channel law $\vtilde \defeq \vhat^{\langle t\rangle}$.
Therefore, we use the superscript \ $\tilde{}$ \ to denote these quantities.

\begin{Definition}{$\Ttilde^{(N)}_1(\bhat)$, $\Ttilde^{(N)}_2(\bhat,y)$, and
    $\Ttilde^{(N)}_2(\bhat,\vyD)$}
  \label{def:ttilde:1:2}

  \mbox{}

  \begin{itemize}

  \item In the case of the upper bound and difference function minimization,
    define
    \begin{align}
      \label{eq:ttilde1}
      \Ttilde^{(N)}_1(\bhat)
        &\defeq
           \sum_{\vy}
             (QW)(\vy \condvs)
             \left(
               \frac{1}{2N}
               \ellsum
                 \sum_{\bhat_{\ell}}
                   \Ptilde(\bhat_{\ell} | \vy \condcs)
                   \left[
                     \bhat_{\ell} = \bhat
                   \right]
             \right), \\
      \label{eq:ttilde2}
      \Ttilde^{(N)}_2(\bhat,y)
        &
         \defeq
           \sum_{\vy}
             (QW)(\vy \condvs)
             \left(
               \frac{1}{2N}
               \ellsum
                 \sum_{\bhat_{\ell}}
                   \Ptilde(\bhat_{\ell} | \vy \condcs)
                   \left[
                     \bhat_{\ell} = \bhat
                   \right]
                   \left[
                     y_{\ell} = y
                   \right]
             \right),
    \end{align}
    with asymptotic versions $\Ttilde_1(\bhat) \defeq \lim_{N \to \infty}
    \Ttilde^{(N)}_1(\bhat)$ and $\Ttilde_2(\bhat,y) \defeq \lim_{N \to \infty}
    \Ttilde^{(N)}_2(\bhat,y)$. Here, $\Ptilde(\bhat_{\ell} | \vy)$ is defined in
    the spirit of Rem.~\ref{remark:induced:pmfs:1}, and $[\bhat_{\ell} = \bhat]$
    is defined to equal one if $\bhat_{\ell} = \bhat$ and to equal zero
    otherwise, etc. Note that $\Ttilde^{(N)}_1(\bhat) = \sum_y
    \Ttilde^{(N)}_2(\bhat,y)$ for all $\bhat$.

  \item In the case of the lower bound maximization, define
    \begin{align}
      \label{eq:ttilde2:mod}
      \Ttilde^{(N)}_2(\bhat,\vyD)
        &
         \defeq
           \sum_{\vy}
             (QW)(\vy \condvs)
             \left(
               \frac{1}{2N}
               \ellsum
                 \sum_{\bhat_{\ell}}
                   \Vtilde(\bhat_{\ell} | \vy \condcs)
                   \left[
                     \bhat_{\ell} = \bhat
                   \right]
                   \left[
                     \vy_{\ell-D_1}^{\ell+D_1} = \vyD
                   \right]
             \right),
    \end{align}
    with asymptotic version $\Ttilde_2(\bhat,\vyD) \defeq \lim_{N \to \infty}
    \Ttilde^{(N)}_2(\bhat,\vyD)$.

  \end{itemize}
\end{Definition}

\begin{Algorithm}{Numerical/Stochastic Computation of
    $\Ttilde^{(N)}_1(\bhat)$, $\Ttilde^{(N)}_2(\bhat,y)$, and
    $\Ttilde^{(N)}_2(\bhat,\vyD)$}
  \label{alg:numerical:Ttilde:1:2}

  Direct evaluation of the above quantities for all possible realizations of
  channel output $\vy$ is prohibitive. However, there are efficient stochastic
  procedures to numerically approximate them to a precision that is good
  enough for practical purposes. Namely, we replace the ensemble average with
  the time average and proceed as follows:

\begin{itemize}

\item Generate a sequence of channel output $\vy$ of length $2N$ in the
  \emph{original} channel according to its pmf $(QW)(\vy \condvs)$.

\item For the generated output sequence $\vy$ in the original channel, compute
  $\Ptilde(\bhat_{\ell} | \vy \condcs) = \Ptilde(\shat_{\ell-1}, x_{\ell}, \shat_{\ell}| \vy, \shat_{-N})$ for all $\ell \in \sIN$ by applying the BCJR
  algorithm~\cite{BCJR1974} to the \emph{AF-FSMC model}. (Note that
  $\bhat_\ell$ includes the channel input $x_\ell$.)

\item Whenever $\bhat_{\ell} = \bhat$, add the computed $\Ptilde(\bhat_{\ell}
  | \vy \condcs)$ to $\Ttilde^{(N)}_1(\bhat)$. Similarly, whenever
  $\bhat_{\ell} = \bhat$ and $y_{\ell} = y$, add the computed
  $\Ptilde(\bhat_{\ell} | \vy \condcs)$ to $\Ttilde^{(N)}_2(\bhat,y)$.

\end{itemize}
(Of course, in the case of the lower bound optimization, $\Ptilde(\bhat_{\ell}
| \vy \condcs)$, $\Ttilde^{(N)}_2(\bhat,y)$, and ``AF-FSMC'' have to be
replaced by $\Vtilde(\bhat_{\ell} | \vy \condcs)$,
$\Ttilde^{(N)}_2(\bhat,\vyD)$, and ``AB-FSMC'', respectively, in the above
sentences.) Because of the assumptions that we made in
\secref{sec:channel:models:1}, the resulting estimates are equal to
$\Ttilde^{(N)}_1(\bhat)$, $\Ttilde^{(N)}_2(\bhat,y)$, and
$\Ttilde^{(N)}_2(\bhat,\vyD)$ with probability $1$ as $N\to \infty$.

\end{Algorithm}

\begin{Definition}{$\Ttilde^{(N)}_3(\bhat)$, $\Ttilde^{(N)}_4(\bhat,y)$, and
    $\Ttilde^{(N)}_4(\bhat,\vyD)$}
  \label{def:ttilde:3:4}

  \mbox{}

  \begin{itemize}

  \item In the case of the upper bound and difference function minimization,
    define
    \begin{align}
      \label{eq:ttilde3}
      \Ttilde^{(N)}_3(\bhat)
        &\defeq
          \sumxy
            Q(\vx)
            W(\vy | \vx \condcs)
            \left(
              \frac{1}{2N}
                \ellsum
                \sum_{\bhat_{\ell}}
                  \Ptilde(\bhat_{\ell}|\vx,\vy \condcs)
                  \left[
                    \bhat_{\ell} = \bhat
                  \right]
                  \left[
                    x_{\ell} = x(\bhat)
                  \right]
            \right), \\
      \label{eq:ttilde4}
      \Ttilde^{(N)}_4(\bhat, y)
        &\defeq
             \sumxy
               Q(\vx)
               W(\vy | \vx \condcs)
               \left(
                 \frac{1}{2N}
                 \ellsum
                   \sum_{\bhat_{\ell}}
                     \Ptilde(\bhat_{\ell}|\vx,\vy \condcs)
                     \left[
                       \bhat_{\ell} = \bhat
                     \right]
                     \left[
                       x_{\ell} = x(\bhat)
                     \right]
                     \left[
                       y_{\ell} = y
                     \right]
               \right),
    \end{align}
    with asymptotic versions $\Ttilde_3(\bhat) \defeq \lim_{N \to \infty}
    \Ttilde^{(N)}_3(\bhat)$ and $\Ttilde_4(\bhat,y) \defeq \lim_{N \to \infty}
    \Ttilde^{(N)}_4(\bhat,y)$. Here, $\Ptilde(\bhat_{\ell} | \vx, \vy)$ is
    defined in the spirit of Rem.~\ref{remark:induced:pmfs:1} and
    $\left[x_{\ell} = x(\bhat)\right]$ is used to emphasize that the summands
    of $\Ttilde^{(N)}_3(\bhat)$ and $\Ttilde^{(N)}_4(\bhat,y)$ are non-zero
    only when the $\ell$-th channel input $x_{\ell}$ in $\vx$ is compatible
    with the input symbol of the AF-FSMC branch $\bhat$, denoted by
    $x(\bhat)$.  Note that $\Ttilde^{(N)}_3(\bhat) = \sum_y
    \Ttilde^{(N)}_4(\bhat,y)$ for all $\bhat$.

  \item In the case of the lower bound maximization, define
    \begin{align}
      \label{eq:ttilde4:mod}
      \Ttilde^{(N)}_4(\bhat, \vyD)
        &\defeq
             \sumxy
               Q(\vx)
               W(\vy | \vx \condcs)
               \left(
                 \frac{1}{2N}
                 \ellsum
                   \sum_{\bhat_{\ell}}
                     \Vtilde(\bhat_{\ell}|\vx,\vy \condcs)
                     \left[
                       \bhat_{\ell} = \bhat
                     \right]
                     \left[
                       x_{\ell} = x(\bhat)
                     \right]
                     \left[
                       \vy_{\ell-D_1}^{\ell+D_2} = \vyD
                     \right]
               \right),
    \end{align}
    with asymptotic version $\Ttilde_4(\bhat,\vyD) \defeq \lim_{N \to \infty}
    \Ttilde^{(N)}_4(\bhat,\vyD)$.

  \end{itemize}
\end{Definition}

\begin{Algorithm}{Numerical/Stochastic Computation of $\Ttilde^{(N)}_3(\bhat)$,
    $\Ttilde^{(N)}_4(\bhat,y)$, and $\Ttilde^{(N)}_4(\bhat,\vyD)$}
  \label{alg:numerical:Ttilde:3:4}

  We proceed as follows:
  \begin{itemize}

  \item Generate a sequence of channel input $\vx$ of length $2N$ according to
    the source distribution $Q(\vx \condvs)$.

  \item Generate a sequence of channel output $\vy$ of length $2N$ in the
    \emph{original} channel according to its channel law $W(\vy|\vx \condcs)$.

  \item For the generated input and output sequences $\vx$ and $\vy$, compute
    $\Ptilde(\bhat_{\ell} | \vx, \vy \condcs) = \Ptilde(\shat_{\ell-1},
    x_{\ell},\shat_{\ell}| \vx, \vy \condcs)$ by applying the BCJR
    algorithm~\cite{BCJR1974} to the \emph{AF-FSMC} channel. Obviously,
    $\Ptilde(\bhat_{\ell} | \vx, \vy \condcs)$ will be zero if the $\ell$-th
    element of $\vx$ does not equal $x(\bhat_\ell)$.

  \item Whenever $\bhat_{\ell} = \bhat$ and $x_{\ell} = x(\bhat)$, add the
    computed $\Ptilde(\bhat_{\ell} | \vx, \vy \condcs)$ to
    $\Ttilde^{(N)}_3(\bhat)$. Similarly, whenever $\bhat_{\ell} = \bhat$,
    $x_{\ell} = x(\bhat)$, and $y_{\ell} = y$, add the computed
    $\Ptilde(\bhat_{\ell} | \vx, \vy \condcs)$ to
    $\Ttilde^{(N)}_4(\bhat,y)$,

  \end{itemize}
  (Of course, in the case of the lower bound optimization,
  $\Ptilde(\bhat_{\ell} | \vx, \vy \condcs)$, $\Ttilde^{(N)}_4(\bhat,y)$, and
  ``AF-FSMC'' have to be replaced by $\Vtilde(\bhat_{\ell} | \vx, \vy
  \condcs)$, $\Ttilde^{(N)}_4(\bhat,\vyD)$, and ``AB-FSMC'', respectively, in
  the above sentences.) Because of the assumptions that we made in
  \secref{sec:channel:models:1}, the resulting estimates are equal to
  $\Ttilde^{(N)}_3(\bhat)$, $\Ttilde^{(N)}_4(\bhat,y)$, and
  $\Ttilde^{(N)}_4(\bhat,\vyD)$ with probability $1$ as $N\to \infty$.
\end{Algorithm}

%% file: optimize_upper_bound1.tex

\section{Optimizing the Information Rate Upper Bound}
\label{sec:optimize:upper:bound:1}

Assume that at the current iteration we have found an AF-FSMC model with
channel law $\What^{\langle t\rangle}$ and the corresponding information rate
upper bound $\Iupper(\What^{\langle t\rangle})$. In order to simplify
notation, we will use $\Wtilde$ instead of $\What^{\langle t\rangle}$. Let
\begin{align}
  \Psiupper^{(N)}(\Wtilde, \What \condcs)
    &\defeq
       \Iupper^{(N)}(\What \condcs)
     + \frac{1}{2N}
       \sum_{\vy}
         (QW)(\vy \condvs) \,
         D_{\vbhat}
           \left(
               \Ptilde(\vbhat | \vy \condcs)
             \left\lVert
               \Phat(\vbhat | \vy \condcs)
             \right.
           \right)
\end{align}
be the surrogate function for the upper bound $\Iupper^{(N)}(\What \condcs)$
and let its asymptotic version be $\Psiupper(\Wtilde, \What) \defeq \lim_{N
  \to \infty} \Psiupper^{(N)}(\Wtilde, \What \condcs)$. In the surrogate
function, the conditional pmfs $\Ptilde(\vbhat | \vy \condcs)$ and
$\Phat(\vbhat | \vy \condcs)$ are induced by the channel laws $\Wtilde$ and
$\What$, respectively (cf.~Rem.~\ref{remark:induced:pmfs:1}).

\begin{Lemma}{Important Properties of $\Psiupper$}
  \label{lemma:properties:upper}

  We recognize the following properties of the surrogate function:
  \begin{enumerate}

  \item For any $\What$, the function $\Psiupper^{(N)}(\Wtilde, \What
    \condcs)$ is never below $\Iupper^{(N)}(\What)$. Moreover, for $\What=
    \Wtilde$, $\Psiupper^{(N)}(\Wtilde, \What \condcs)$ equals
    $\Iupper^{(N)}(\Wtilde)$.

  \item The function $\Psiupper^{(N)}(\Wtilde, \What \condcs)$ can be
    simplified to
    \begin{align}
      \Psiupper^{(N)}(\Wtilde, \What \condcs)
        &= \overline{c}^{(N)}(\Wtilde \condcs)
           -
           \sum_{\bhat}
             \log
             \left(
               \What(\shat | \shatL, x)
             \right)
             \Ttilde^{(N)}_1(\bhat)
           -
           \sum_{\bhat}
           \sum_{y}
             \log
               \left(
                 \What(y | \bhat)
               \right)
             \Ttilde^{(N)}_2(\bhat,y),
               \label{eq:upper:surrogate:3}
    \end{align}
    where $\overline{c}^{(N)}(\Wtilde \condcs)$ is independent of $\What$ and
    where $\Ttilde^{(N)}_1(\bhat)$ and $\Ttilde^{(N)}_2(\bhat,y)$ were defined
    in \eqref{eq:ttilde1} and \eqref{eq:ttilde2}, respectively. Note that
    $\shatL$ denotes the previous (or left) state of $\bhat = (\shatL, x,
    \shat)$.

  \item The function $\Psiupper^{(N)}(\Wtilde, \What \condcs)$ is convex in
    $\What$, \ie, in $\bigl\{ \What(\shat | \shatL, x) \bigr\} \cup \bigl\{
    \What(y | \bhat) \bigr\}$.

  \end{enumerate}
\end{Lemma}

\begin{Proof}
  See Appendix~\ref{app:upper:property}.
\end{Proof}

\begin{Lemma}{Minimizing the Surrogate Function $\Psiupper$}
  \label{lemma:minimize:upper}

  Assume that we are at iteration $t$ and that $\Wtilde = \What^{\langle
    t\rangle}$. The $\What$ that minimizes $\Psiupper(\Wtilde, \What)$ at the
  next iteration is given by
  \begin{alignat}{2}
  \label{eq:update:What:upper:1}
    \What^{*}(\shat | \shatL, x)
      &\ \propto \
          \Ttilde_1(\bhat)
            &\quad
            &(\text{for all $\bhat = (\shatL, x, \shat) \in \setBhat$}), \\
  \label{eq:update:What:upper:2}
    \What^{*}(y | \bhat)
      &\ \propto \
          \Ttilde_2(\bhat,y)
            &\quad
            &(\text{for all $\bhat \in \setBhat$ and all $y \in \setY$}),
  \end{alignat}
  where the proportionality
  constants are chosen such that the constraints
  \begin{align}
    \label{eq:upper:const1}
    \sum_{\shat}
      \What(\shat | \shatL, x)
      &= 1
           \quad (\text{for all $(\shatL, x) \in \setShat \times \setX$}), \\
    \label{eq:upper:const2}
    \sum_{y}
      \What(y | \bhat)
      &= 1
           \quad (\text{for all $\bhat \in \setBhat$})
  \end{align}
  are fulfilled.
\end{Lemma}

\begin{Proof}
  See Appendix~\ref{app:upper:minimize}.
\end{Proof}

We observe that the update equations in Lemma~\ref{lemma:minimize:upper} are
not dissimilar to the update equations for the Baum-Welch
algorithm~\cite{Baum1972}, which was proposed for parameter estimation in
hidden Markov models (HMMs).\footnote{The Baum-Welch algorithm is an early
  instance of the EM algorithm. The EM theory was later generalized in 1977 by
  Dempster, Laird, and Rubin \cite{dempster:laird:rubin:1977}.} In contrast to
the Baum-Welch algorithm, here we are also averaging over $\vy$. Note that
some simplifications in the update equations arise in the case where $\shatL$
and $x$ determine the next state $\shat$.

With these results in our hand, the proposed iterative optimization of the
upper bound is given as follows.

\begin{Algorithm}{Iterative Optimization of the Information Rate Upper Bound}
  \label{alg:iterative:upper}

  \mbox{}

  \begin{enumerate}

  \item For all $\bhat \in \setBhat$ and all $y \in \setY$, set
    $\What^{\langle 0\rangle}(\shat | \shatL, x)$ and $\What^{\langle
      0\rangle}(y | \bhat)$ to some positive initial values.

  \item Set $t = 0$.

  \item \label{item:upper:bound:loop} As long as desired or until convergence,
    repeat the following steps.
    \begin{enumerate}

    \item \label{item:upper:bound:inner:loop} Set $\Wtilde \defeq
      \What^{\langle t\rangle}$ and use
      Algorithm~\ref{alg:numerical:Ttilde:1:2} to estimate $\bigl\{
      \Ttilde^{(N)}_1(\bhat) \bigr\}$ and $\bigl\{ \Ttilde^{(N)}_2(\bhat,y)
      \bigr\}$.

  \item Update $\bigl\{ \What^{\langle t+1\rangle}(\shat | \shatL, x) \bigr\}$
    and $\bigl\{ \What^{\langle t+1\rangle}(y | \bhat) \bigr\}$ according to
    \eqref{eq:update:What:upper:1}-\eqref{eq:upper:const2}.

    \item Increase $t$ by one.

    \item Go back to step~\ref{item:upper:bound:inner:loop}.

    \end{enumerate}

  \end{enumerate}
\end{Algorithm}

\begin{Lemma}{Convergence Properties of the Iterative Optimization}
  \label{lemma:convergence:upper}

  For any finite $N$, all limit points of an instance of
  Algorithm~\ref{alg:iterative:upper} are stationary points of
  $\Iupper^{(N)}(\What)$. Similarly, for $N \to \infty$, all limit points of
  an instance of Algorithm~\ref{alg:iterative:upper} are stationary points of
  $\Iupper(\What)$.

\end{Lemma}

\begin{Proof}
  This can be proven by using Wu's convergence results for the
  EM algorithm~\cite{Wu:83:1}. In the case $N \to \infty$, the required
  continuity of $\Psiupper(\Wtilde, \What)$ in $\Wtilde$ and $\What$ follows
  from results in~\cite{LeGland:Mevel:00:2, Mevel:Finesso:04:1}.
\end{Proof}

Note that in general, $\Iupper^{(N)}(\What)$ and $\Iupper(\What)$ can have
local maxima where the algorithm can get stuck. However, because of the
stochastic evaluation of $\bigl\{ \Ttilde^{(N)}_1(\bhat) \bigr\}$ and $\bigl\{
\Ttilde^{(N)}_2(\bhat,y) \bigr\}$, these local maxima tend to be instable.

%% file: optimize_diff_bound1.tex

\section{Optimizing the Difference Function}
\label{sec:optimize:diff:bound:1}

Assume that at the current iteration we have found an AF-FSMC with channel law
$\What^{\langle t\rangle}$ and with corresponding difference function value
$\Iuldiff(\What^{\langle t\rangle})$. In order to simplify notation we will
use $\Wtilde$ instead of $\What^{\langle t\rangle}$. Let
\begin{align}
  \Psidiff^{(N)}(\Wtilde, \What \condcs)
    &\defeq
       \Iuldiff^{(N)}(\What \condcs)
     + \frac{1}{2N}
       \sumx
         \sumy
           Q(\vx)
           W(\vy|\vx \condcs) \,
           D_{\vbhat}
             \left(
                 \Ptilde(\vbhat | \vx, \vy \condcs)
               \left\lVert
                 \Phat(\vbhat | \vx, \vy \condcs)
               \right.
             \right)
\end{align}
be the surrogate function for the difference function $\Iuldiff^{(N)}(\What)$
and let its asymptotic version be $\Psidiff(\Wtilde, \What) \defeq \lim_{N \to
  \infty} \Psidiff^{(N)}(\Wtilde, \What \condcs)$. In the surrogate function,
the conditional pmfs $\Ptilde(\vbhat | \vx, \vy \condcs)$ and $\Phat(\vbhat |
\vx, \vy \condcs)$ are induced by the channel laws $\Wtilde$ and $\What$,
respectively (cf.~Rem.~\ref{remark:induced:pmfs:1}).

\begin{Lemma}{Important Properties of $\Psidiff$}\label{lemma:properties:diff}

  We recognize the following properties of the surrogate function:
  \begin{enumerate}

  \item For any $\What$, the function $\Psidiff^{(N)}(\Wtilde, \What
    \condcs)$ is never below $\Iuldiff^{(N)}(\What)$. Moreover, for $\What=
    \Wtilde$, $\Psidiff^{(N)}(\Wtilde, \What \condcs)$ equals
    $\Iuldiff^{(N)}(\Wtilde)$.

  \item The function $\Psidiff^{(N)}(\Wtilde, \What \condcs)$
    can be simplified to
    \begin{align}
      \Psidiff^{(N)}(\Wtilde, \What \condcs)
        &= c^{(N)}_{\Delta}(\Wtilde \condcs)
           -
           \sum_{\bhat}
             \log
               \left(
                 \What(\shat | \shatL, x)
               \right)\Ttilde^{(N)}_3(\bhat)
           -
           \sum_{\bhat}
             \sum_{y}
               \log
                 \left(
                   \What(y | \bhat)
                 \right)
               \Ttilde^{(N)}_4(\bhat,y),
                 \label{eq:diff:surrogate:3}
    \end{align}
    where $c^{(N)}_{\Delta}(\Wtilde \condcs)$ is independent of $\What$ and
    where $\Ttilde^{(N)}_3(\bhat)$ and $\Ttilde^{(N)}_4(\bhat,y)$ were defined
    in \eqref{eq:ttilde3} and \eqref{eq:ttilde4}, respectively. Note that
    $\shatL$ denotes the previous (or left) state of $\bhat = (\shatL, x,
    \shat)$.

  \item The function $\Psidiff^{(N)}(\Wtilde, \What \condcs)$ is convex in
    $\What$, \ie, in $\bigl\{ \What(\shat | \shatL, x) \bigr\} \cup \bigl\{
    \What(y | \bhat) \bigr\}$.

  \end{enumerate}
\end{Lemma}

\begin{Proof}
  See Appendix~\ref{app:diff:property}.
\end{Proof}

\begin{Lemma}{Minimizing the Surrogate Function $\Psidiff$}
  \label{lemma:minimize:diff}

  Assume that we are at iteration $t$ and that $\Wtilde = \What^{\langle
    t\rangle}$. The $\What$ that minimizes $\Psidiff(\Wtilde, \What)$ at the
  next iteration is given by
  \begin{alignat}{2}
    \label{eq:update:What:diff:1}
    \What^{*}(\shat | \shatL, x)
      &\ \propto \
         \Ttilde_3(\bhat),
           &\quad
           &(\text{for all $\bhat = (\shatL, x, \shat) \in \setBhat$}), \\
    \label{eq:update:What:diff:2}
    \What^{*}(y | \bhat)
      &\ \propto \
         \Ttilde_4(\bhat,y),
           &\quad
           &(\text{for all $\bhat \in \setBhat$ and all $y \in \setY$}),
  \end{alignat}
  where the proportionality constants are chosen such that the constraints
  \begin{align}
    \label{eq:diff:const1}
    \sum_{\shat}
      \What(\shat | \shatL, x)
      &= 1
           \quad (\text{for all $(\shatL, x) \in \setShat \times \setX$}), \\
    \label{eq:diff:const2}
    \sum_{y}
      \What(y | \bhat)
      &= 1
           \quad (\text{for all $\bhat \in \setBhat$})
  \end{align}
  are fulfilled.
\end{Lemma}

\begin{Proof}
  The proof is very similar to the proof of Lemma~\ref{lemma:minimize:upper}.
  We leave the details to the reader.
\end{Proof}

\begin{Remark}{Simplification for Data-Controllable
    AF-FSMCs}\label{rem:fir:simplify}

  In case the \emph{AF-FSMC} is a data-controllable FSMC, the input sequence
  $\vx$ determines the state sequence $\vshat$ and therefore also the branch
  sequence $\vbhat$. This leads to simplifications in the computation of
  $\Ttilde_3(\bhat)$ and $\Ttilde_4(\bhat,y)$.

  In particular, if the \emph{AF-FSMC} is a partial response channel with
  memory order $\hat M$, the state at time index $\ell-1$ is $\shat_{\ell-1} =
  \vx_{\ell-\hat M}^{\ell-1}$. With this we obtain $\Ttilde_3(\bhat) = 1$ for
  all branches $\bhat \in \setBhat$ and so $\What^{*}(\shat | \shatL, x) = 1$
  for all $\bhat = (\shatL, x, \shat) \in \setBhat$. Moreover, for any $\bhat
  \in \setBhat$ and $y \in \setY$ we have
  \begin{align*}
    \Ttilde_4(\bhat, y)
      &= \left.
           \sum_{\vx_{-\infty}^{-\hat M - 1}}
             Q\left( \vx_{-\infty}^{0} \right)
             W\left( y_{0} | \vx_{-\infty}^{0} \right)
         \right|_{\left(
                    \vx_{-\hat M}^{-1},
                    x_{0},
                    \vx_{-\hat M+1}^{0}
                  \right)
                   = \bhat, \
                  y_{0} = y},
  \end{align*}
  with the corresponding formula for $\What^{*}(y | \bhat)$. In particular, if
  the \emph{original} channel is a partial response channel with memory order
  $M$ then we can simplify this expression even further. Indeed, if $M > \hat
  M$ then
  \begin{align*}
    \Ttilde_4(\bhat, y)
      &= \left.
           \sum_{\vx_{-M}^{-\hat M - 1}}
             Q\left( \vx_{-M}^{0} \right)
             W\left( y_{0} | \vx_{-M}^{0} \right)
         \right|_{\left(
                    \vx_{-\hat M}^{-1},
                    x_{0},
                    \vx_{-\hat M+1}^{0}
                  \right)
                   = \bhat, \
                  y_{0} = y},
  \end{align*}
  and if $M \leq \hat M$ then
  \begin{align*}
    \Ttilde_4(\bhat, y)
      &= \left.
           Q\left( \vx_{-M}^{0} \right)
           W\left( y_{0} | \vx_{-M}^{0} \right)
         \right|_{\left(
                    \vx_{-\hat M}^{-1},
                    x_{0},
                    \vx_{-\hat M+1}^{0}
                  \right)
                   = \bhat, \
                  y_{0} = y}.
  \end{align*}
\end{Remark}

With these results in our hands, the proposed iterative optimization of the
difference function looks as follows.

\begin{Algorithm}{Iterative Optimization of the Difference Function}
  \label{alg:iterative:diff}

  \mbox{}

  \begin{enumerate}

  \item For all $\bhat \in \setBhat$ and all $y \in \setY$, set
    $\What^{\langle 0\rangle}(\shat | \shatL, x)$ and $\What^{\langle
      0\rangle}(y | \bhat)$ to some positive initial values.

  \item Set $t = 0$.

  \item \label{item:diff:bound:loop} As long as desired or until convergence,
    repeat the following steps.
    \begin{enumerate}

    \item \label{item:diff:bound:inner:loop} Set $\Wtilde \defeq
      \What^{\langle t\rangle}$ and use
      Algorithm~\ref{alg:numerical:Ttilde:3:4} to estimate $\bigl\{
      \Ttilde^{(N)}_3(\bhat) \bigr\}$ and $\bigl\{ \Ttilde^{(N)}_4(\bhat,y)
      \bigr\}$.

    \item Update $\bigl\{ \What^{\langle t+1\rangle}(\shat | \shatL, x)
      \bigr\}$ and $\bigl\{ \What^{\langle t+1\rangle}(y | \bhat) \bigr\}$
      according to \eqref{eq:update:What:diff:1}-\eqref{eq:diff:const2}.

    \item Increase $t$ by one.

    \item Go back to step~\ref{item:diff:bound:inner:loop}.

    \end{enumerate}

  \end{enumerate}
\end{Algorithm}

\begin{Lemma}{Convergence Properties of the Iterative Optimization}
  \label{lemma:convergence:diff}

  For any finite $N$, all limit points of an instance of
  Algorithm~\ref{alg:iterative:diff} are stationary points of
  $\Iuldiff^{(N)}(\What)$. Similarly, for $N \to \infty$, all limit points of
  an instance of Algorithm~\ref{alg:iterative:diff} are stationary points of
  $\Iuldiff(\What)$.

\end{Lemma}

\begin{Proof}
  This can be proven by using Wu's convergence results for the
  EM algorithm~\cite{Wu:83:1}. In the case $N \to \infty$, the required
  continuity of $\Psidiff(\Wtilde, \What)$ in $\Wtilde$ and $\What$ follows
  from results in~\cite{LeGland:Mevel:00:2, Mevel:Finesso:04:1}.
\end{Proof}

%% file: optimize_lower_bound1.tex

\section{Optimizing the Information Rate Lower Bound}
\label{sec:optimize:lower:bound:1}

Assume that at the current iteration we have found an AB-FSMC model with
channel law $\vhat^{\langle t\rangle}$ and the corresponding information rate
lower bound $\Ilower(\vhat^{\langle t\rangle})$. In order to simplify
notation, we will use $\vtilde$ instead of $\vhat^{\langle t\rangle}$. The
derivation of a suitable surrogate function will be somewhat longer than the
corresponding derivations in the case of the upper bound and the difference
function.

We start by plugging the
definition~\eqref{eq:backward:auxiliary:channel:law:1}-\eqref{eq:backward:auxiliary:channel:law:1:part:3}
of $\Vhat$ in terms of $\vhat$ into the information rate lower bound
\eqref{eq:information:rate:lower:1} and obtain
\begin{align}
  \Ilower^{(N)}(\vhat \condcs)
    &\defeq
       \frac{1}{2N}
       \sumxy
         Q(\vx \condvs)
           W(\vy|\vx \condcs)
           \log
             \left(
               \frac{\sum_{\vshat} \vhat(\vshat, \vx, \vy)}
                    {Q(\vx \condvs) \,
                     \sum_{\vshat', \vx'}
                       \vhat(\vshat', \vx', \vy)}
             \right).
    \label{eq:information:rate:lower:part:0:reformulated}
\end{align}
For the following consideration, it will be useful to split this expression
into three parts, \ie, $\Ilower^{(N)}(\vhat \condcs) = \Ilower^{(N)}_0(\vhat
\condcs) + \Ilower^{(N)}_1(\vhat \condcs) + \Ilower^{(N)}_2(\vhat \condcs)$,
where
\begin{align}
  \label{eq:information:rate:lower:part:0}
  \Ilower^{(N)}_0(\vhat \condcs)
    &\defeq
       -
       \frac{1}{2N}
       \sumx
         Q(\vx)
           \log
             \left(
               Q(\vx)
             \right), \\
  \label{eq:information:rate:lower:part:1}
  \Ilower^{(N)}_1(\vhat \condcs)
    &\defeq
       +
       \frac{1}{2N}
       \sumxy
         Q(\vx)
           W(\vy|\vx)
           \log
             \left(
               \sum_{\vshat}
                 \vhat(\vshat, \vx, \vy)
             \right), \\
  \label{eq:information:rate:lower:part:2}
  \Ilower^{(N)}_2(\vhat \condcs)
    &\defeq
       -
       \frac{1}{2N}
       \sumy
         (QW)(\vy)
         \log
           \left(
             \sum_{\vshat', \vx'}
                     \vhat(\vshat', \vx', \vy)
           \right).
\end{align}
It is clear that $\Ilower^{(N)}_0(\vhat \condcs)$ is independent of $\vhat$,
therefore, in order to derive a suitable surrogate function
$\Psilower^{(N)}(\vhat \condcs)$ for $\Ilower^{(N)}(\vhat \condcs)$, it is
sufficient to focus on $\Ilower^{(N)}_1(\vhat \condcs)$ and
$\Ilower^{(N)}_2(\vhat \condcs)$. Our approach will be to derive surrogate
functions $\Psilower^{(N)}_1(\vhat \condcs)$ and $\Psilower^{(N)}_2(\vhat
\condcs)$ for $\Ilower^{(N)}_1(\vhat \condcs)$ and $\Ilower^{(N)}_2(\vhat
\condcs)$, respectively, which we will then combine. We define the first and
second surrogate functions to be, respectively,
\begin{align}
  \Psilower^{(N)}_1(\vtilde, \vhat)
    &\defeq
       \Ilower^{(N)}_1(\vhat \condcs)
       -
       \frac{1}{2N}
       \sumxy
         Q(\vx)
         W(\vy|\vx)
         D_{\vshat}
           \left(
               \frac{\vtilde(\vshat, \vx, \vy)
                    }
                    {\sum_{\vshat'}
                       \vtilde(\vshat', \vx, \vy)
                    }
           \ \left\lVert \
               \frac{\vhat(\vshat, \vx, \vy)
                    }
                    {\sum_{\vshat'}
                       \vhat(\vshat', \vx, \vy)
                    }
           \right.
           \right),
    \label{eq:information:rate:lower:surrogate:function:part:1} \\
  \Psilower^{(N)}_2(\vtilde, \vhat)
    &\defeq
       \underline{c}^{(N)}_2(\vtilde)
       -
       \sum_{\bhat}
         \sum_{\vyD}
           \frac{1}{\gamma}
            \left(
              \frac{\vhat(\bhat,\vyD)}
                   {\vtilde(\bhat,\vyD)}
            \right)^{\gamma}
            \Ttilde^{(N)}_2(\bhat,\vyD),
  \label{eq:information:rate:lower:surrogate:function:part:2}
\end{align}
where $\gamma$ is some arbitrary positive real number, where
$\underline{c}^{(N)}_2(\vtilde)$ is chosen such that
$\Psilower^{(N)}_2(\vtilde, \vtilde) = \Ilower^{(N)}_2(\vtilde)$, and where
$\Ttilde^{(N)}_2(\bhat,\vyD)$ was defined in~\eqref{eq:ttilde2:mod}.

\begin{Lemma}{Important Properties of $\Psilower_1$}
  \label{lemma:properties:lower:surrogate:function:part:1}

  We recognize the following properties of the first surrogate function:
  \begin{enumerate}

  \item For any $\vhat$, the function $\Psilower^{(N)}_1(\vtilde, \vhat
    \condcs)$ is never above $\Ilower^{(N)}_1(\vhat)$. Moreover, for $\vhat =
    \vtilde$, $\Psilower^{(N)}_1(\vtilde, \vhat \condcs)$ equals
    $\Ilower^{(N)}_1(\vtilde)$.

  \item The function $\Psilower^{(N)}_1(\vtilde, \vhat)$ can be
    simplified to
    \begin{align}
      \Psilower^{(N)}_1(\vtilde, \vhat)
        &= \underline{c}^{(N)}_1(\vtilde)
           +
           \sum_{\bhat}
             \sum_{\vyD}
               \log
                 \left(
                   \vhat(\bhat,\vyD)
                 \right)
               \Ttilde^{(N)}_4(\bhat,\vyD)
    \label{eq:information:rate:lower:surrogate:function:part:1:reformulation}
    \end{align}
    where $\underline{c}^{(N)}_1(\vtilde)$ is independent of $\vhat$ and where
    $\Ttilde^{(N)}_4(\bhat,\vyD)$ was defined in~\eqref{eq:ttilde4:mod}.

  \item The function $\Psilower^{(N)}_1(\vtilde, \vhat)$ is concave in $\vhat$,
    \ie, in $\bigl\{ \vhat(\bhat,\vyD) \bigr\}$.

  \end{enumerate}
\end{Lemma}

\begin{Proof}
  See Appendix~\ref{app:lower:property:part:1}.
\end{Proof}

\begin{Lemma}{Important Properties of $\Psilower_2$}
  \label{lemma:properties:lower:surrogate:function:part:2}

  We recognize the following properties of the second surrogate function:
  \begin{enumerate}

  \item For $\vhat = \vtilde$, the function value $\Psilower^{(N)}_2(\vtilde,
    \vhat)$ equals the function value $\Ilower^{(N)}_2(\vtilde)$.

  \item For $\vhat = \vtilde$, the gradient of $\Psilower^{(N)}_2(\vtilde,
    \vhat)$ w.r.t.~$\vhat$ equals the gradient of $\Ilower^{(N)}_2(\vhat)$
    w.r.t.~$\vhat$.

  \item The function $\Psilower^{(N)}_2(\vtilde, \vhat)$ is concave in
    $\vhat$, \ie, in $\bigl\{ \vhat( \bhat, \vyD) \bigr\}$.

  \end{enumerate}
\end{Lemma}

\begin{Proof}
  See Appendix~\ref{app:lower:property:part:2}.
\end{Proof}

Based on the above definitions, we are ready to introduce our surrogate
function for the information rate lower bound
\begin{align}
  \Psilower^{(N)}(\vtilde, \vhat)
    &\defeq
       \underline{c}^{(N)}(\vtilde)
       +
       \Psilower^{(N)}_1(\vtilde, \vhat)
       +
       \Psilower^{(N)}_2(\vtilde, \vhat),
\end{align}
with its asymptotic version be $\Psilower(\vtilde, \vhat) \defeq \lim_{N \to
  \infty} \Psilower^{(N)}(\vtilde, \vhat \condcs)$, where
$\underline{c}^{(N)}(\vtilde)$ is chosen such that $\Psilower^{(N)}(\vtilde,
\vtilde) = \Ilower^{(N)}(\vtilde)$.

\begin{Lemma}{Important Properties of $\Psilower$}
  \label{lemma:properties:lower:surrogate:function}

  We recognize the following properties of the surrogate function:
  \begin{enumerate}

  \item For $\vhat = \vtilde$, the function value $\Psilower^{(N)}(\vtilde,
    \vhat)$ equals the function value $\Ilower^{(N)}(\vtilde)$.

  \item For $\vhat = \vtilde$, the gradient of $\Psilower^{(N)}(\vtilde,
    \vhat)$ w.r.t.~$\vhat$ equals the gradient of $\Ilower^{(N)}(\vhat)$
    w.r.t.~$\vhat$.

  \item The function $\Psilower^{(N)}(\vtilde, \vhat)$ is concave in
    $\vhat$, \ie, in $\bigl\{ \vhat( \bhat, \vyD) \bigr\}$.

  \end{enumerate}
\end{Lemma}

\begin{Proof}
  Follows from Lemmas~\ref{lemma:properties:lower:surrogate:function:part:1}
  and~\ref{lemma:properties:lower:surrogate:function:part:2}.
\end{Proof}

\begin{Lemma}{Maximizing the Surrogate Function $\Psilower$}
  \label{lemma:maximize:lower}

  Assume that we are at iteration $t$ and that $\vtilde = \vhat^{\langle
    t\rangle}$. The $\vhat$ that minimizes $\Psilower(\vtilde, \vhat)$ at the
  next iteration is given by
  \begin{alignat}{2}
    \label{eq:vhat:update:1}
    \vhat^{*}(\bhat,\vyD)
      &= \left(
           \frac{\Ttilde_4(\bhat,\vyD)}
                {\Ttilde_2(\bhat,\vyD)}
         \right)^{\frac{1}{\gamma}}
         \vtilde(\bhat,\vyD)
           \quad
           &&(\text{for all $\bhat \in \setBhat$
              and all $\vyD \in \setY^{D_1+D_2+1}$}),
  \end{alignat}

\end{Lemma}

\begin{Proof}
  Follows easily from the definition of $\Psilower^{(N)}(\vtilde, \vhat)$, the
  reformulation of $\Psilower^{(N)}_1(\vtilde, \vhat)$ in
  Lemma~\ref{lemma:properties:lower:surrogate:function:part:1}, and the
  gradient expressions for $\Psilower^{(N)}_2(\vtilde, \vhat)$ in
  App.~\ref{app:lower:property:part:2}.
\end{Proof}

With these results in our hand, the proposed iterative optimization of the
lower bound looks as follows.

\begin{Algorithm}{Iterative Optimization of the Information Rate Lower Bound}
  \label{alg:iterative:lower}

  \mbox{}

  \begin{enumerate}

  \item For all $\bhat \in \setBhat$ and all $\vyD \in \setY^{D_1+D_2+1}$, set
    $\vhat^{\langle 0\rangle}(\bhat, \vyD)$ to some positive initial values.
    Fix some $\gamma > 0$. (In the simulation sections we will talk about
    suitable strategies for (adaptively) choosing the value of $\gamma$.)

  \item Set $t = 0$.

  \item \label{item:lower:bound:loop} As long as desired or until convergence,
    repeat the following steps.
    \begin{enumerate}

    \item \label{item:lower:bound:inner:loop} Set $\vtilde \defeq
      \vhat^{\langle t\rangle}$ and use
      Algorithms~\ref{alg:numerical:Ttilde:1:2}
      and~\ref{alg:numerical:Ttilde:3:4} to estimate $\bigl\{
      \Ttilde^{(N)}_2(\bhat,\vyD) \bigr\}$ and $\bigl\{ \Ttilde^{(N)}_4(\bhat,\vyD)
      \bigr\}$.

    \item Update $\bigl\{ \vhat^{\langle t+1\rangle}(\bhat,\vyD) \bigr\}$
      according to~\eqref{eq:vhat:update:1}.

    \item Increase $t$ by one.

    \item Go back to step~\ref{item:lower:bound:inner:loop}.

    \end{enumerate}

  \end{enumerate}
\end{Algorithm}

\begin{Remark}{Normalization}
  \label{remark:lower:bound:parameter:normalization:1}

  Note that no normalization is involved in the update
  equation~\eqref{eq:vhat:update:1}. However, because of the way that $\vhat$
  appears in the numerator and denominator of the logarithm
  in~\eqref{eq:information:rate:lower:part:0:reformulated}, the following
  normalization can be applied to $\vhat(\bhat,\vyD)$ (if desired): all
  $\vhat(\bhat,\vyD)$ can be multiplied by some positive constant that is
  independent of $\bhat$ and $\vyD$. It is important to notice that this
  normalization is different from the normalizations that were applied in
  Lemmas~\ref{lemma:minimize:upper} and~\ref{lemma:minimize:diff}.
\end{Remark}

\begin{Remark}{Convergence}

  Because we cannot prove that our $\Psilower^{(N)}(\vtilde, \vhat)$ is a
  never-above surrogate function, we cannot make convergence statements
  like those in Lemmas~\ref{lemma:convergence:upper} and~\ref{lemma:convergence:diff}.
\end{Remark}

\begin{Remark}{Simplification for Data-Controllable Auxiliary
    FSMCs}\label{rem:fir:simplify:lower}

  In data-controllable AB-FSMCs, $\Ttilde^{(N)}_4(\bhat,\vyD)$ has a
  simplified closed-form and can be pre-computed before optimization
  iterations. Please refer to Remark~\ref{rem:fir:simplify} for more details.
  This means that for the optimization of the lower bound for
  data-controllable channels, one only needs to evaluate
  $\Ttilde^{(N)}_2(\bhat,\vyD)$ at each iteration.
\end{Remark}

\begin{Remark}{Optimization of the Alternative Information Rate Lower Bound}

  Assuming the additional constraint on $\vhat(\bhat,\vyD)$ that we imposed in
  Rem.~\ref{remark:special:case:information:rate:lower:bound:2}, we have
  $\sum_{\vshat, \vx} \vhat(\vshat, \vx, \vy) = 1$ for all $\vy$ and
  so $\Ilower^{(N)}_2(\vhat) = 0$.  Therefore, the surrogate function can be
  chosen to be $\Psilower^{(N)}(\vtilde, \vhat) = \underline{c}^{(N)}(\vtilde) +
  \Psilower^{(N)}_1(\vtilde, \vhat)$, where $\underline{c}^{(N)}(\vtilde)$ is
  chosen such that $\Psilower^{(N)}(\vtilde, \vtilde) =
  \Ilower^{(N)}(\vtilde)$ and where $\Psilower^{(N)}_1(\vtilde, \vhat)$ is
  defined as in~\eqref{eq:information:rate:lower:surrogate:function:part:1}.
  It follows from Lemma~\ref{lemma:properties:lower:surrogate:function:part:1}
  that $\Psilower^{(N)}(\vtilde, \vtilde)$ is a never-above surrogate
  function. We leave the details to the reader to derive an optimization
  algorithm where at each step the value of the lower bound is non-decreasing.

  As already alluded to at the end of Sec.~\ref{sec:information:rate:bound:1},
  we have just seen that the optimization of the
  Rem.~\ref{remark:special:case:information:rate:lower:bound:2} lower bound
  has nicer analytical properties. However, we remind the reader of our
  previous remark that usually strictly positive choices for $D_1$ and $D_2$
  are needed to yield good lower bounds. We leave it as an open problem to
  find suitable modifications of this variant of the lower bound such that the
  desired analytical optimization properties are retained yet less parameters
  are needed.
\end{Remark}

%% file: results_partial1.tex

\section{Numerical Results for
             Partial Response Channels}
\label{sec:num:partial:response}

In this section, we provide numerical results for the optimization of the
upper and the lower bound for (output-quantized) partial response channels,
which are connected to a binary, independent and uniformly distributed (i.u.d)
source. In \secref{sec:partial:model} we present the channel model and in
\secref{sec:partial:initialize} we discuss three different initialization
methods for starting the iterative optimization methods.  \secref{sec:soblex}
describes the Soblex optimization method for a later comparison with the
proposed techniques. In \secref{sec:num:partial:compelxity} we discuss how the
proposed algorithms may provide tight information rate bounds with less
computational complexity, when compared to computing the information rate in
the original channel with a large memory length. \secref{sec:num:partial:M:2}
and the following subsections provide an investigation of the convergence
properties of the optimization algorithms. Simple channels are used for this
purpose to reduce the computational investment.

\subsection{\textbf{Source, Channel, and Auxiliary Channel Models}}
\label{sec:partial:model}

As mentioned earlier in this section, we assume a binary i.u.d source with
alphabet $\set{X} = \{ -1, +1 \}$. The original channel is an
(output-quantized) memory-length-$M$ partial response channel with input
alphabet $\set{X}$, with discrete output alphabet $\set{Y}$ (that will be
specified later), and which is described by
\begin{align}
  \label{eq:fir:2}
  y_{\ell}
    &= \mu
       \left(
         \sum_{m=0}^{M}
           h_m x_{\ell-m}
           +
           n_{\ell}
       \right).
\end{align}
Here, $\mu$ is a quantization function that maps elements of $\mathbb{R}$ to
$\setY$, and $\{ x_{\ell} \}$, $\{ y_{\ell} \}$, $\{ n_{\ell} \}$, $\{ h_m \}$
represent the channel input process, the channel output process, an additive
white Gaussian noise (AWGN) process, and the filter coefficients,
respectively. We assume that the coefficients $\vect{h} \defeq [h_0, h_1,
\ldots, h_M]$ are known. Such knowledge is not necessary for the optimization
procedure in its strict sense, because one only needs to know a realization of
an input/output process pair $\bigl( \{ x_{\ell} \}, \{ y_{\ell} \} \bigr)$.
However, knowing the channel coefficients helps us start optimizations at more
appropriate initial points (see the three initialization algorithms described
in \secref{sec:partial:initialize}). As usual, this channel can be described
by an FSMC with $|\setS| = 2^M$ states, whereby each state can be labeled by
the $M$ previous channel inputs $\vx_{\ell-M}^{\ell-1}$, and where there are
two valid outgoing branches for each state, hence, $|\setB| = 2^{M+1}$. The
conditional pmfs $W(s_{\ell} | s_{\ell-1}, x_{\ell})$ and $W(y_{\ell} |
b_{\ell})$ are then defined in the obvious way.

The AF-FSMC (and similarly the AB-FSMC) is chosen to have a trellis
structure that is the same as the trellis structure of an
(output-quantized) partial response channel with memory $\hat M$,
with input alphabet $\setX$, output alphabet $\setY$, and the same
quantization function $\mu$ as for the original channel. In the case
of the upper bound and difference function, we will optimize over
the AF-FSMC parameters $\bigl\{ \What(\hat s_{\ell} | \hat
s_{\ell-1}, \hat x_{\ell}) \bigr\}$ and $\bigl\{ \What(y | \bhat)
\bigr\}$, whereas in the case of the lower bound, we will optimize
over the AB-FSMC parameters $\bigl\{ \vhat(\bhat, \vyD) \bigr\}$.

\subsection{\textbf{Initialization Methods}}
\label{sec:partial:initialize}

The parameters $\bigl\{ \What(\hat s_{\ell} | \hat s_{\ell-1}, x_{\ell})
\bigr\}$ of the AF-FSMC are initialized to the natural settings based on the
above AF-FSMC trellis definition. For the parameters $\bigl\{ \What(y | \bhat)
\bigr\}$ of the AF-FSMC we consider three different initialization methods.
\begin{enumerate}

\item \emph{Initialization based on channel coefficient truncation.} In the
  first method, the initial parameters of the AF-FSMC are derived from
  truncating the original channel into its latest $\Mhat+1$ coefficients
  $\vect{\hat h} = [h_{0}, \ldots, h_{\Mhat}]$. If the memory length $\Mhat$
  of the AF-FSMC is greater than the original channel memory length $M$, then
  we fill the remaining $\Mhat-M$ coefficients with zeros to obtain
  $\vect{\hat h} = [\underbrace{h_{0}, \ldots, h_M}_{M+1}, \underbrace{0,
    \ldots, 0}_{\Mhat-M}]$. In summary, we assume an AF-FSMC with channel
  response
  \begin{align}
    \label{eq:fir:aux}
    y_{\ell}
      &= \mu
           \left(
             \sum_{m=0}^{\Mhat}
               \hat{h}_m x_{\ell-m}
               +
               n_{\ell}
           \right),
  \end{align}
  and find its output probability $\What(y|\bhat)$.

\item \emph{Initialization based on optimized difference function.} In the
  second method, we select $\{ \What(y|\bhat) \}$ to minimize the difference
  function. For partial response channels, $\What(y|\bhat)$ has a closed form
  and can be pre-computed as discussed in \secref{sec:optimize:diff:bound:1},
  Remark~\ref{rem:fir:simplify}.

\item \emph{Initialization based on averaging.} In the third method and for
  each channel output $y \in \setY$, we average the original channel law
  $W(y|b)$ over all original FSMC branches and assign it to all the AF-FSMC
  branches. That is,
  \begin{align}
    \label{eq:mean}
    \What(y|\bhat)
      &= \frac{1}{|\setB|}
         \sum_{b\in\setB}
           W(y|b)
           \quad\quad \text{(for all $y \in \setY$ and
                                all $\bhat \in \setBhat$)}.
  \end{align}
\end{enumerate}
In all numerical analyses, we use the information rate lower bound as defined
in Def.~\ref{def:information:rate:lower:1} with the setting $D_1 = 0$ and $D_2
= 0$ (and \emph{not} the specialized information rate lower bound in
Rem.~\ref{remark:special:case:information:rate:lower:bound:2}). Then, the
parameters $\bigl\{ \vhat(\bhat, y) \bigr\}$ of the AB-FSMC are initialized
based on the initialization of $\bigl\{ \What(\hat s_{\ell} | \hat s_{\ell-1},
x_{\ell}) \bigr\}$ and $\bigl\{ \What(y | \bhat) \bigr\}$. Namely, for any of
the above three initialization methods we set $\vhat(\bhat, y) \defeq
\What(\hat s_{\ell} | \hat s_{\ell-1}, x_{\ell}) \cdot \What(y | \bhat)$.
(Note that because $D_1 =0$ and $D_2 = 0$, we have $\vyD = y$.)

\subsection{\textbf{Soblex Optimization Algorithm}}
\label{sec:soblex}

In the following subsections, we will compare our proposed optimization
algorithms for the upper and the lower bound with an improved variation of the
\emph{simplex} algorithm
(see~\cite{press:flannery:teukolsky:vetterling:numerical:Recipes:in:c} for the
definition of the standard simplex optimization method). Standard optimization
algorithms such as \emph{Powell} or
simplex~\cite{press:flannery:teukolsky:vetterling:numerical:Recipes:in:c} can
easily be trapped by local minima. Instead, we will use a robust optimization
method by combining the standard simplex algorithm with an initial sampling of
the parameter space with the \emph{Sobol} quasi-random
sequences~\cite{press:flannery:teukolsky:vetterling:numerical:Recipes:in:c}.
This method was originally proposed
in~\cite{shams:kennedy:sadeghi:hartley:iccv:07} and was called \emph{Soblex}
optimization.

Here is a brief explanation about how the Soblex algorithm works. The Soblex
optimization is initially given a budget, in terms of the number of cost
function calls. Within the initial budget, Soblex evaluates the cost function
using the Sobol sequence and initializes a simplex-shaped subspace, which is
constructed from points with the lowest costs. The Sobol sequence ensures that
we can progressively sample the parameter space in a virtually uniform
fashion. Unlike most optimization methods that start with a single point in
space, the simplex algorithm starts from a region of space that can be made
arbitrarily close to the global minimum by increasing the Soblex budget.
Intuitively, if the budget is large enough, the simplex subspace can
sufficiently close in on the global minimum to allow successful execution of
the optimization algorithm. The Soblex algorithm has had promising performance
results for efficiently finding the global optimum of problems with low
dimensionality (three-dimensional rotation space
in~\cite{shams:kennedy:sadeghi:hartley:iccv:07}). However, the dimensionality
of the AF-FSMC / AB-FSMC parameter space is much higher and requires a large
Soblex budget. We will use Soblex as a reference to demonstrate the superior
efficiency and performance of our algorithms, compared to general-purpose
optimization methods.

\subsection{\textbf{Reduction in the Computational Complexity}}
\label{sec:num:partial:compelxity}

As mentioned in the Introduction and from a practical viewpoint, the capacity
and the capacity-achieving input distribution of finite-state channels with
not too many states (short memory lengths, $M$) is numerically computable.
Therefore, for partial response channels with small memory lengths, the
proposed optimization techniques may not provide an immediate advantage. The
true benefit of the optimization techniques is for original partial response
channels with large memory lengths, where we are able to obtain lower and
upper bounds on information rates of such channels with less computational
complexity. This is further clarified through the following
argument.

For binary signaling in a partial response channel with memory length $M$, the
computational complexity for running the ``Forward'' or the ``Backward'' part
of the BCJR algorithm is proportional to $2^M \cdot (2N)$, which exponentially
increases in the channel memory length $M$ and is linear in the input/output
window length $2N$. Numerical computation of the information rate only
requires the ``Forward'' part of the BCJR algorithm. Let us assume that we use
an AF-FSMC / AB-FSMC with memory length $\Mhat = \lceil M/2\rceil$ to
iteratively optimize the bounds ($\lceil \, \cdot \, \rceil$ is the integer
ceiling function) and we use a fixed $\gamma$ (such as $\gamma = 1$)
throughout the lower bound iterations presented in
\secref{sec:optimize:lower:bound:1} (see \eqref{eq:vhat:update:1}). As
discussed in Sections~\ref{sec:optimize:upper:bound:1}
and~\ref{sec:optimize:lower:bound:1}, for partial response channels one only
needs to iteratively compute $\Ttilde^{(N)}_2(\bhat,y)$ and
$\Ttilde^{(N)}_2(\bhat,\vyD)$, which requires one ``Forward run'' of the BCJR
algorithm on the input/output sequence to compute the forward messages or
$\alpha$'s, one ``Backward run'' to compute the backward messages or
$\beta$'s, and one final ``Forward run'' to compute combinations of forward
and backward messages or $\sigma$'s~\cite{BCJR1974,
  Kschischang:Frey:Loeliger:01}.

\begin{figure}
  \begin{center}
    \includegraphics[width=0.7\columnwidth]{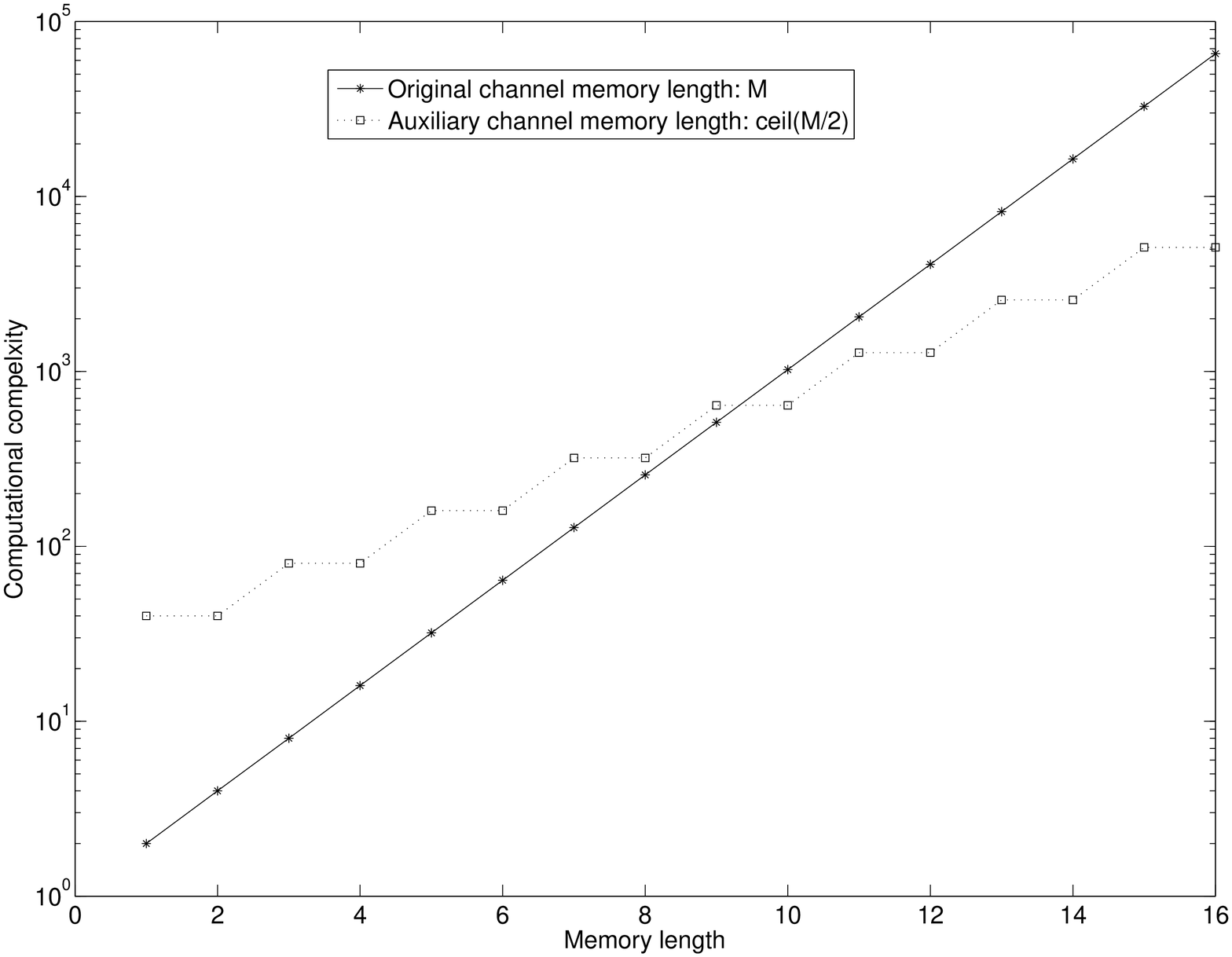}
  \end{center}
  \caption{Comparison of the relative computational complexity (in logarithmic
    scale) for numerical evaluation of the original information rate and for
    optimization of the upper and the lower bounds using simpler AF-FSMCs /
    AB-FSMCs. We have assumed that $3$ iterations are needed for the
    optimization of each bound. For original channel memory lengths greater
    than or equal to $M = 10$, the optimization techniques clearly become
    computationally advantageous.}
  \label{fig:complexity}
\end{figure}

Therefore, the computational cost of each iteration is $3 \cdot 2^{\Mhat}
\cdot (2N)$. For example, the total cost of 3 iterations for each upper and
lower bound plus one final ``Forward run'' to actually compute each bound
becomes $20 \cdot 2^{\Mhat} \cdot (2N)$. This is shown in
\figref{fig:complexity}, where the original partial response channel memory
length $M$ ranges from $1$ to $16$. The vertical axis shows the relative
computational complexity in logarithmic scale by assuming the same
input/output window length. The optimization techniques become computationally
advantageous for original channel memory lengths greater than or equal to $M =
10$.

\begin{figure}
  \begin{center}
    \includegraphics[width=0.7\columnwidth]{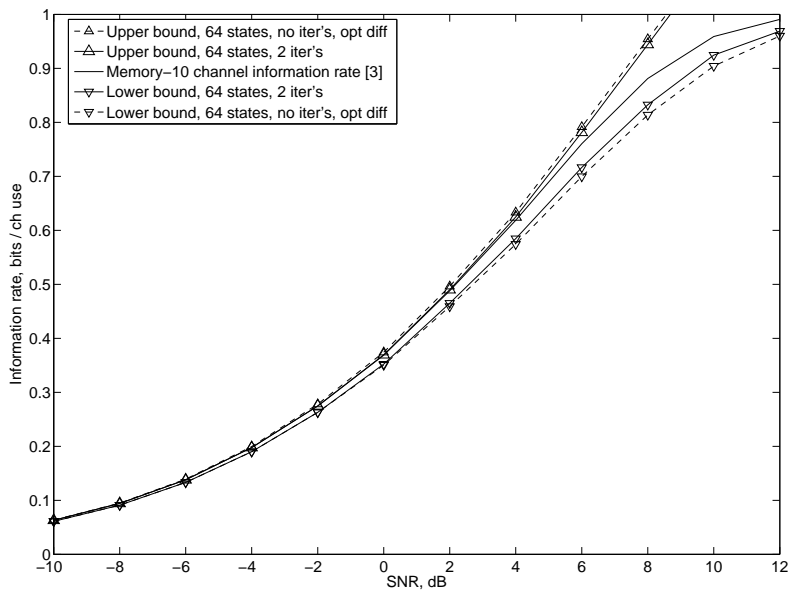}
  \end{center}
  \caption{Information rate bounds for the quantized version of the memory-10
    channel introduced in~\cite{Arnold:Loeliger:Vontobel:Kavcic:Zeng:06:1}
    using AF-FSMCs / AB-FSMCs with memory $\Mhat = 6$. It is clear from this
    figure that at low SNR, even a closed-form and non-iterative optimization
    of the difference function is enough to provide tight bounds at a much
    lower computational complexity.}
  \label{fig:ch11:ch6}
\end{figure}

To verify that the proposed technique can provide tight bounds for partial
response channels with large memory lengths, we consider the $11$-tap original
channel with memory length $M=10$ studied
in~\cite{Arnold:Loeliger:Vontobel:Kavcic:Zeng:06:1}. The channel coefficients
are defined as $h_i = 1/(1+(i-5)^2)$ for $0 \le i \le 10$. In contrast
to~\cite{Arnold:Loeliger:Vontobel:Kavcic:Zeng:06:1} we also apply a
quantization function $\mu$ that is based on partition points at
\begin{align}
  -\infty, \,
  -2.5\sigma_y, \,
  -2.45\sigma_y, \,
  -2.4\sigma_y, \,
  \ldots,  \,
  +2.4\sigma_y, \,
  +2.45\sigma_y \,
  +2.5\sigma_y, \,
  +\infty,
\end{align}
where $\sigma_y$ is the output standard deviation. The results are shown in
\figref{fig:ch11:ch6}, where the SNR is defined according
to~\cite{Arnold:Loeliger:Vontobel:Kavcic:Zeng:06:1}. The first curve from the
bottom shows the lower bound using an AB-FSMC with memory length $\Mhat = 6$
or with $64$ states. The AB-FSMC is initialized based on the ``difference
function optimization initialization method'' (as explained in
Remark~\ref{rem:fir:simplify}, the AB-FSMC parameters can be pre-computed for
such an initialization) and no lower bound optimization steps are performed.
Similarly, the first curve from the top shows the upper bound where the
AF-FSMC is initialized based on the ``difference function optimization
initialization method'' and no upper bound optimization steps are performed.
If we allow two more iterations, we obtain the next set of inner curves which
slightly tighten the bounds, especially for higher SNRs. For a fair comparison
of the computational complexities of information rates and their bounds, we
used a fixed $\gamma = 1$ in the lower bound iterations (see
\eqref{eq:vhat:update:1}) and did not spend extra time to find a better
$\gamma$ that would potentially result in higher lower bounds. It is clear
from this figure that at low SNR, even a closed-form and non-iterative
optimization of the difference function is enough to provide tight bounds at a
much lower computational complexity (there are only $64$ states in the AF-FSMC
/ AB-FSMC as opposed to $1024$ states in the original channel).  Even with 2
iterations, the complexity of computing bounds is still lower than computing
the information rate in the original channel. It is also noted that at low
SNR, the proposed upper bound is tighter than those studied
in~\cite{Arnold:Loeliger:Vontobel:Kavcic:Zeng:06:1} using the reduced-state
upper bound (RSUB) method with $100$ states.\footnote{It is reminded that the
  upper bounds in~\cite{Arnold:Loeliger:Vontobel:Kavcic:Zeng:06:1}, Fig 8, and
  those given here are not directly comparable, because of the quantized
  output used in our set-up. However, our quantization is fine enough, so that
  the information rates do not change much even by increasing the quantization
  levels. Therefore, the general conclusion applies.}

\subsection{\textbf{Original Partial Response Channel
                     with Memory Length $M=2$}}
\label{sec:num:partial:M:2}

In this subsection we study the convergence properties of the upper and lower
bound optimizations for a simple partial response original channel with memory
length $M = 2$. We also compare tightness of the optimized bounds with those
obtained via the Soblex numerical optimization described in
\secref{sec:soblex}.

Because this original channel has memory length $M = 2$, there are three
normalized coefficients in \eqref{eq:fir:2} given as
\begin{align}
  \big[
    h_0, \ h_1, \ h_2
  \big]
    &= \big[
         0.5774, \ -0.5774, \ -0.5774
       \big].
\end{align}
We refer to this partial response channel as CH3. The quantization function
is based on partition points at
\begin{align}
  -\infty, \,
  -2.25, \,
  -1.5, \,
  -0.75, \,
  0, \,
  +0.75, \,
  +1.5, \,
  +2.25, \,
  +\infty,
\end{align}
resulting in $8$ quantized outputs and $\setY = \{1, \ldots, 8\}$. The SNR is
equal to $0$ dB as defined
in~\cite{Arnold:Loeliger:Vontobel:Kavcic:Zeng:06:1}.

The AF-FSMC / AB-FSMC used for this channel are output-quantized
partial response channels with memory length $\Mhat = 1$. That is,
the AF-FSMC / AB-FSMC channel will have $|\setShat| = 2$ states and
$|\setBhat| = 4$ branches. Since partial response channels are
controllable, optimization over $\What(\shat | \shatL, x)$ is not
required (current state $\shatL$ and input $x$ immediately determine
next state $\shat$). Therefore, we only need to optimize over
$\What(y | \bhat)$ or $\vhat(\bhat, y)$. Since there are $4$
possible branches and $8$ output levels, optimization of $\What(y |
\bhat)$ is over 32 real-valued and non-negative parameters. For the
upper bound, $\sum_{y}\What(y | \bhat) = 1$ for all branches and
this will further reduce the optimization to an optimization over 28
parameters. However, for the lower bound, there are no such
normalization constraints on $\vhat(\bhat, y)$,
cf.~Remark~\ref{remark:lower:bound:parameter:normalization:1}.

\subsubsection{Upper Bound Tests}

\begin{figure}
  \mbox{} \\
  \begin{center}
    \includegraphics[width=\columnwidth]{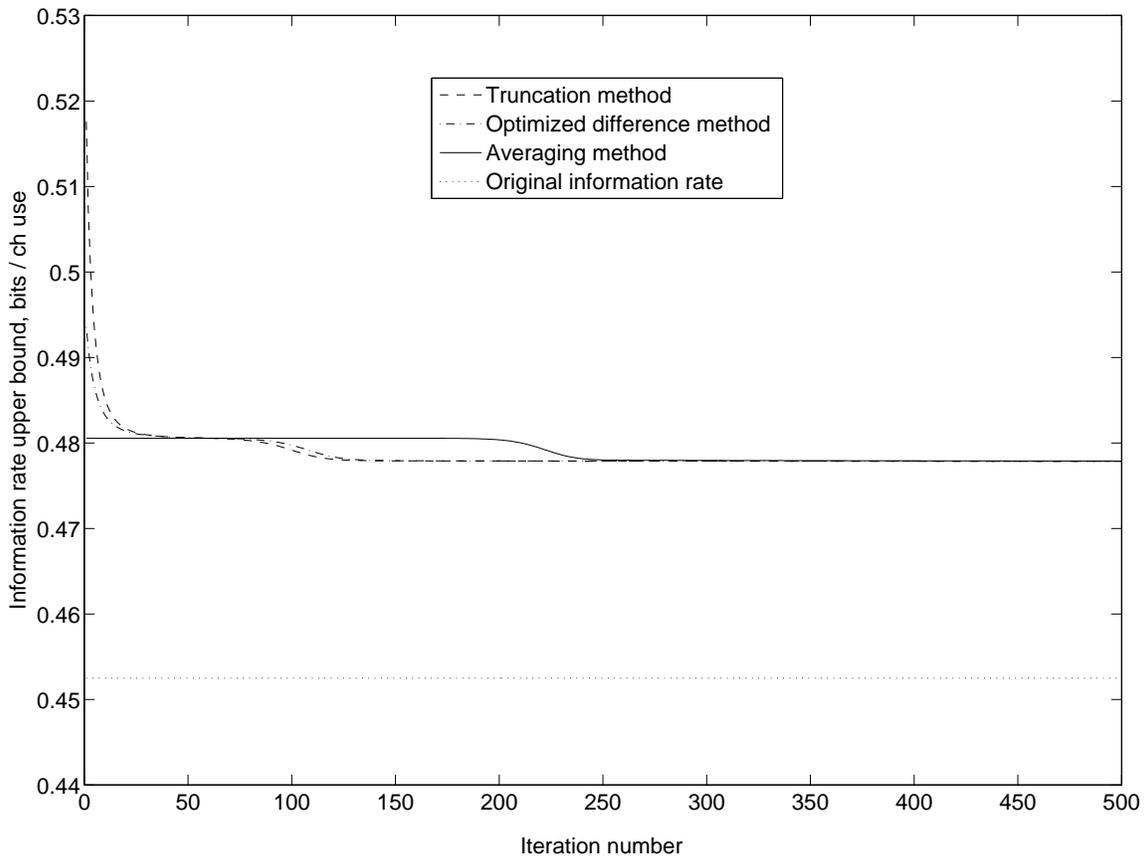}
  \end{center}
  \caption{Comparison of three initialization methods for the optimization of
    the upper bound. The original partial response channel has a memory length
    of $M = 2$ and the AF-FSMC has a memory length of $\Mhat = 1$. The SNR is
    equal to $0$ dB. Initializing the optimization algorithm by the
    ``truncation initialization method'' or the ``optimized difference
    function initialization method'' leads to faster convergence than
    initializing by the ``averaging initialization method''. It is noted
    that after the first $10$ iterations, the improvements is at most about
    $7.3\times 10^{-3}$ bits~/~channel use (independently of which
    initialization method is used).}
  \label{fig:upper:ch3:init:methods}
\end{figure}

\figref{fig:upper:ch3:init:methods} shows the effect of the three
initialization methods discussed above upon the upper bound optimizations. The
first $500$ iterations of $3000$ iterations performed are shown in this
figure, where the remaining $2500$ iterations improved the upper bound only in
the order of $10^{-5}$ bits~/~channel use. To avoid the small fluctuations in
stochastic evaluation of $\Ttilde^{(N)}_2(\bhat, y)$ in \eqref{eq:ttilde2} and
to solely study the convergence behavior of the optimization algorithm, we
used a single window of the channel output $\vy$ with length $2N_1 = 5\times
10^5$ for all iterations. Three main observations are made from this figure.
First it is noted that after the first $10$ iterations, the improvements in
for all initialization methods are at most about $7.3\times 10^{-3}$
bits~/~channel use.  Therefore, for all practical purposes, we may stop the
optimizations after only a few iterations.  Second, for all three
initialization methods the upper bound optimization algorithm virtually
converges to the same number, where the difference between the upper bounds in
the $3000$-th iteration is in the order $10^{-8}$ bits~/~channel use. Third,
we observe that initialization based on the ``difference function optimization
initialization method'' does not necessarily result in the lowest upper bound
in the first iteration (as can be seen from the figure, the lowest upper bound
in the first iteration belongs to initialization based on the ``averaging
initialization method'').

We also note that all three methods experience a \emph{flat} period
over which the improvement in the upper bound is almost negligible.
However, this flat period is shorter when the optimization algorithm
is initialized by the ``difference function optimization
initialization method'' and its convergence is quicker than when the
optimization algorithm is initialized by the ``averaging
initialization method''. In fact, when using the ``averaging
initialization method'', the upper bound improves only by $1.9\times
10^{-5}$ bits~/~channel use during the first $174$ iterations,
whereas both the ``truncation initialization method'' and the
``difference function optimization initialization method'' virtually
reach the final value within about $125$ iterations (the difference
between the bounds at iteration $125$ and $3000$ is less than
$3.8\times 10^{-4}$ bits / channel use).

The optimization of the upper bound has a low sensitivity to the numerical
technique used for computing $\Ttilde^{(N)}_2(\bhat, y)$ in
\eqref{eq:ttilde2}. To show this, we performed $100$ runs of the optimization
procedure using the optimum difference method and using $100$ different
(randomly generated) channel output windows $\vy$ of lengths $2N_1 = 5\times
10^5$ and $2N_2 = 2\times 10^6$ for computing the upper bound and
$\Ttilde^{(N)}_2(\bhat, y)$ in \eqref{eq:ttilde2}. The number of iterations in
each run was limited to $300$. With the relatively small window length $2N_1$,
the absolute and normalized standard deviation of the optimized upper bound
over $100$ runs\footnote{The normalized standard deviation is unit-less and is
  obtained by dividing the standard deviation by the mean of the upper bound
  over $100$ runs.} are $1.2 \times 10^{-3}$ bits~/~channel use and $2.5
\times 10^{-3}$, respectively, which are negligible for all practical
purposes. If we increase the window length by a factor of $4$ to $2N_2 =
2\times 10^6$, the absolute and normalized standard deviation of the optimized
upper bound over $100$ runs reduce to $6.3 \times 10^{-4}$ bits~/~channel use
and $1.3 \times 10^{-3}$, respectively.

\tableref{tab:upper:soblex} compares our proposed optimization method with the
Soblex method explained in \secref{sec:soblex}. Even for a very simple AF-FSMC
with memory length $\Mhat = 1$ and $8$ quantized outputs, the dimensionality
of the parameter space is $28$, which is very large. This makes it a difficult
problem for a conventional optimization algorithm without any estimate of the
initial point. Here, we use the Soblex algorithm, which provides a general and
reasonable method for finding initial points and performs better than the
standard simplex or Powell algorithm. In our comparative tests, we used
initial budgets of $1000$, $2000$, and $10000$ and ran the Soblex algorithm
$100$ different times using $100$ non-overlapping initial seeds for the Sobol
sequence. The fractional tolerance (frac.
tol.)~\cite{press:flannery:teukolsky:vetterling:numerical:Recipes:in:c}, which
affects the termination condition of the simplex algorithm, is either
$10^{-5}$ or $10^{-6}$. In the last row of the table, we have shown the
performance of $100$ stochastic runs of our proposed optimization technique
(Algorithm~\ref{alg:iterative:upper} in \secref{sec:optimize:upper:bound:1})
with $300$ iterations. In all tests, the window length for computing the upper
bound and $\Ttilde^{(N)}_2(\bhat, y)$ is $2N_1 = 5\times 10^5$.

The second column in \tableref{tab:upper:soblex} shows the statistical mean of
the upper bound over $100$ runs. It is observed from this column that the mean
upper bound decreases as the Soblex initial budget increases or as the
fractional tolerance decreases. However, the proposed
Algorithm~\ref{alg:iterative:upper} attains the minimum mean upper bound.
Although the difference between the upper bound given by
Algorithm~\ref{alg:iterative:upper} and those obtained by Soblex is small, it
is consistently above the numerical tolerance of
Algorithm~\ref{alg:iterative:upper}, which was found to be $1.2 \times
10^{-3}$ bits~/~channel use for $2N_1 = 5\times 10^5$ in the previous
paragraph. In the third column of \tableref{tab:upper:soblex}, we have used
the worst-case upper bound in the $300$-th iteration of
Algorithm~\ref{alg:iterative:upper} over $100$ runs as the reference, which is
found to be $\Iupper = 0.4789$ bits~/~channel use. Then, we have counted the
percentage of Soblex runs where the optimized upper bound is worse than our
worst-case performance. As can be seen from this column, the Soblex method
does not perform as well as Algorithm~\ref{alg:iterative:upper} in terms of
robustness in finding the lowest upper bound. The fourth column of this table
shows the number of cost calls per run in the Soblex algorithm (computation of
information rate bound) or in Algorithm~\ref{alg:iterative:upper} (computation
of information rate bound and $\Ttilde^{(N)}_2(\bhat, y)$). As discussed in
Sections~\ref{sec:num:partial:compelxity}, computing $\Ttilde^{(N)}_2(\bhat,
y)$ plus a new upper bound in each iteration is four times more complex than
computing the upper bound only. Hence, we have $300 \times 4$ in the last
column. Finally, the actual average time spent in each run is shown in the
last column. It is observed that with the initial budget of $10000$, Soblex
has $10-11$ times higher computational complexity than
Algorithm~\ref{alg:iterative:upper}. It is noted that the real advantage of
Algorithm~\ref{alg:iterative:upper} is for more complicated AF-FSMCs with
larger memory lengths and output levels, where the Soblex method becomes less
reliable and computationally more challenging.

After finding the optimum upper bound in one of the stochastic runs of
Algorithm~\ref{alg:iterative:upper} with $300$ iterations and window length of
$2N_1 = 5 \times 10 ^5$, we passed the optimized AF-FSMC parameters to the
standard simplex algorithm (with no initial budget to randomly sample space)
to see if simplex method can further improve the upper bound. The fractional
tolerance for the simplex algorithm was set to $10^{-7}$. It turns out that
simplex algorithm can improve the upper bound found by
Algorithm~\ref{alg:iterative:upper} by only $6.4\times 10^{-5}$ bits~/~channel
use, which is negligible for all practical purposes.

\begin{table}
  \caption{Comparison of the Soblex optimization
    Algorithm~\cite{shams:kennedy:sadeghi:hartley:iccv:07} with the proposed
    Algorithm~\ref{alg:iterative:upper} in
    \secref{sec:optimize:upper:bound:1}. Algorithm~\ref{alg:iterative:upper}
    is superior to Soblex method both in terms of reliably finding the
    lowest upper bound and computational
    efficiency.}\label{Tab2}
  \vspace{1em}
  \begin{center}
    \begin{tabular}{|c||c|c|c|c|}
      \hline
        \scriptsize{Algorithm Specifications} &
        \scriptsize{Mean upper bound,} &
        \scriptsize{Percentage of runs with} &
        \scriptsize{Number of cost calls / run} &
        \scriptsize{Average time} \\
        &
        \scriptsize{bits~/~channel use}&
        \scriptsize{a bound above $0.4797$ b/ch use} &
        \scriptsize{or iteration complexity} &
        \scriptsize{per run, sec.} \\
      \hline
      \hline
        \scriptsize{Soblex budget$=1000$, frac. tol. $= 10^{-5}$} &
        $0.4791$ & $53\%$ & $2180$ & $218$ \\
      \hline
        \scriptsize{Soblex budget$=2000$, frac. tol. $= 10^{-5}$} &
        $0.4789$ & $37\%$ & $3244$ & $325$ \\
      \hline
        \scriptsize{Soblex budget$=10000$, frac. tol. $= 10^{-5}$} &
        $0.4788$ & $32\%$ & $1111$ & $1117$ \\
      \hline
        \scriptsize{Soblex budget$=10000$, frac. tol. $= 10^{-6}$} &
        $0.4783$ & $6\%$ & $12741$ & $1278$ \\
      \hline
        \scriptsize{\textbf{Algorithm~\ref{alg:iterative:upper},
        \secref{sec:optimize:upper:bound:1}}} &
        $\mathbf{0.4764}$ & $\mathbf{0\%}$ & $\mathbf{1200 (300\times 4)}$ & $\mathbf{110}$ \\
      \hline
    \end{tabular}
    \label{tab:upper:soblex}
  \end{center}
\end{table}

\subsubsection{Lower Bound Tests}

\begin{figure}
  \mbox{} \\
  \begin{center}
    \includegraphics[width=\columnwidth]{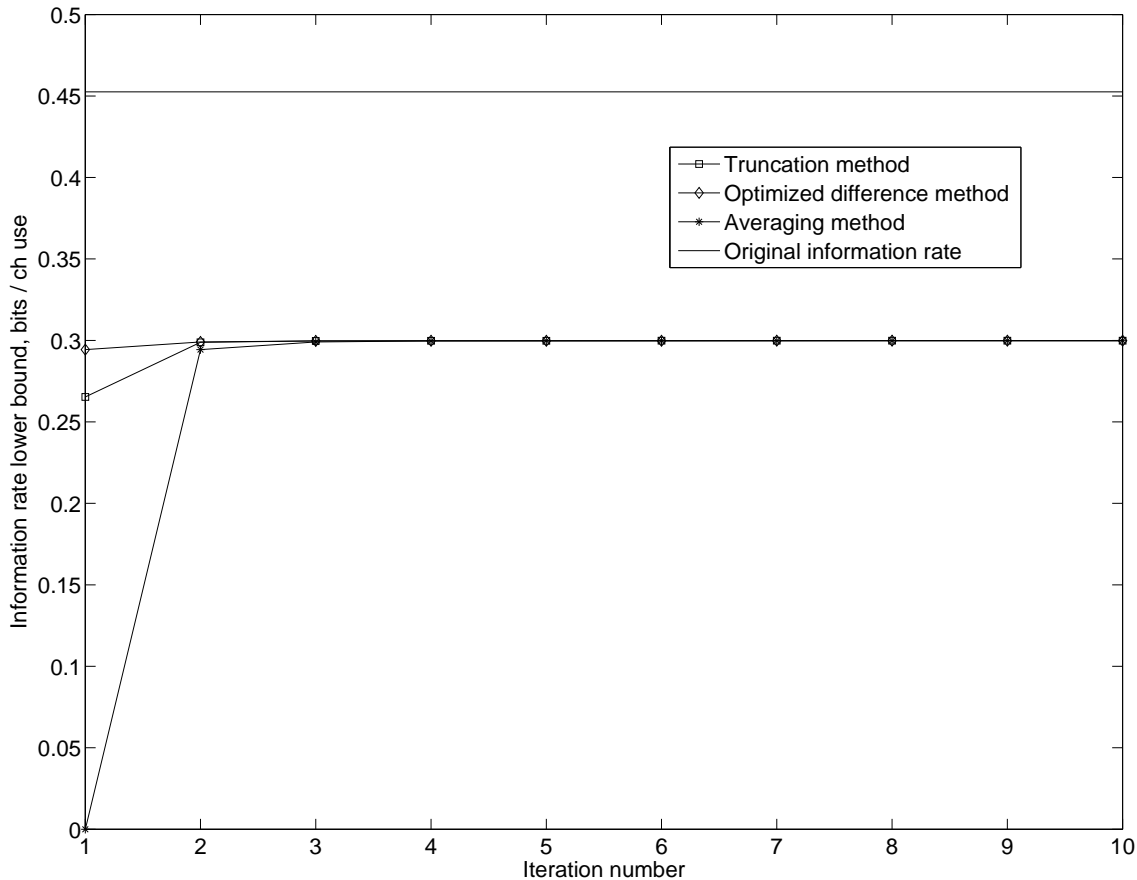}
  \end{center}
  \caption{Comparison of three initialization methods for the optimization of
    the lower bound. The original partial response channel has a memory length
    of $M = 2$ and the AB-FSMC has a memory length of $\Mhat = 1$. The SNR is
    equal to $0$ dB. Optimization with initialization based on the ``optimized
    difference function initialization method'' has the highest lower bound
    value in the first iteration. For all three initialization methods, the
    optimization algorithm virtually converges to the final lower bound value
    within $3$ to $4$ iterations.}
  \label{fig:lower:ch3:init:methods}
\end{figure}

\figref{fig:lower:ch3:init:methods} shows the effect of the three
initialization methods discussed in \secref{sec:partial:initialize}
on the lower bound optimization. The first $10$ iterations of $3000$
iterations performed are shown in this figure, where the remaining
$2990$ iterations improved the lower bound only by a maximum of
$2.5\times 10^{-6}$ bits / channel use. To avoid the small
fluctuations in stochastic evaluation of $\Ttilde^{(N)}_2(\bhat,
y)$\footnote{Note that because we use $D_1 =0$ and $D_2 = 0$ in
lower bound optimizations, we have $\vyD = y$ in $\Ttilde^{(N)}_2$.}
used in \eqref{eq:vhat:update:1} and to solely study the convergence
behavior of the optimization algorithm, we used a single window of
channel output $\vy$ with length $2N_1 = 5\times 10^5$ for all
iterations. All other parameters are the same as for the upper bound
tests. At each iteration, the parameter $\gamma$ in
\eqref{eq:vhat:update:1} was varied in the set $\Gamma = \{1,2,
\ldots, 10\}$. A small $\gamma$ corresponds to more freedom in
updating $\vhat$ with respect to $\vtilde$, whereas a large $\gamma$
corresponds to more conservative update of $\vhat$ relative to
$\vtilde$. In the limit as $\gamma \to \infty$, $\vhat = \vtilde$.
Our simple algorithm for choosing $\gamma$ is summarized as follows.

\begin{Algorithm}{An algorithm for choosing $\gamma$ in
                  \eqref{eq:vhat:update:1}}
\label{alg:ghat:update}

\mbox{}

\begin{enumerate}

\item Choose a finite set for varying $\gamma$ such as $\Gamma$.

\item At iteration $t$, update $\vhat$ according to each individual $\gamma$
  in the set $\Gamma$ using \eqref{eq:vhat:update:1} and compute the
  corresponding information rate lower bound.

\item For the next iteration, choose the $\gamma$ and its corresponding
  $\vhat$ that results in the highest lower bound in the set considered.

\item Increase $t$ by one. Compute $\Ttilde^{(N)}_2(\bhat, y)$ (and
  $\Ttilde^{(N)}_4(\bhat, y)$ if needed) and go back to step 1.

\end{enumerate}
\end{Algorithm}

Obviously, the larger the size of the set $\Gamma$ is the more time we spend
in optimizing the lower bound in each iteration. So the practical size of
$\Gamma$ is chosen by taking this factor into account. However, we do not
concern ourselves with this issue here, because the purpose of the following
analysis is understanding the convergence behavior of the algorithm and also
its comparison with other local optimization methods such as Soblex.

Two main observations are made from \figref{fig:lower:ch3:init:methods}.
First, the lower bound optimization has a fast-converging behavior.  While
optimization with initialization based on the ``averaging initialization
method'' yields the worst initial lower bound, viz.~$\Ilower = 0$, it
converges very quickly and within $3$ to $4$ iterations. On the other hand,
optimization with initialization based on the ``optimized difference function
initialization method'' results in the highest lower bound in the first
iteration and virtually converges within $2$ to $3$ iterations. Second, for
all three initialization methods the optimization algorithm converges
virtually to the same lower bound, where the difference between the lower
bounds for the $3000$-th iteration is in the order $10^{-11}$ bits~/~channel
use.

The optimization of the lower bound has a low sensitivity to numerical
techniques used for computing $\Ttilde^{(N)}_2(\bhat, y)$ in
\eqref{eq:ttilde2:mod}. To show this, we performed $100$ runs of the
optimization procedure with initialization based on the ``optimized difference
function initialization method'' and using $100$ different (randomly
generated) channel output windows $\vy$ of lengths $2N_1 = 5\times 10^5$ and
$2N_2 = 2\times 10^6$ for computing the lower bound and
$\Ttilde^{(N)}_2(\bhat, y)$. The number of iterations in each run was limited
to $50$, due to the fast-converging behavior of the lower bound optimizations.
We used $\Gamma = \{1, 2, \ldots, 10\}$ in Algorithm~\ref{alg:ghat:update}.
With a relatively small window length $2N_1$, the absolute and normalized
standard deviation of the optimized lower bound over $100$ runs are $1.0
\times 10^{-3}$ bits~/~channel use and $3.4 \times 10^{-3}$, respectively,
which are negligible for all practical purposes. If we increase the window
length by a factor of $4$ to $2N_2 = 2\times 10^6$, the absolute and
normalized standard deviation of the optimized lower bound over $100$ runs
reduce to $5.7 \times 10^{-4}$ bits~/~channel use and $1.9 \times 10^{-3}$,
respectively.

\tableref{tab:lower:soblex} compares our proposed optimization method with
Soblex method explained in \secref{sec:soblex}. Even for a very simple AB-FSMC
with memory length $\Mhat = 1$ and $8$ quantized outputs, the dimensionality
of the unconstrained parameter space is $32$, which is very large. This makes
it a difficult problem for a conventional optimization algorithm without any
estimate of the initial point. Here, we use the Soblex algorithm, which
provides a general and reasonable method for finding initial points and
performs better than the standard simplex or Powell. In our comparative tests,
we used initial budgets of $1000$, $2000$, $5000$, $20000$, and $50000$ and
ran the Soblex algorithm $100$ different times using $100$ non-overlapping
initial seeds for the Sobol sequence. The fractional tolerance (frac.
tol.)~\cite{press:flannery:teukolsky:vetterling:numerical:Recipes:in:c} is
$10^{-4}$. In the last row of the table, we have shown the performance of
$100$ stochastic runs of our proposed optimization technique
(Algorithm~\ref{alg:iterative:lower} in \secref{sec:optimize:lower:bound:1})
with $50$ iterations. In all tests, the window length of $\vy$ for computing
the upper bound and $\Ttilde^{(N)}_2(\bhat, y)$ is $2N_1 = 5\times 10^5$.

The second column in \tableref{tab:lower:soblex} shows the statistical mean of
the lower bound over $100$ runs. It is observed from this column that the mean
lower bound increases as the number of initial budget increases. However, the
proposed Algorithm~\ref{alg:iterative:lower} attains the maximum mean lower
bound by a large margin, which is beyond the numerical tolerance of the
algorithm. In the third column of \tableref{tab:lower:soblex}, we have used
the worst-case lower bound in the $50$-th iteration of
Algorithm~\ref{alg:iterative:lower} over $100$ runs as the reference. The
worst-case lower bound is $\Ilower = 0.2970$ bits~/~channel use. Then, we have
counted the percentage of Soblex runs where the optimized lower bound is above
(or better than) our worst-case performance. As can be seen from this column,
Algorithm~\ref{alg:iterative:lower} is noticeably superior in terms of
robustness in finding the highest lower bound. The fourth column of this table
shows the number of cost calls per run in the Soblex algorithm (computation of
information rate lower bound) or in Algorithm~\ref{alg:iterative:lower}
(computation of information rate lower bound and $\Ttilde^{(N)}_2(\bhat, y)$).
We used $\Gamma = \{1,2, \ldots, 10\}$ in Algorithm~\ref{alg:ghat:update} for
choosing $\gamma$. Therefore, at each iteration of our algorithm, we computed
$\Ttilde^{(N)}_2(\bhat, y)$ once plus ten lower bounds. According to the
discussions in \secref{sec:num:partial:compelxity}, this is $13$ times more
complex than computing the lower bound only. Finally, the actual average time
spent in each run is shown in the last column. For example, with the initial
budget of $20000$, Soblex is about $37$ times more computationally complex
than Algorithm~\ref{alg:iterative:lower}. The real advantage of
Algorithm~\ref{alg:iterative:lower} is, however, for more complicated AB-FSMC
with larger memory lengths and output levels, where the Soblex method becomes
less reliable and computationally more challenging.

After finding the best lower bound in one of the stochastic runs of
Algorithm~\ref{alg:iterative:lower} with $50$ iterations and window length of
$2N_1 = 5 \times 10 ^5$, we passed the optimized AB-FSMC parameters to the
standard Powell algorithm to see if it can further improve the lower bound.
The fractional tolerance for Powell algorithm was set to $10^{-4}$. We
observed that Powell algorithm can improve the lower bound found by
Algorithm~\ref{alg:iterative:lower} by only $2.8\times 10^{-5}$ bits~/~channel
use, which is negligible for all practical purposes.

\begin{table}
  \caption{Comparison of the Soblex optimization
    algorithm~\cite{shams:kennedy:sadeghi:hartley:iccv:07} with the proposed
    Algorithm~\ref{alg:iterative:lower} in
    \secref{sec:optimize:lower:bound:1}. Algorithm~\ref{alg:iterative:lower}
    is superior to Soblex method both in terms of reliably finding the highest
    lower bound and computational efficiency.}
  \label{Tab3}
  \vspace{1em}
  \begin{center}
    \begin{tabular}{|c||c|c|c|c|}
      \hline
        \scriptsize{Algorithm Specifications} &
        \scriptsize{Mean lower bound,} &
        \scriptsize{Percentage of runs with} &
        \scriptsize{Number of cost calls / run} &
        \scriptsize{Average time} \\
        &
        \scriptsize{bits~/~channel use}&
        \scriptsize{a bound above $0.2970$ b/ch use} &
        \scriptsize{or iteration complexity} &
        \scriptsize{per run, sec.} \\
      \hline
      \hline
        \scriptsize{Soblex budget$=1000$}&$0.2803$&$3\%$&$5189$&$875$ \\
      \hline
        \scriptsize{Soblex budget$=2000$}&$0.2826$&$3\%$&$6149$&$1037$ \\
      \hline
        \scriptsize{Soblex budget$=5000$}&$0.2845$&$5\%$&$9127$&$1539$ \\
      \hline
        \scriptsize{Soblex budget$=20000$}&$0.2848$&$9\%$&$23916$&$4032$ \\
      \hline
        \scriptsize{Soblex budget$=50000$}&$0.2883$&$11\%$&$53939$&$9096$ \\
      \hline
        \scriptsize{\textbf{Algorithm~\ref{alg:iterative:lower},
        \secref{sec:optimize:lower:bound:1}}} &
        $\mathbf{0.3000}$&$\mathbf{100\%}$&$\mathbf{650 (50\times 13)}$&$\mathbf{109}$ \\
      \hline
    \end{tabular}
    \label{tab:lower:soblex}
  \end{center}
\end{table}

\subsection{\textbf{Original Partial Response Channel
                     with Memory Length $M=3$}}
\label{sec:num:partial:M:3}

In this subsection we study the convergence properties of the upper and lower
bound optimizations for a simple partial response original channel with memory
length $M = 3$. Because the original channel has memory length $M = 3$, there
are four normalized coefficients in \eqref{eq:fir:2} given as
\begin{align}
  [h_0, h_1, h_2, h_3]
    &= [0.5, 0.5, -0.5, -0.5].
\end{align}
This partial response channel is referred to as EPR4 in the
literature~\cite{Arnold:Loeliger:Vontobel:Kavcic:Zeng:06:1}. The quantization
function is based on partition points at
\begin{align}
  -\infty, \,
  -2.5, \,
  -2, \,
  -1.5, \,
  -1, \,
  -0.5, \,
  0, \,
  +0.5, \,
  +1, \,
  +1.5, \,
  +2, \,
  +2.5, \,
  +\infty,
\end{align}
resulting in $12$ quantized outputs and $\setY = \{1, \ldots, 12\}$. The SNR
is equal to $0$ dB.

The AF-FSMC / AB-FSMC chosen for this channel are partial response channels
with memory length $\Mhat = 2$ with $|\setShat| = 4$ states and $|\setBhat| =
8$ branches.

\subsubsection{Upper Bound Tests}

We studied the effect of three initialization methods discussed in
\secref{sec:partial:initialize} on the upper bound optimizations. We observed
that the convergence behavior of the upper bound in the EPR4 channel is
similar to that of CH3 in \figref{fig:upper:ch3:init:methods}. In particular,
very fast convergence behavior with initialization based on the ``truncation
initialization method'' or the ``optimized difference function initialization
method'' is observed, where the optimization algorithm virtually converges
within the first $10$ iterations. For example, when the ``optimized difference
function initialization method'' is used, the upper bound minimization
algorithm starts at $0.4536$ bits~/~channel use and reaches $0.4377$ bits /
channel use at the $10$-th iteration and does not decrease noticeably after
that. On the other hand, when the ``averaging initialization method'' is used,
the convergence of the minimization algorithm is poor and it is not recommended
for optimizations. For brevity purposes, these observations are not shown in a
separate figure.

\subsubsection{Lower Bound Tests}

\begin{figure}
  \begin{center}
    \includegraphics[width=\columnwidth]{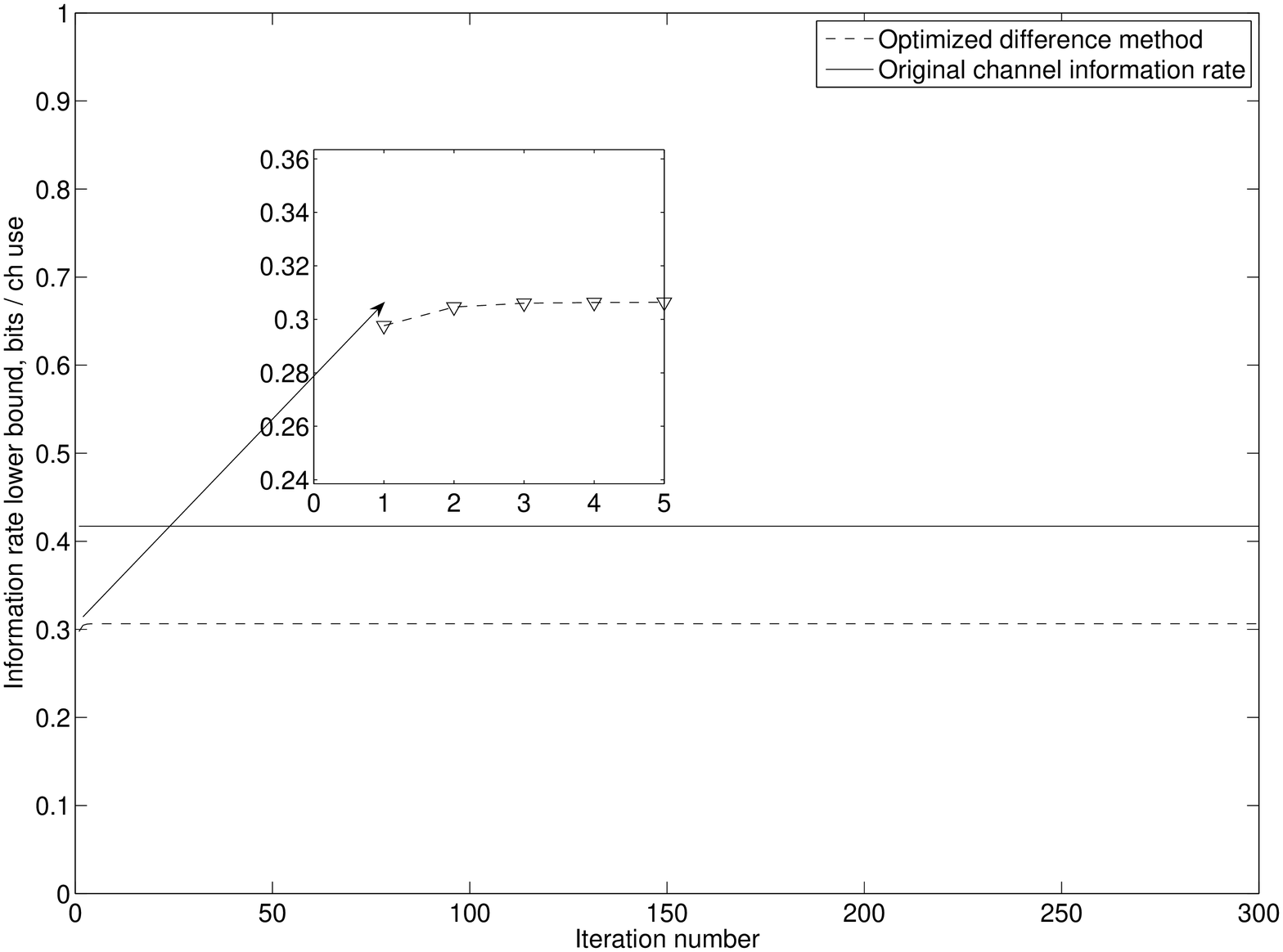}
  \end{center}
  \caption{Optimization of the lower bound for the EPR4 channel with $300$
    iterations at an SNR equal to $0$ dB. The initialization is based on the
    ``optimized difference function initialization method''. We used $\Gamma =
    \{1, 1.5, 2, 2.5, \ldots, 9.5, 10, 100\}$ in
    Algorithm~\ref{alg:ghat:update} for choosing $\gamma$ at each iteration,
    where $\gamma = 100$ has a stabilizing effect for $\vhat$.  After only a
    few iterations the lower bound reaches its maximum.}
  \label{fig:lower:epr4}
\end{figure}

\figref{fig:lower:epr4} shows the lower bound optimizations for the EPR4
channel for $300$ iterations, where initialization is based on the ``optimized
difference function initialization method''. As can be seen from this figure,
after only $2-3$ iterations the lower bound reaches its maximum. For this
test, we used $\Gamma = \{1, 1.5, 2, 2.5, \ldots, 9.5, 10, 100\}$ in
Algorithm~\ref{alg:ghat:update} for choosing $\gamma$ at each iteration, where
$\gamma = 100$ has a stabilizing effect for $\vhat$. In our analysis, we
observed that lower values of $\gamma$ result in drastically \emph{decreasing}
lower bounds over some iterations. In such cases, a large $\gamma$ prevents
this situation by choosing a $\vhat$ that is very close to the previous
$\vtilde$.  As an example, if we fix $\gamma = 1$ for all iterations, we
observe that the lower bound would decrease/increase over iterations in
oscillatory manner and the maximum lower bound obtained is slightly below the
lower bound with variable $\gamma$ (the difference is $7.2 \times 10^{-4}$
bits~/~channel use).  Therefore, using a variable $\gamma$ is beneficial for
both stabilizing and potentially providing higher lower bounds.

%% file: results_fading1.tex

\section{Numerical Results for Fading Channels}
\label{sec:num:fading}

In this section, we provide numerical results for the optimization of the
upper bound, the lower bound, and the difference function for fading channels.

\subsection{\textbf{Source, Channel, and Auxiliary Channel Models}}

Throughout this section, we assume that the source is characterized by
i.u.d.~binary constant power signaling (BPSK). The original channel is an
(output-quantized) correlated and flat-fading channel (FFC), also known as
frequency non-selective fading channel, and defined as
\begin{align}
  \label{eq:fading}
  y_{\ell}
    &= \mu
         \big(
           g_{\ell} x_{\ell}
           +
           n_{\ell}
         \big).
\end{align}
Here, $\mu$ is a quantization function that maps elements of $\mathbb{R}$ to
some discrete set $\setY$, $\{ n_{\ell} \}$ is a complex-valued AWGN process
with variance per dimension $N_{0}/2$ and independent real and imaginary
components, and $g_{\ell} \defeq a_{\ell} e^{j\theta_{\ell}}$ is the
complex-valued FFC gain, with the FFC amplitude $a_{\ell}$ and the FFC phase
$\theta_{\ell}$. The fading power is normalized to $\sigma^2_g = 0.5$ per
complex dimension. The average power of the channel input is
$\mathcal{E}_{\mathrm{s}}$. Since the fading power is normalized, the average
SNR per symbol is $\mathcal{E}_{\mathrm{s}} / N_{0}$.

The actual realization of the time-varying FFC gain $g_{\ell}$ is unknown to
the receiver and to the transmitter \emph{a priori}. It is assumed, however,
that the statistics of the time-varying FFC gain $g_{\ell}$ are known and do
not change over time. Hence, the fading process is stationary. We assume that
the fading process is a Gauss-Markov process as follows
\begin{align}
  \label{eq:gauss:markov}
  g_{\ell}
    &= \alpha g_{\ell-1}
       +
       w_{\ell}.
\end{align}
In this model, the FFC gain is complex-valued Gaussian distributed
with Rayleigh-distributed FFC channel amplitude $a_\ell$ and
uniformly distributed FFC channel phase $\theta_\ell$. In
\eqref{eq:gauss:markov}, $w_\ell$ is the complex-valued and white
Gaussian process noise. The Gauss-Markovian assumption is not
absolutely necessary for optimizing the information bounds,
especially the lower bound. However, it will facilitate obtaining
tight lower bounds for $H$ ($ = H(Y|X)$) for the original fading
channel, which is required for the computation of the upper bound
(see \eqref{eq:upper:analytical},
\secref{sec:information:rate:bound:1}). $\alpha$ in
\eqref{eq:gauss:markov} determines the correlation of the fading
channel gain at two successive time indices. We assume that this
correlation coefficient is given as $\alpha \triangleq
E\big\{G_{\ell}G^{*}_{\ell-1}\big\}/2\sigma^2_g = J_{0}\big(2\pi
f_{\mathrm{D}}T\big)$, which is the first correlation coefficient in
Clarke's model~\cite{Clarke1968,Proakis:00:1}. $J_{0}$ is the
zero-order Bessel function of the first kind, $f_{\mathrm{D}}$ is
the Doppler frequency shift, $T$ is the transmitted symbol period,
and the superscript $*$ denotes complex conjugate. Moreover,
according to Clarke's model, we note that the real and imaginary
parts of the fading process $\{ g_\ell \}$ and the process noise $\{
w_\ell \}$ are independent of each other.

Direct evaluation of information rates in correlated fading channels with no
channel state information is still an open problem. Therefore, having access
to tight upper and lower bound for such channels is valuable. It is noted that
fading channels are not naturally finite-state channels. Nevertheless, the
technique first proposed in~\cite{Arnold:Loeliger:Vontobel:02:1} (see
also~\cite{Arnold:Loeliger:Vontobel:Kavcic:Zeng:06:1}) is still able to
provide upper and lower bounds for non-finite state fading channels using
auxiliary FSMCs. The numerical results in this section will show that using
the optimization techniques introduced in this paper, we are able to
noticeably tighten these bounds on information rates of fading channels.

The AF-FSMC (and similarly the AB-FSMC) is chosen to have a trellis structure
with data-independent transitions (similar to the trellis that describes the
Gilbert-Elliott channel, cf.~Example~\ref{example:gilbert:elliott:channel:1}),
with input alphabet $\setX$, output alphabet $\setY$, and the same
quantization function $\mu$ as for the original channel. In particular, the
number of states will be $\hat K_a \cdot \hat K_{\theta}$, where $\hat K_a$
and $\hat K_{\theta}$ represent the number of auxiliary channel fading
amplitude quantization intervals and auxiliary channel phase quantization
intervals, respectively. Moreover, the transition between any two states is
possible. In our experiments, we will look at the effect of using different
values of $\hat K_a$ and $\hat K_{\theta}$ for the AF-FSMC / AB-FSMC.

\subsection{\textbf{Initialization Methods}}

For fading channels, we mainly use two methods for initializing the AF-FSMC
parameters.
\begin{enumerate}

\item \emph{Natural initialization method (based on FSMC modeling of the
    fading channel amplitude and phase)}. FSMC modeling of fading channel
  amplitude and phase has been used in the literature for a variety of
  purposes, including receiver design,
  see~\cite{Sadeghi_TCOM_2006,Sadeghi_SPM_2007} and the references therein for
  more details. In this initialization method, we model the fading channel
  amplitude $a_\ell$ with $\hat K_a$ states and the fading channel phase with
  $\hat K_\theta$ states and assume a first-order Markov transition between
  the states. This will result in a non-data-controllable AF-FSMC model with
  $\hat K_a \cdot \hat K_\theta$ states, where the initial AF-FSMC state
  transition probability $\What(\shat | \shatL, x)$ and channel law $\What(y |
  \bhat)$ are readily computable.

\item \emph{Initialization based on optimized difference function.} In this
  method, we select an initial $\{ \What(y|\bhat) \}$ that (locally)
  minimizes the difference function. Unlike the case of partial response
  channels, there is no closed-form expression for the minimum of the
  difference function. Therefore, we use Algorithm~\ref{alg:iterative:diff} to
  iteratively minimize the difference function, which in turn is initialized
  by the ``natural initialization method''.

\end{enumerate}
Similarly to \secref{sec:num:partial:response}, in all numerical
analyses, we use the information rate lower bound as defined in
Def.~\ref{def:information:rate:lower:1} with the setting $D_1 = 0$
and $D_2 = 0$ (and \emph{not} the specialized information rate lower
bound in
Rem.~\ref{remark:special:case:information:rate:lower:bound:2}).
Moreover, the parameters $\bigl\{ \vhat(\bhat, y) \bigr\}$ of the
AB-FSMC are initialized based on the initialization of $\bigl\{
\What(\hat s_{\ell} | \hat s_{\ell-1}, x_{\ell}) \bigr\}$ and
$\bigl\{ \What(y | \bhat) \bigr\}$. Namely, for any of the above two
initialization methods we set $\vhat(\bhat, y) \defeq \What(\hat
s_{\ell} | \hat s_{\ell-1}, x_{\ell}) \cdot \What(y | \bhat)$.

\subsection{\textbf{A Lower Bound on $H$}}

As mentioned in \eqref{eq:upper:analytical} in
\secref{sec:information:rate:bound:1}, computation of the upper bound requires
knowledge of the conditional entropy rate
\begin{align}
  H &\defeq
       \lim_{N\to \infty}\frac{1}{2N}H^{(N)}
     \defeq
       \lim_{N\to \infty}-
         \frac{1}{2N}
           \sumxy
             Q(\vx)
             W(\vy|\vx)
             \log
               \bigl(
                 W(\vy|\vx)
               \bigr)
\end{align}
for the original channel, which cannot be obtained using any auxiliary
channel. (A similar statement holds for the computation of the difference
function.) For partial response channels, this conditional entropy was readily
computable for the original channel, as it was the conditional entropy with
perfect CSI and only depended on the distribution of the AWGN. However, in
fading channels, $H$ is not easily computable. It should be mentioned that $H$
has a closed-form expression in the case that the above channel model is such
that the receiver does not use output-quantization and such that the sender
uses constant power signaling, e.g., BPSK~\cite{Haimovich2004}.  However, this
expression is not readily applicable here, because we need to quantize the
channel output $y_{\ell}$ for the optimization of the bounds.\footnote{The
  closed-form expression of $H$ given in~\cite{Haimovich2004} may provide an
  estimate of the conditional entropy for very finely quantized channel output
  and using the relation between discrete entropy and differential entropy
  given in~\cite[pp.  228-229]{Cover:Thomas:91}.}

In this subsection, we provide a lower bound on $H$, which is applicable for
any quantized channel output and can be made as tight as required (at the
expense of some computational complexity). Referring to~\cite[pp.
63-65]{Cover:Thomas:91}, we note that, using stationarity of the involved
processes, the conditional entropy rate can also be written as
\begin{align}
  H &= \lim_{N\to \infty}
         H(Y_0|\vY_{-N+1}^{-1},\vX_{-N+1}^{N}).
\end{align}
Using the fact that there is no feedback from the receiver to the transmitter
and the fact that, given the channel input up to time index $\ell$, the
channel output up to time index $\ell$ is independent of the channel input
from time $\ell + 1$ on, this can be written as
\begin{align}
  H &= \lim_{N\to \infty}
         H(Y_0|\vY_{-N+1}^{-1},\vX_{-N+1}^{0}).
\end{align}
Using the fact that conditioning cannot increase entropy, we can obtain lower
bounds on $H$. A good candidate for conditioning is a past fading channel gain
such as $G_{-D}$ for $D \ge 1$. To see this, let us elaborate on
$H(Y_0|\vY_{-N+1}^{-1},\vX_{-N+1}^{0})$ as follows
\begin{align}
  H(Y_0|\vY_{-N+1}^{-1},\vX_{-N+1}^{0})
    &\ge H(Y_0|\vY_{-N+1}^{-1},\vX_{-N+1}^{0}, G_{-D}) \\
    &=   H(Y_0|\vY_{-D+1}^{-1},\vX_{-D+1}^{0}, G_{-D})
           \label{eq:gauss:markov:independence}\\
    &=   H(Y_D|\vY_{1}^{D-1},\vX_{1}^{D}, G_{0}).
\end{align}
In \eqref{eq:gauss:markov:independence} we have used the fact that $Y_0$ given
$G_{-D}$ is independent of all channel inputs and outputs up to and including
time index $-D$, \ie, independent of $\vY_{-N+1}^{-D}$ and $\vX_{-N+1}^{-D}$.
Finally, the last equality follows form the channel being stationary. It can
also be shown that by increasing the delay $D$ in the above, tighter lower
bounds on $H$ will be obtained.

As an example of how the above lower bound may be computed, let us consider
the case where the conditioning delay is $D=2$. Then, we need to find
\begin{align}
  \label{eq:D:2}
  H(Y_2|Y_{1},\vX_{1}^{2}, G_{0})
    &= H(\vY_1^2|\vX_{1}^{2}, G_{0})-H(Y_1|\vX_{1}^{2}, G_{0}) \\
    &= H(\vY_1^2|\vX_{1}^{2}, G_{0})-H(Y_1|X_{1}, G_{0}),
\end{align}
which requires knowledge about $P(\vy_1^2|\vx_{1}^{2}, g_{0})$ and
$P(y_{1}|x_{1}, g_{0})$. These quantities can be computed by integrating
over the \emph{missing} channel gains. Let us first consider $P(y_{1}|x_{1},
g_{0})$ which can be written as
\begin{align}\label{eq:D:2:expand}
  P(y_{1}|x_{1}, g_{0})
    &= \int
         P(y_{1}, g_{1}|x_{1}, g_{0})
       \, \dint{g_{1}} \\
    &= \int
         P(y_{1}|x_{1}, g_{1}) \, f(g_{1}|g_{0})
       \, \dint{g_{1}},
         \label{eq:D:2:expand:2}
\end{align}
where we have used the Gauss-Markovian property of the fading process and the
fact that according to \eqref{eq:fading}, $y_\ell$ given $x_\ell$ and $g_\ell$
is independent of all other previous channel gains. In
\eqref{eq:D:2:expand:2}, $P_{Y_1|X_1, G_1}(k|x_{1}, g_{1})$ is the area under
the AWGN distribution between the two quantization points
$y^{\mathrm{q}}_{k-1}$ and $y^{\mathrm{q}}_{k}$ with a mean of $x_{1}g_{1}$
and variance of $N_0/2$ per complex dimension. Also, $f(g_{1}|g_{0})$ is the
probability density function of $g_1$ given $g_{0}$, which according to
\eqref{eq:gauss:markov} is Gaussian with mean $\alpha g_{0}$ and variance
$\sigma_g^2-\sigma_g^2\alpha^2 = 0.5(1-\alpha^2)$ per complex dimension.

Similarly, $P(\vy_1^2|\vx_{1}^{2}, g_{0})$ is expanded as
\begin{align*}
  P(\vy_1^2|\vx_{1}^{2}, g_{0})
    &= \int
         \int
           P(\vy_1^2, g_2, g_{1}|\vx_{1}^{2}, g_{0})
         \, \dint{g_{2}}
       \, \dint{g_{1}} \\
    &= \int
         P(y_1|g_1,x_{1}) \, f(g_1|g_0)
         \,
         \left(
           \int
             P(y_2|g_2,x_{2}) \, f(g_2|g_1)
           \, \dint{g_{2}}
         \right)
       \, \dint{g_{1}}.
\end{align*}
In principle, computations of the above integrals should be carried out in the
complex domain, because both channel output $y_{\ell}$ and channel gain
$g_{\ell}$ are complex-valued. However, according to the independence property
of the real and the imaginary parts of the AWGN and the fading process, one
can simplify computations greatly by only evaluating the probabilities either
for the real or for the imaginary dimension, computing the corresponding
entropies, and multiplying the obtained result by a factor of $2$. Evaluation
of the lower bound can be similarly extended to larger conditioning delays
$D$, which result in tighter lower bounds, but at the cost of more
computational complexity.

\begin{figure}
  \mbox{} \\
  \begin{center}
    \includegraphics[width=\columnwidth]{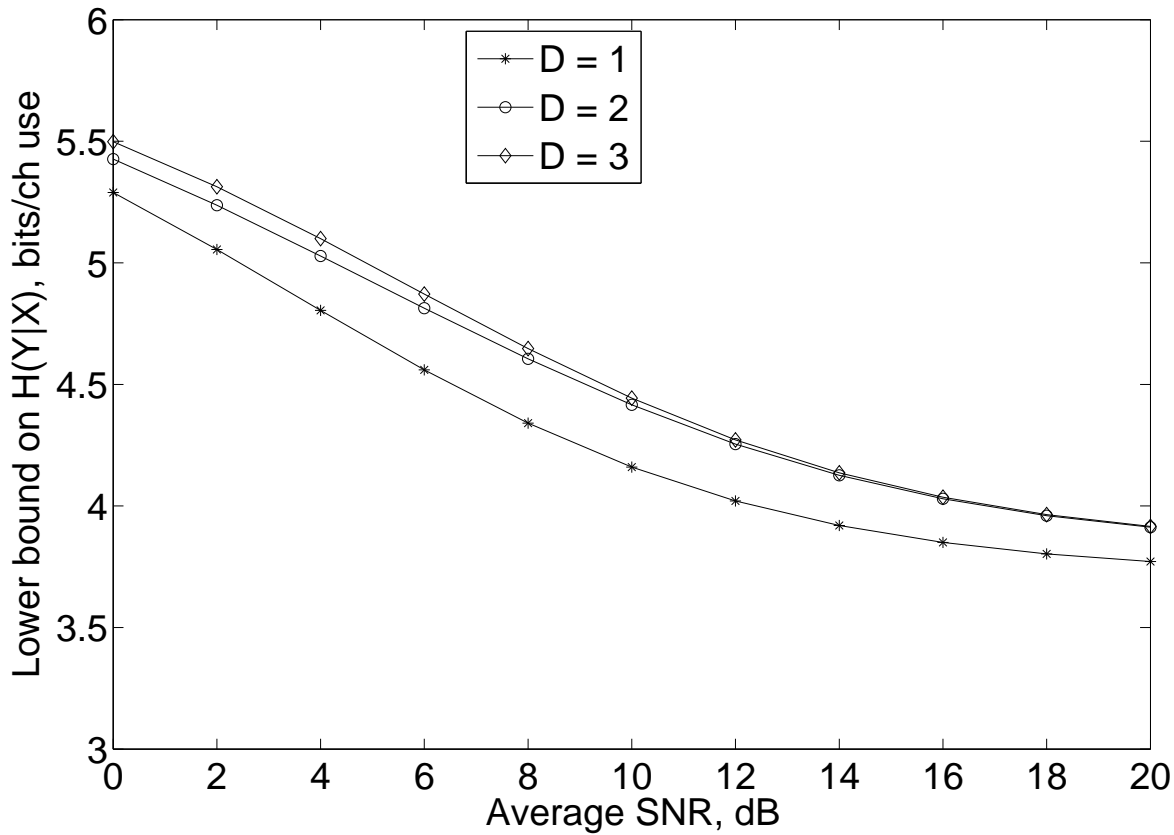}
  \end{center}
  \caption{Lower bounds on the conditional entropy rate $H = H(Y|X)$ in the
    original Gauss-Markov fading channel with various conditioning delays from $D=1$ to $D=3$ as a function of the average SNR
    $\mathcal{E}_{s}/N_{0}$. The Gauss-Markov fading channel has $\alpha =
    J_{0}\big(2\pi f_{\mathrm{D}}T\big)$ and the normalized Doppler frequency shift is
    $f_{\mathrm{D}}T = 0.1$.}
  \label{fig:lower:hyx:fading}
\end{figure}

We have computed lower bounds on $H$ using delays from $D=1$ to
$D=3$ with a reasonable complexity. \figref{fig:lower:hyx:fading}
shows the lower bound for conditioning delays from $D=1$ to $D=3$
for the Gauss-Markov fading channel with $\alpha = J_{0}\big(2\pi
f_{\mathrm{D}}T\big)$ and the normalized Doppler frequency shift
$f_{\mathrm{D}}T = 0.1$. The horizontal axis shows the average SNR.
Each real and imaginary component of the channel output $y_{\ell}$
is quantized into $10$ intervals using a quantization function that
is based on partition points at
\begin{align}
  -\infty, \
  -2.0\sigma_y, \
  -1.5\sigma_y, \
  -1.0\sigma_y, \
  -0.5\sigma_y, \
   0, \
  +0.5\sigma_y, \
  +1.0\sigma_y, \
  +1.5\sigma_y, \
  +2.0\sigma_y, \
  +\infty
\end{align}
where $\sigma^2_y$ is the average output variance per dimension and defined as
$\sigma^2_y = \sigma^2_g\mathcal{E}_{s}+N_0/2$. We note that for high SNR
conditions, the entropy rate $H$ visually reaches its limit with the
conditioning delay of $D=3$. In the following subsections, we will see that
even with such modest conditioning delays, we can obtain good upper bounds for
fading channels.

\subsection{\textbf{Optimizing the Difference Function}}
\label{sec:num:fading:diff}

\begin{figure}
  \begin{center}
    \includegraphics[width=\columnwidth]{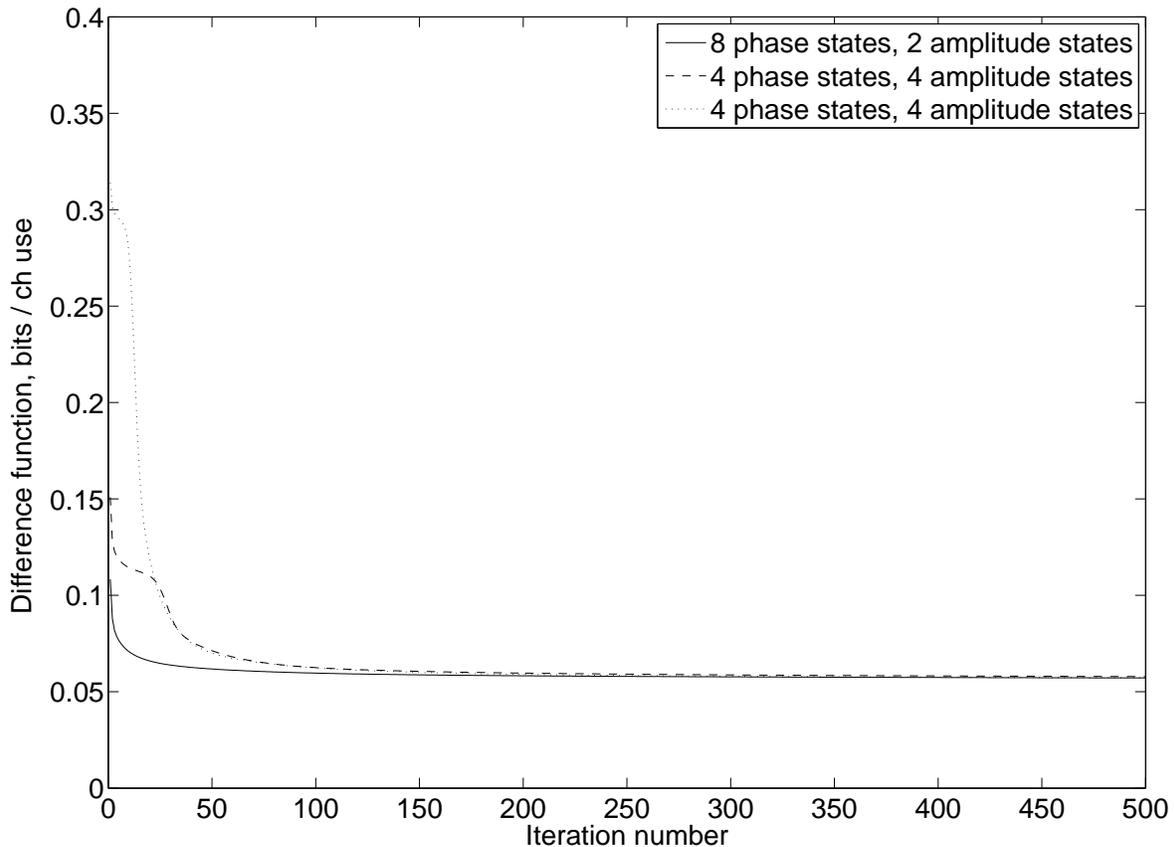}
  \end{center}
  \caption{Comparison of the difference function optimization algorithm with
    initialization by the ``natural initialization method''. The original
    fading channel has a normalized Doppler frequency shift of
    $f_{\mathrm{D}}T = 0.1$ and the average SNR is $0$ dB. The AF-FSMC model
    has $16$ states. The ``natural initialization method'' is applied with
    three different mappings of the fading channel amplitude and phase into
    the AF-FSMC model.}
  \label{fig:fading:diff:init:methods}
\end{figure}

\figref{fig:fading:diff:init:methods} shows the iterative optimization of the
difference function. The original fading channel has a normalized Doppler
frequency shift of $f_{\mathrm{D}}T = 0.1$ and the average SNR is $0$ dB. The
window length of the channel output $\vy$ for optimizations is $2N = 1\times
10^6$.  The number of states in the AF-FSMC model is set to $16$ states.
Initialization is done according to the ``natural initialization method''
using three different mappings of the fading channel amplitude and phase into
the FSMC model. The difference function is computed using the lower bound for
$H$ with $D = 3$ as discussed above. Therefore, the values shown are, in fact,
upper bounds on the difference function values. The figure only shows the
first $500$ iterations of a total of $3000$ iterations, because the
improvement in the difference function after the first $500$ iterations was
less than $1.9 \times 10^{-3}$ bits~/~channel use. Although the optimization
algorithm eventually converge to the same difference value (independently of
the initialization method), the optimization run with an AF-FSMC with more
phase states ($\hat K_{\theta} = 8$) and less amplitude states ($\hat K_a =
2$) converges faster. Also note that after a few iterations, the value of the
difference function for this SNR is, at most, only about $0.06$ bits~/~channel
use, which shows the effectiveness of the optimization technique.

\subsection{\textbf{Upper Bound Tests}}
\label{sec:num:fading:upper}

\begin{figure}
  \begin{center}
    \includegraphics[width=\columnwidth]
                    {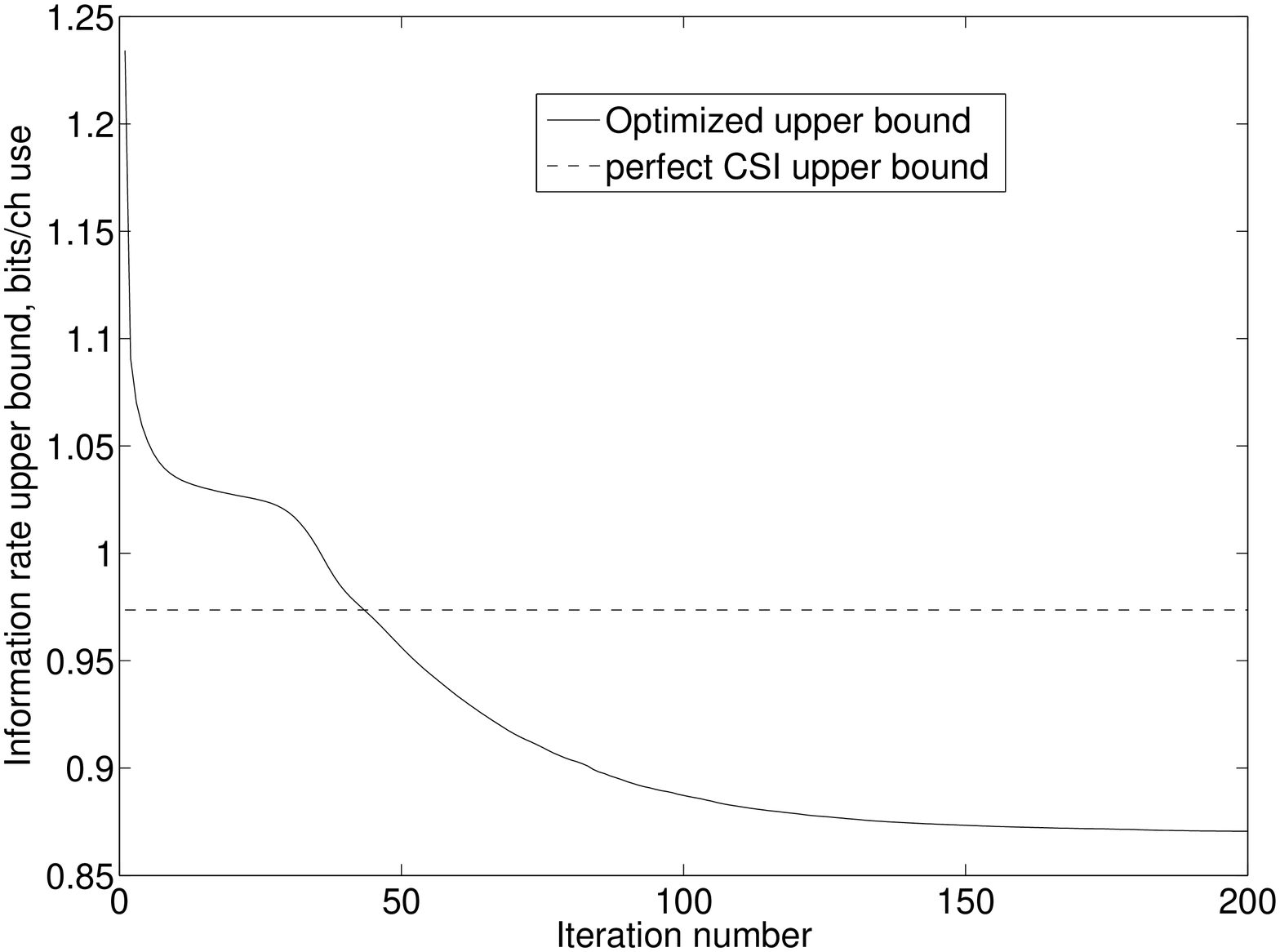}
  \end{center}
  \caption{Convergence of the upper bound optimizations using the ``natural
    initialization method'' of the Gauss-Markov fading channel into $\hat
    K_{\theta} = 8$ fading phase states and $\hat K_{a} = 2$ fading amplitude
    states. The average SNR is $16$ dB. From first to last iteration, one is
    able to reduce the upper bound significantly (by $0.36$ bits~/~channel
    use), well below the CSI upper bound, if enough number of iterations are
    allowed.}
  \label{fig:fading:upper:init:methods:16dB}
\end{figure}

\figref{fig:fading:upper:init:methods:16dB} shows $200$ iterations for the
optimizing of the upper bound. The original fading channel has a normalized
Doppler frequency shift of $f_{\mathrm{D}}T = 0.1$ and the average SNR is $16$
dB. We have used the ``natural initialization method'' with $\hat K_{\theta} =
8$ phase states $\hat K_{a} = 2$ amplitude states. It is observed from this
figure that if enough number of iterations are allowed, one is able to
optimize the upper bound well below the CSI upper bound (the new upper bound
is $0.8643$ bits~/~channel use, whereas the CSI upper bound is $0.9736$
bits~/~channel use.)

Furthermore, we observed that initializing the upper bound optimization
algorithm by the ``optimized difference function initialization method'' may
provide higher initial upper bounds that do not improve with further
iterations. This may be due to the fact that this initialization method yields
parameters where the upper bound has a local minimum, whereas the ``natural
initialization method'' yields parameters at a non-stationary point of the
upper bound, which can subsequently be improved iteratively. So for the upper
bound, the ``natural initialization method'' with about $100$ to $200$
iterations is recommended.

\subsection{\textbf{Lower Bound Tests}}
\label{sec:num:fading:lower}

\tableref{tab:lower:fading} shows examples of obtained lower bounds based on a
variety of initialization methods. Namely, the ``natural initialization
methods'' and the ``optimized difference function initialization method'' are
used with four and five, respectively, different settings of amplitude and
phase states.

The original Gauss-Markov fading channel has a normalized Doppler frequency
shift of $f_{\mathrm{D}}T = 0.1$ and the average SNR is $0$ dB. The number of
states in the AB-FSMC model is set to $16$ states. We used $\Gamma = \{1, 1.5,
2, 2.5, \ldots, 9.5, 10, 100\}$ in Algorithm~\ref{alg:ghat:update} for
choosing $\gamma$ at each iteration, where $\gamma = 100$ has a stabilizing
effect for $\vhat$ and does not let the lower bound decrease over iterations.
It is clear from this table that initialization based on the ``optimized
difference function initialization method'' can result in a higher information
rate lower bound. For this initialization method we also observe that
increasing the number of iterations from $50$ to $3000$ yields an improvement
of only $0.0028$ bits~/~channel use, which is negligible for all practical
purposes.

Finally, we would like to emphasize the importance of optimizing the lower
bound from the mismatched decoding perspective. FSMC models have been proposed
in the literature~\cite{Goldsmith1996, Goldsmith2002, Wesel2001,
  Collins_ISIT2006, Shwedyk1999, Shwedyk2003} for channel estimation and
decoding in time-varying, continuous-valued, and continuous state-space fading
channels (see~\cite{Sadeghi_SPM_2007} for a review of FSMC-based decoding in
fading channels). However, the fact that such decoding is mismatched to the
physical fading channel often goes unnoticed. In particular, the choice of
FSMC parameters (type of FSMC, number of amplitude and phase states, the FSMC
state transition probabilities, and the FSMC output probabilities) are often
chosen on an ad-hoc basis and mostly using the natural mapping of fading
channel amplitude and phase into FSMC models. On the other hand, the results
in this section showed that by optimizing the AB-FSMC parameters, we can
achieve higher lower bounds on mismatched decoding rates and therefore achieve
higher information rates in AB-FSMC-based decoding for fading channels.

\begin{table}
  \caption{The effect of different initialization methods on the lower
    bound. Better lower bounds are often obtained if the ``optimized
    difference function initialization method'' is used,
    which was observed to be a
    typical behavior of the lower bound across SNR.}
  \label{Tab4}
  \vspace{1em}
  \begin{center}
    \begin{tabular}{|c||c|}
      \hline
        \scriptsize{Initialization Method} &
        \scriptsize{Best Lower Bound Found, bits~/~channel use} \\
      \hline
      \hline
        \scriptsize{Natural, $\hat K_{\theta} = 16$, $\hat K_a = 1$}& $0.3357$ \\
      \hline
        \scriptsize{Natural, $\hat K_{\theta} = 8$, $\hat K_a = 2$} & $0.3501$ \\
      \hline
        \scriptsize{Natural, $\hat K_{\theta} = 4$, $\hat K_a = 4$} & $0.3345$ \\
      \hline
        \scriptsize{Natural, $\hat K_{\theta} = 2$, $\hat K_a = 8$} & $0.3056$ \\
      \hline
        \scriptsize{$50$ Iterations for Optimized Diff,
                    $\hat K_{\theta} = 16$, $\hat K_a = 1$} &
        $0.3562$ \\
      \hline
        \scriptsize{$\mathbf{50}$ \textbf{Iterations for Optimized Diff,}
                    $\mathbf{\hat K_{\theta} = 8}$, $\mathbf{\hat K_a = 2}$} &
        $\mathbf{0.3580^{*}}$ \\
      \hline
        \scriptsize{$\mathbf{3000}$ \textbf{Iterations for Optimized Diff,}
                    $\mathbf{\hat K_{\theta} = 8}$, $\mathbf{\hat K_a = 2}$} &
        $\mathbf{0.3608^{**}}$ \\
      \hline
        \scriptsize{$50$ Iterations for Optimized Diff, $\hat K_{\theta} = 4$,
                    $\hat K_a = 4$} &
        $0.3539$ \\
      \hline
        \scriptsize{$50$ Iterations for Optimized Diff, $\hat K_{\theta} = 2$,
                    $\hat K_a = 8$}&$0.3529$ \\
      \hline
    \end{tabular}
    \label{tab:lower:fading}
  \end{center}
\end{table}

\subsection{\textbf{Optimized Bounds Versus SNR}}
\label{sec:num:fading:snr}

\begin{figure}
  \begin{center}
    \includegraphics[width=\columnwidth]{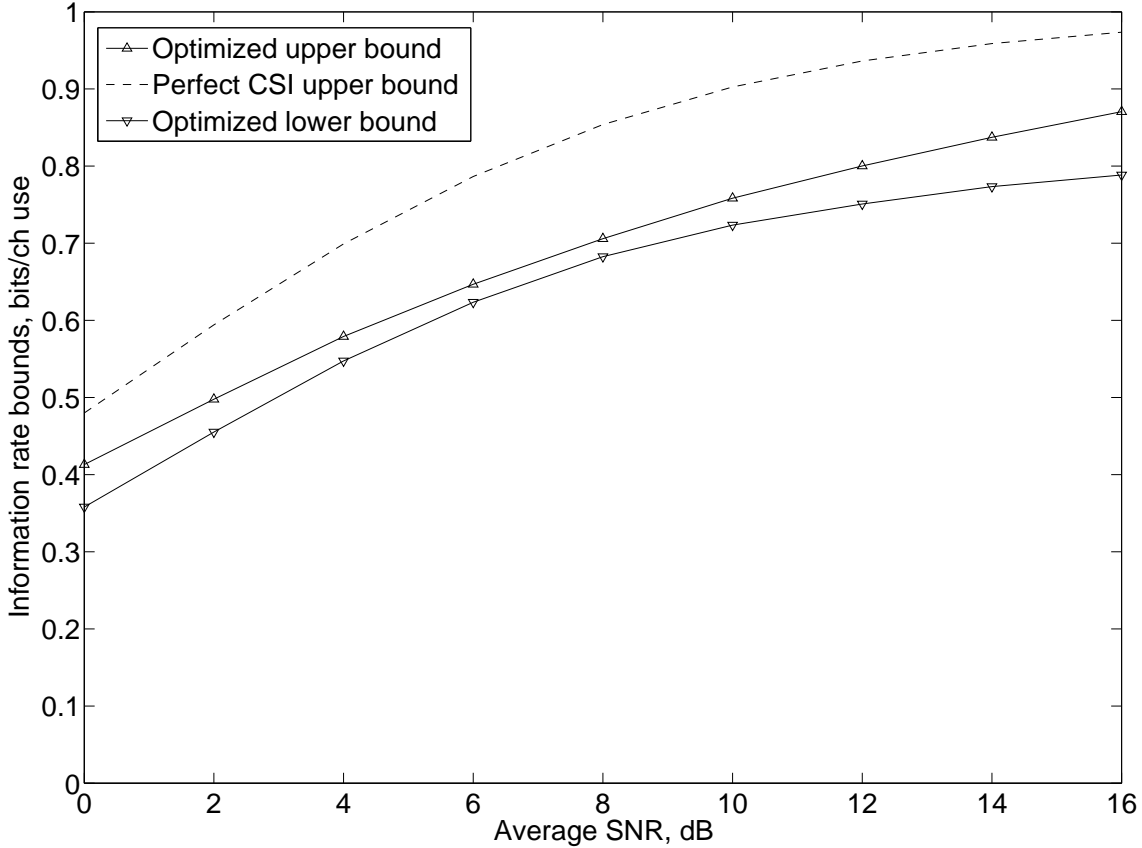}
  \end{center}
  \caption{Optimized upper and lower bounds for a practical range of SNR and
    for the Gauss-Markov fading channel with the normalized fading rate of
    $f_{\mathrm{D}}T = 0.1$. The upper bound is considerably tighter than the
    CSI upper bound and together with the lower bound, can successfully limit
    the range of possible information rates.}
  \label{fig:fading:snr}
\end{figure}

\figref{fig:fading:snr} shows optimized upper and lower bounds for a practical
range of SNRs and for the considered Gauss-Markov fading channel as in
previous figures with a relatively fast fading rate of $f_{\mathrm{D}}T =
0.1$. We used $\Gamma = \{1, 1.5, 2, 2.5, \ldots, 9.5, 10, 100\}$ in
Algorithm~\ref{alg:ghat:update} for choosing $\gamma$ in the lower bound
optimizations.  The upper bound is considerably tighter than the CSI upper
bound and together with the lower bound, can successfully bound the range of
possible information rates.  To the best of authors' knowledge, no such good
bounds have previously been shown in the literature.

%% file: conclusions1.tex

\section{Conclusions}
\label{conclusion}

In this paper, we devised iterative algorithms for the minimization of an
information rate upper bound and the maximization of an information rate lower
bound for communication channels with memory. Moreover, we also discussed the
minimization of the so-called difference function which represents the
difference between the upper bound and a specialized version of the lower
bound.

Our proposed optimization techniques are EM-type algorithms and optimize
auxiliary FSMC models under the constraint that the (time-invariant) trellis
is fixed. We applied the optimization techniques to original channels that are
finite state (such as partial response channels) and non-finite state (such as
fading channels). In both cases, we observed that the optimization techniques
considerably tighten the bounds with a reasonable computational complexity.
This is particularly important for finite-state channels with a large memory
length ($M \ge 10$), where the numerical computation of the information rate
is complex, or for non-finite channels, where direct or numerical evaluation
of the information rate is not possible. Using the proposed techniques, we
improved the upper bound of the considered fading channel by as much as 0.11
bits / channel use below the CSI upper bound in binary signaling, which
together with the optimized lower bound, successfully provided bounds for the
fading channel information rates.  Optimizing the lower bound is also
significant from a mismatched decoding viewpoint for receivers that are
equipped with the decoding metric for the auxiliary FSMC model, which is
mismatched to the original communication channel.

%% file: appendices.tex

\appendices

\section{Proof of the Statement in Rem.~\ref{remark:special:case:information:rate:lower:bound:2}}
\label{app:proof:remark:special:case:information:rate:lower:bound:2}

In this appendix we verify that $\sum_{\vshat, \vx} \vhat(\vshat, \vx, \vy) =
1$ if $\sum_{\shat_{\ell}} \vhat( \shat_{\ell-1}, x_{\ell}, \shat_{\ell},
\vy_{\ell-D_1}^{\ell+D_2}) = 1$ for all $\shat_{\ell-1}$, all $x_{\ell}$, and
all $\vy_{\ell-D_1}^{\ell+D_2}$. Indeed,
\begin{align}
  \sum_{\vshat, \vx}
    \vhat(\vshat, \vx, \vy)
    &= \sum_{\vshat, \vx}
         \prod_{\ell \in \sIN}
           Q
           \left(
             x_{\ell}
             \left|
             x_{-N+1}^{\ell-1}
             \right.
           \right)
           \vhat
             \left(
               \shat_{\ell-1}, x_{\ell}, \shat_{\ell}, \vy_{\ell-D_1}^{\ell+D_2}
             \right) \\
    &= \sum_{\vshat_{-N+1}^{N-1}, \, \vx_{-N+1}^{N-1}}
         \prod_{\ell = -N+1}^{N-1}
           Q
           \left(
             x_{\ell}
             \left|
             x_{-N+1}^{\ell-1}
             \right.
           \right)
           \vhat
             \left(
               \shat_{\ell-1}, x_{\ell}, \shat_{\ell}, \vy_{\ell-D_1}^{\ell+D_2}
             \right) \\
    &\hspace{5cm}
           \cdot \
           \underbrace{
             \sum_{x_N}
               Q
               \left(
                 x_N
                 \left|
                 x_{-N+1}^{N-1}
                 \right.
               \right)
               \underbrace{
                 \sum_{\shat_N}
                   \vhat
                     \left(
                       \shat_{N-1}, x_{N}, \shat_{N}, \vy_{N-D_1}^{N+D_2}
                     \right)
                 }_{= 1}
           }_{= 1} \\
    &= \sum_{\vshat_{-N+1}^{N-1}, \vx_{-N+1}^{N-1}}
         \prod_{\ell = -N+1}^{N-1}
           Q
           \left(
             x_{\ell}
             \left|
             x_{-\ell+1}^{N-1}
             \right.
           \right)
           \vhat
             \left(
               \shat_{\ell-1}, x_{\ell}, \shat_{\ell}, \vy_{\ell-D_1}^{\ell+D_2}
             \right) \\
    &= \cdots \\
    &= 1.
\end{align}

\section{Proof of Lemma~\ref{lemma:properties:upper}}
\label{app:upper:property}

For ease
of reference, let us repeat here the relevant surrogate function:
\begin{align}\nonumber
    \Psiupper^{(N)}(\Wtilde, \What \condcs)
    &\defeq
       \Iupper^{(N)}(\What \condcs)
       +
       \frac{1}{2N}
       \sum_{\vy}
         (QW)(\vy \condvs) \,
         D_{\vbhat}
           \left(
               \Ptilde(\vbhat | \vy \condcs)
             \left\lVert
               \Phat(\vbhat | \vy \condcs)
             \right.
           \right).
\end{align}

\subsection{Property 1}

This follows from standard KL divergence properties.

\subsection{Property 2}

Using the definition of $\Iupper^{(N)}(\What \condcs)$ in
\eqref{eq:information:rate:upper:1}, we obtain
\begin{align}\nonumber
  \Psiupper^{(N)}(\Wtilde, \What \condcs)
    &= \frac{1}{2N}
       \sumxy
         Q(\vx)
         W(\vy | \vx \condvs)
         \log
           \left(
             W(\vy|\vx \condcs)
           \right)\\\nonumber
    &\quad\
       -
       \frac{1}{2N}\sum_{\vy}
         (QW)(\vy \condvs)
         \sum_{\vbhat}
           \Ptilde(\vbhat|\vy \condcs)
           \log
             \left(
               (Q\What)(\vy \condvs)
             \right)\\ \nonumber
    &\quad\
       + \frac{1}{2N}\sum_{\vy}
         (QW)(\vy \condvs)
         \sum_{\vbhat}
           \Ptilde(\vbhat | \vy \condcs)
           \log
             \left(
               \Ptilde(\vbhat | \vy \condcs)
             \right)\\
    &\quad\
       - \frac{1}{2N}\sum_{\vy}
         (QW)(\vy \condvs)
         \sum_{\vbhat}
           \Ptilde(\vbhat|\vy \condcs)
           \log
             \left(
               \Phat(\vbhat | \vy \condcs)
             \right).
\end{align}
We note that the first and third terms on the right-hand side are only
functions of $Q$, $W$, $\Wtilde$, and are independent of $\What$. By combining
these two terms as $\overline{c}^{(N)}_1(\Wtilde \condcs)$ and by combining
the second and fourth terms, we can write
\begin{align}\nonumber
     \Psiupper^{(N)}(\Wtilde, \What \condcs)
    &= \overline{c}^{(N)}_1(\Wtilde \condcs)
       -\frac{1}{2N}\sum_{\vy}
         (QW)(\vy \condvs)
         \sum_{\vbhat}
         \Ptilde(\vbhat|\vy \condcs)
         \log
           \left(
             \Phat(\vbhat,\vy  \condvs)
           \right).
\end{align}
Now using the AF-FSMC's decomposition property~\eqref{eq:aux:FSMC:decompose:2}
for $\Phat(\vbhat,\vy \condvs) = \Phat(\vx,\vshat,\vy \condvs)$, we can
simplify the surrogate function further to
\begin{align}\nonumber
     \Psiupper^{(N)}(\Wtilde, \What \condcs)
    &= \overline{c}^{(N)}_1(\Wtilde \condcs) \\ \nonumber
    &\quad\
       -
       \frac{1}{2N}\sum_{\vy}
         (QW)(\vy \condvs)
         \sum_{\vbhat}
         \Ptilde(\vbhat|\vy \condcs)
         \log
           \left(
             Q(\vx  \condvs)
           \right)\\\nonumber
    &\quad\
       -
       \frac{1}{2N}\sum_{\vy}
         (QW)(\vy \condvs)
         \sum_{\vbhat}
         \Ptilde(\vbhat|\vy \condcs)
         \ellsum\log
           \left(
             \What(\shat_{\ell} | \shat_{\ell-1}, x_{\ell})
           \right)\\\label{eq:upper:What:2}
    &\quad\
       -
       \frac{1}{2N}\sum_{\vy}
         (QW)(\vy \condvs)
         \sum_{\vbhat}
         \Ptilde(\vbhat|\vy \condcs)
         \ellsum\log
           \left(
             \What(y_{\ell} | \bhat_{\ell})
           \right),
\end{align}
where we notice that the second term on the right-hand side is independent of
$\What$. Therefore, by combining $\overline{c}^{(N)}_1(\Wtilde \condcs)$ and
this term into $\overline{c}^{(N)}(\Wtilde \condcs)$ and after a few
manipulations of the third and fourth terms, we obtain
\begin{align}
     \Psiupper^{(N)}(\Wtilde, \What \condcs)
    &= \overline{c}^{(N)}(\Wtilde \condcs)
       -
       \sum_{\bhat}
         \log
         \left(
           \What(\shat | \shatL, x)
         \right)\Ttilde^{(N)}_1(\bhat)
       -
       \sum_{\bhat}
       \sum_{y}
         \log
           \left(
             \What(y | \bhat)
           \right)
       \Ttilde^{(N)}_2(\bhat,y),\label{eq:app:upper:surrogate:3}
\end{align}
where $\Ttilde^{(N)}_1(\bhat)$ and $\Ttilde^{(N)}_2(\bhat,y)$ were defined
in~\eqref{eq:ttilde1} and~\eqref{eq:ttilde2}, respectively.

\subsection{Property 3}

In order to prove convexity of $\Psiupper^{(N)}(\Wtilde, \What \condcs)$
w.r.t.\ $\bigl\{ \What(\shat | \shatL, x) \bigr\} \cup \bigl\{ \What(y |
\bhat) \bigr\}$, we need to show that the corresponding Hessian matrix is
positive semi-definite. This can be done as follows. First, we use the
expression in~\eqref{eq:app:upper:surrogate:3} for $\Psiupper^{(N)}(\Wtilde,
\What \condcs)$ to obtain the following second-order partial derivatives:
\begin{align}
  &
  \ddiff{\What(\shat | \shatL, x)}\Psiupper^{(N)}
     = \frac{\Ttilde^{(N)}_1(\bhat)}{(\What(\shat | \shatL, x))^2},
  \quad\quad
  \ddiff{\What(y | \bhat)}\Psiupper^{(N)}
     = \frac{\Ttilde^{(N)}_2(\bhat)}{(\What(y | \bhat))^2}, \\
  &
  \fddiff{\What(\shat | \shatL, x)}{\What(\shat' | \shatL', x')}
    \Psiupper^{(N)}
     = 0,
  \
  \fddiff{\What(y | \bhat)}{\What(y' | \bhat')}
    \Psiupper^{(N)}
     = 0,
  \
  \fddiff{\What(\shat | \shatL, x)}{\What(y' | \bhat')}
    \Psiupper^{(N)}
     = 0.
\end{align}
Secondly, we note that $\Ttilde^{(N)}_1(\bhat)$ and $\Ttilde^{(N)}_2(\bhat,y)$
are non-negative, cf.~\eqref{eq:ttilde1} and~\eqref{eq:ttilde2}. Thirdly, we
combine these results and see that the Hessian matrix is diagonal with
non-negative diagonal elements, \ie, the Hessian matrix is indeed positive
semi-definite.

Let us remark that in the above computations of derivatives we did not impose
the constraint that $\bigl\{ \What(\shat | \shatL, x) \bigr\}$ and $\bigl\{
\What(y | \bhat) \bigr\}$ lie in the corresponding probability simplices for
every $(\shatL, x)$ and every $\bhat$, respectively, and therefore we actually
proved a stronger convexity result than really needed.  Note that this
approach of proving convexity worked well because the surrogate function is
well-behaved also outside these probability simplices. This is in contrast
to~\cite{Vontobel:Kavcic:Arnold:Loeliger:04:1:subm} where information rates
where optimized as a function of source probabilities: for $N \to \infty$, the
information rate was not well defined outside the corresponding simplices and
so it was important to take directional derivatives within the simplices. For
more details we refer the reader
to~\cite{Vontobel:Kavcic:Arnold:Loeliger:04:1:subm}.

\section{Proof of Lemma~\ref{lemma:minimize:upper}}
\label{app:upper:minimize}

In this appendix, we minimize the surrogate function $\Psiupper^{(N)}(\Wtilde,
\What \condcs)$ subject to the following constraints
\begin{align}
  \label{eq:app:upper:const1}
  \sum_{\shat}
    \What(\shat | \shatL, x)
    &= 1
         \quad (\text{for all $(\shatL, x) \in \setShat \times \setX$}), \\
  \label{eq:app:upper:const2}
  \sum_{y}
    \What(y | \bhat)
    &= 1
         \quad (\text{for all $\bhat \in \setBhat$}).
\end{align}
(For the moment we omit the non-negativity constraints; we will see at the end
that they are automatically satisfied.) Clearly, the Lagrangian is
\begin{align}
  \label{eq:upper:Lagrange}
  L
  &\defeq
     \Psiupper^{(N)}(\Wtilde,\What \condcs)
     -
     \sum_{\shatL, x}
       \mu_1(\shatL, x)
       \left(
         \sum_{\shat}\What(\shat | \shatL, x)- 1
       \right)
     -
     \sum_{\bhat}
       \mu_2(\bhat)
       \left(
         \sum_{y}\What(y | \bhat)- 1
       \right),
\end{align}
where $\bigl\{ \mu_1(\shatL, x) \bigr\}$ and $\bigl\{ \mu_2(\bhat) \bigr\}$
are Lagrange multipliers. Using the expression in~\eqref{eq:upper:surrogate:3}
for $\Psiupper^{(N)}(\Wtilde,\What \condcs)$ and setting the derivative of $L$
w.r.t.~$\What(\shat | \shatL, x)$ equal to zero yields
\begin{align}
  \label{eq:upper:Lagrange:derivative1}
  \diff{\What(\shat | \shatL, x)} L
  &= \frac{-1}{\What(\shat | \shatL, x)}
     \Ttilde^{(N)}_1(\bhat)
     -
     \mu_1(\shatL, x)
   \overset{!}{=}
     0,
\end{align}
which results in
\begin{align}
  \label{eq:app:update:What:upper:1}
  \What^{*}(\shat | \shatL, x)
    &= \frac{\Ttilde_1(\bhat)}{-\mu_1(\shatL, x)}.
\end{align}
Therefore, the Lagrange multiplier $\mu_1(\shatL, x)$ is just a normalization
constant so that $\sum_{\shat} \What^{*}(\shat | \shatL, x) = 1$. Since
$\Ttilde_1(\bhat)$ is non-negative, cf.~\eqref{eq:ttilde1}, non-negativity of
the pmf $\What^{*}(\shat | \shatL, x)$ is automatically satisfied. The optimum
setting of $\What^{*}(y | \bhat)$ is similarly found.

\section{Proof of Lemma~\ref{lemma:properties:diff}}
\label{app:diff:property}

For ease of reference, let us repeat here the relevant surrogate function:
\begin{align}
    \Psidiff^{(N)}(\Wtilde, \What \condcs)
    &\defeq
       \Iuldiff^{(N)}(\What \condcs)
       +
       \frac{1}{2N}
       \sumx
       Q(\vx)
       \sumy
         W(\vy|\vx \condcs) \,
         D_{\vbhat}
           \left(
               \Ptilde(\vbhat | \vx, \vy \condcs)
             \left\lVert
               \Phat(\vbhat | \vx, \vy \condcs)
             \right.
           \right).
             \label{eq:app:surrogate:diff:2}
\end{align}

\subsection{Property 1}

This follows from standard KL divergence properties.

\subsection{Property 2}
Using the definition~\eqref{eq:information:rate:diff:1} of difference
function, the surrogate function in \eqref{eq:app:surrogate:diff:2} is written
as
\begin{align}\nonumber
    \Psidiff^{(N)}(\Wtilde,\What \condcs)
    &= \frac{1}{2N}
       \sumxy
         Q(\vx \condvs)
           W(\vy | \vx \condcs)
           \log
             \left(
               W(\vy|\vx \condcs)
             \right)\\\nonumber
       &\quad\
        - \frac{1}{2N}
          \sumxy
            Q(\vx \condvs)
            W(\vy | \vx \condcs)
            \sumbhat
              \Ptilde(\vbhat|\vx,\vy \condcs)
              \log
                \left(
                  \What(\vy|\vx \condcs)\right)
       \\ \nonumber
       &\quad\
        + \frac{1}{2N}
          \sumxy
            Q(\vx \condvs)
            W(\vy | \vx \condcs)
            \sumbhat
              \Ptilde(\vbhat|\vx,\vy \condcs)
                \log
                  \left(
                    \Ptilde(\vbhat|\vx,\vy \condcs)
                  \right) \\
       &\quad\
        - \frac{1}{2N}
          \sumxy
            Q(\vx \condvs)
            W(\vy | \vx \condcs)
            \sumbhat
              \Ptilde(\vbhat|\vx,\vy \condcs)
                \log
                  \left(
                    \Phat(\vbhat|\vx,\vy \condcs)
                  \right).
\end{align}
The first and third terms on the right-hand side are only functions of $Q$,
$W$, and $\Wtilde$ and are independent of $\What$. Therefore, by combining
them as $c^{(N)}_{\Delta}(\Wtilde \condcs)$, we can write
\begin{align}\nonumber
  \Psidiff^{(N)}(\Wtilde,\What \condcs)
    &\defeq
       c^{(N)}_{\Delta}(\Wtilde \condcs)
       -
       \frac{1}{2N}
       \sumxy
         Q(\vx \condvs)
         W(\vy | \vx \condcs)
         \sumbhat
           \Ptilde(\vbhat|\vx,\vy \condcs)
             \log
               \left(
                 \Phat(\vbhat,\vy|\vx \condcs)\right).
\end{align}
Now using the AF-FSMC's decomposition property~\eqref{eq:aux:FSMC:decompose:2}
for $\What(\vbhat,\vy |\vx \condcs)$, we can simplify the surrogate function
further to
\begin{align}\nonumber
  \Psidiff^{(N)}(\Wtilde,\What \condcs)
    &= c^{(N)}_{\Delta}(\Wtilde \condcs) \\
    &\quad\
       -
       \frac{1}{2N}
       \sumxy
         Q(\vx \condvs)
         W(\vy | \vx \condcs)
           \sumbhat
             \Ptilde(\vbhat|\vx,\vy \condcs)
         \ellsum\log
           \left(
             \What(\shat_{\ell} | \shat_{\ell-1}, x_{\ell})
           \right)\\\nonumber
     &\quad\
       -
       \frac{1}{2N}
       \sumxy
         Q(\vx \condvs)
         W(\vy | \vx \condcs)
           \sumbhat
            \Ptilde(\vbhat|\vx,\vy \condcs)
         \ellsum\log
           \left(
             \What(y_{\ell} | \bhat_{\ell})
           \right).
\end{align}
After a few manipulations, we obtain
\begin{align}
  \Psidiff^{(N)}(\Wtilde,\What \condcs)
    &= c^{(N)}_{\Delta}(\Wtilde \condcs)
       -
       \sum_{\bhat}
         \log
           \left(
             \What(\shat | \shatL, x)
           \right)\Ttilde^{(N)}_3(\bhat)
       -
       \sum_{\bhat}
         \sum_{y}
           \log
             \left(
               \What(y | \bhat)
             \right)
           \Ttilde^{(N)}_4(\bhat,y).
             \label{eq:app:diff:surrogate:3}
\end{align}
where $\Ttilde^{(N)}_3(\bhat)$ and $\Ttilde^{(N)}_4(\bhat,y)$ were defined
in~\eqref{eq:ttilde3} and~\eqref{eq:ttilde4}, respectively.

\subsection{Property 3}

Similarly to the proof of Property 3 of Lemma~\ref{lemma:properties:upper},
the convexity of $\Psidiff^{(N)}(\Wtilde, \What \condcs)$ w.r.t.\ $W$ is
established by looking at the corresponding Hessian matrix and by verifying
that it is positive semi-definite. We leave the details to the reader.

\section{Proof of
  Lemma~\ref{lemma:properties:lower:surrogate:function:part:1}}
\label{app:lower:property:part:1}

For ease of reference, let us repeat here the relevant surrogate function:
\begin{align}
  \label{eq:information:rate:lower:surrogate:function:part:1:app}
  \Psilower^{(N)}_1(\vtilde, \vhat)
    &\defeq
       \Ilower^{(N)}_1(\vhat \condcs)
       -
       \frac{1}{2N}
       \sumxy
         Q(\vx)
         W(\vy|\vx)
         D_{\vshat}
           \left(
               \frac{\vtilde(\vshat, \vx, \vy)
                    }
                    {\sum_{\vshat'}
                       \vtilde(\vshat', \vx, \vy)
                    }
           \ \left\lVert \
               \frac{\vhat(\vshat, \vx, \vy)
                    }
                    {\sum_{\vshat'}
                       \vhat(\vshat', \vx, \vy)
                    }
           \right.
           \right).
\end{align}

\subsection{Property 1}

This follows from standard KL divergence properties.

\subsection{Property 2}

Inserting the definition~\eqref{eq:information:rate:lower:part:1} of
$\Ilower^{(N)}_1(\What)$
into~\eqref{eq:information:rate:lower:surrogate:function:part:1:app}, we
obtain
\begin{align}
  \Psilower^{(N)}_1(\vhat \condcs)
    &= \underline{c}^{(N)}_1(\vtilde)
       +
       \frac{1}{2N}
       \sumxy
         Q(\vx)
         W(\vy|\vx)
         \sum_{\vshat}
           \frac{\vtilde(\vshat, \vx, \vy)
                }
                {\sum_{\vshat'}
                   \vtilde(\vshat', \vx, \vy)
                }
           \log
             \big(
               \vhat(\vshat, \vx, \vy)
             \big),
\end{align}
where $\underline{c}^{(N)}_1(\vtilde)$ is independent of $\vhat$.  Applying
the product decomposition~\eqref{eq:backward:auxiliary:channel:law:1:part:2}
of $\vhat(\vshat, \vx, \vy)$, and the
relationship~\eqref{eq:capital:v:small:v:1}, \ie,
\begin{align}
  \frac{\vtilde(\vshat, \vx, \vy)}
       {\sum_{\vshat'}
          \vtilde(\vshat', \vx, \vy)
       }
   = \Vtilde(\vshat | \vx, \vy),
\end{align}
we get, after some simplifying steps,
\begin{align}
  \Psilower^{(N)}_1(\vtilde, \vhat)
    &= \underline{c}^{(N)}_1(\vtilde)
       +
       \sum_{\bhat}
         \sum_{\vyD}
           \log
             \left(
               \vhat(\bhat,\vyD)
             \right)
           \Ttilde^{(N)}_4(\bhat,\vyD).
\end{align}

\subsection{Property 3}

Similarly to the proof of Property 3 of Lemma~\ref{lemma:properties:upper},
the concavity of $\Psilower^{(N)}_1(\vtilde, \vhat \condcs)$ w.r.t.~$\vhat$ is
established by looking at the corresponding Hessian matrix and by verifying
that it is negative semi-definite. We leave the details to the reader.

\section{Proof of Lemma~\ref{lemma:properties:lower:surrogate:function:part:2}}
\label{app:lower:property:part:2}

For ease of reference, let us repeat here the relevant part of the information
rate lower bound and the corresponding surrogate function:
\begin{align}
  \Ilower^{(N)}_2(\vhat \condcs)
    &\defeq
       -
       \frac{1}{2N}
       \sumy
         (QW)(\vy)
         \log
           \left(
             \sum_{\vshat', \vx'}
                     \vhat(\vshat', \vx', \vy)
           \right),
    \label{eq:information:rate:lower:part:2:app} \\
  \Psilower^{(N)}_2(\vtilde, \vhat)
    &\defeq
       \underline{c}^{(N)}_2(\vtilde)
       -
       \sum_{\bhat}
         \sum_{\vyD}
           \frac{1}{\gamma}
            \left(
              \frac{\vhat(\bhat,\vyD)}
                   {\vtilde(\bhat,\vyD)}
            \right)^{\gamma}
            \Ttilde^{(N)}_2(\bhat,\vyD).
  \label{eq:information:rate:lower:surrogate:function:part:2:app}
\end{align}

\subsection{Property 1}

This statement follows trivially from the definition of
$\underline{c}_2(\vtilde)$.

\subsection{Property 2}

On one hand, taking derivatives of $\Psilower^{(N)}_2(\vtilde, \vhat)$
w.r.t.~$\vhat(\bhat,\vyD)$ we obtain
\begin{align}
  \left.
    \diff{\vhat(\bhat, \vyD)}
      \Psilower^{(N)}_2(\vhat)
  \right|_{\vhat = \vtilde}
    &= \left.
         -
         \frac{\gamma}
              {\gamma}
         \cdot
         \frac{\vhat(\bhat,\vyD)^{\gamma-1}}
              {\vtilde(\bhat,\vyD)^{\gamma}}
         \cdot
         \Ttilde^{(N)}_2(\bhat,\vyD)
       \right|_{\vhat = \vtilde}
     = -
       \frac{1}
            {\vtilde(\bhat,\vyD)}
       \cdot
       \Ttilde^{(N)}_2(\bhat,\vyD).
         \label{app:lower:property:part:2:gradient:1}
\end{align}
On the other hand, $\Ilower^{(N)}_2(\vhat)$ can be expanded to the
following expression
\begin{align}
  \Ilower^{(N)}_2(\vhat)
    &= \underline{c}^{(N)}_2(\vtilde)
       -
       \frac{1}{2N}
       \sumy
         (QW)(\vy)
         \sum_{\vshat, \vx}
           \frac{\vtilde(\vshat, \vx, \vy)
                }
                {\sum_{\vshat', \vx'}
                   \vtilde(\vshat', \vx', \vy)
                }
           \log
             \left(
               \vhat(\vshat, \vx, \vy)
             \right) \nonumber \\
     &\quad\
       -
       \frac{1}{2N}
       \sumy
         (QW)(\vy)
         D_{\vshat, \vx}
           \left(
               \frac{\vtilde(\vshat, \vx, \vy)
                    }
                    {\sum_{\vshat', \vx'}
                       \vtilde(\vshat', \vx', \vy)
                    }
           \ \left\lVert \
               \frac{\vhat(\vshat, \vx, \vy)
                    }
                    {\sum_{\vshat', \vx'}
                       \vhat(\vshat', \vx', \vy)
                    }
           \right.
           \right),
             \label{eq:information:rate:lower:part:2:app:mod}
\end{align}
where $\underline{c}^{(N)}_2(\vtilde)$ is independent of $\vhat$. At first
sight, expression~\eqref{eq:information:rate:lower:part:2:app:mod} looks more
complicated than expression~\eqref{eq:information:rate:lower:part:2:app},
however, the relevant gradient of
expression~\eqref{eq:information:rate:lower:part:2:app:mod} is easier to find.
Namely, the gradient of the first term with respect to $\vhat$ is the zero
vector and so we do not have to worry about it. Similarly, the third term is a
KL divergence and so the gradient with respect to $\vhat$ at $\vhat = \vtilde$
is also the zero vector. Therefore, we only need to have a close look at the
second term. Applying the product
decomposition~\eqref{eq:backward:auxiliary:channel:law:1:part:2} of
$\vhat(\vshat, \vx, \vy)$, and
using~\eqref{eq:backward:auxiliary:channel:law:1}, \ie,
\begin{align}
  \frac{\vtilde(\vshat, \vx, \vy)}
       {\sum_{\vshat', \vx'}
          \vtilde(\vshat', \vx', \vy)
       }
  &= \Vtilde(\vshat, \vx | \vy),
\end{align}
we get
\begin{align}
  \left.
    \diff{\vhat(\bhat, \vyD)}
      \Ilower^{(N)}_2(\vhat)
  \right|_{\vhat = \vtilde}
    &= \left.
         -
         \diff{\vhat(\bhat, \vyD)}
           \sum_{\bhat'}
             \sum_{(\vyD)'}
               \log
                 \left(
                   \vhat(\bhat',(\vyD)')
                 \right)
               \Ttilde^{(N)}_2(\bhat',(\vyD)')
       \right|_{\vhat = \vtilde} \\
    &= \left.
         -
         \frac{1}{\vhat(\bhat,\vyD)}
         \cdot
         \Ttilde^{(N)}_2(\bhat,\vyD)
       \right|_{\vhat = \vtilde}
     = -
       \frac{1}{\vtilde(\bhat,\vyD)}
       \cdot
       \Ttilde^{(N)}_2(\bhat,\vyD),
         \label{app:lower:property:part:2:gradient:2}
\end{align}
which indeed agrees with the expression
in~\eqref{app:lower:property:part:2:gradient:1}.

\subsection{Property 3}

Similarly to the proof of Property 3 of Lemma~\ref{lemma:properties:upper},
the concavity of $\Psilower^{(N)}_2(\vtilde, \vhat)$ w.r.t.~$\vhat$ is
established by looking at the corresponding Hessian matrix and by verifying
that it is negative semi-definite. We leave the details to the reader.

%% file: acknowledgments1.tex

\section*{Acknowledgments}

The work of Parastoo Sadeghi was partly supported under Australian Research
Council's Discovery Projects funding scheme (project number DP0773898). The
authors acknowledge correspondence with Fredrik Rusek which motivated them to
generate \figref{fig:ch11:ch6}. They are also grateful to Henry Pfister for
mentioning to them the papers by Mevel et al.